\documentclass[conference]{IEEEtran}
\IEEEoverridecommandlockouts

\usepackage{amsmath,amssymb,amsfonts}
\usepackage{algorithmic}
\usepackage[T1]{fontenc}

\usepackage{textcomp}
\usepackage{refcount}

\usepackage{xspace}
\usepackage{graphicx}
\usepackage{color}
\usepackage[sort,compress]{cite}
\usepackage{subcaption}
\definecolor{accessbl}{cmyk}{1, 0.4, 0.3, 0}
\DeclareCaptionFont{bl}{\color{accessbl}}
\usepackage{hyperref}
\usepackage[normalem]{ulem}

\usepackage{epsfig,endnotes,paralist,multirow}
\usepackage{wrapfig}
\usepackage{epsfig}
\usepackage[normalem]{ulem}
\usepackage{url}
\usepackage{soul}
\usepackage{multirow}
\usepackage{array}
\usepackage{bbm}
\usepackage{makecell}

\def\includeComments{include}
\def\includ{include}
\def\comm[#1]{\ifx\includeComments\includ  \bl \texttt{\textbf{\textit{Note: #1}}} \par \fi}
\def\inlinecomm[#1]{\ifx\includeComments\includ  \textit{Note: #1} \fi}

\newcommand{\blue}[1]{\textcolor{blue}{#1}} %

\newcommand{\para}[1]{\smallskip \noindent {\bf #1}}
\newcommand{\softpara}[1]{\smallskip \noindent \underline{#1}}
\newcommand{\vsp}{\vspace{0.00in}}

\def\blackbox{\hfill {\vrule height6pt width6pt depth0pt}}

\newcounter{theorem}
\newtheorem{thm}{Theorem}

\newtheorem{corollary}{Corollary}

\newtheorem{defin}{Definition}
\newtheorem{ex}{Example}
\newtheorem{lem}{Lemma}
\newtheorem{ob}{Observation}

\newenvironment{theorem}{\vsp \begin{thm} \nopagebreak}{{\hfill$\blackbox$} \end{thm} \vsp}
\newenvironment{thm-prf}{\begin{thm} \nopagebreak}{\end{thm}}

\newenvironment{lem-wo-prf-box}{\vsp \begin{lem} \nopagebreak}{\end{lem}}
\newenvironment{lem-prf}{\vsp \begin{lem} \nopagebreak}{\end{lem}}
\newenvironment{prf}{{\em Proof:} \nopagebreak }{{\hfill$\blackbox$}}

\newenvironment{cor-prf}{\vsp \begin{corollary} \nopagebreak}{\end{corollary}}

\newcounter{packednmbr}

\newcommand{\id}[1]{\mbox{\em #1\xspace}}

\newcommand{\eps}{\mbox{EP}\xspace}
\newcommand{\epss}{\mbox{EPs}\xspace}
\newcommand{\es}{\mbox{ES}\xspace}

\newcommand{\floor}[1]{\ensuremath{\lfloor #1 \rfloor}}

\newcommand{\eat}[1]{}

\let\emph\textit

\newcommand{\php}{\mbox{$p_{ob}$}\xspace}          %
\newcommand{\bt}{\mbox{$\tau_{b}$}\xspace}         %
\newcommand{\ft}{\mbox{$\tau_{f}$}\xspace}         %
\newcommand{\pt}{\mbox{$\tau_{p}$}\xspace}         %
\newcommand{\bp}{\mbox{$p_{b}$}\xspace}          %
\newcommand{\fp}{\mbox{$p_{f}$}\xspace}          %
\newcommand{\gt}{\mbox{$\tau_g$}\xspace}      %
\newcommand{\gp}{\mbox{$p_g$}\xspace}       %
\newcommand{\ct}{\mbox{$\tau_c$}\xspace}      %
\newcommand{\fth}{\mbox{$f^\tau$}\xspace}      %
\newcommand{\fthi}{\mbox{$f_i^\tau$}\xspace}      %

\newcommand{\swaplatency}{\ensuremath{L_{\mbox{\scriptsize\it swap}}}}

\newcommand{\swapfidelity}{\ensuremath{F_{\mbox{\scriptsize\it swap}}}}
\newcommand{\fusionfidelity}{\ensuremath{F_{\mbox{\scriptsize\it fuse}}}}
\newcommand{\purifyfidelity}{\ensuremath{F_{\mbox{\scriptsize\it pur}}}}

\newcommand{\purifyprob}{\ensuremath{p_{\mbox{\scriptsize\it pur}}}}
\newcommand{\iterpurifyfidelity}[1]{\ensuremath{F_{\mbox{\scriptsize\it pur}, #1}}}
\newcommand{\iterpurifyprob}[1]{\ensuremath{p_{\mbox{\scriptsize\it pur},#1}}}
\newcommand{\discretefid}{\ensuremath{\mathcal{F}}}
\newcommand{\discreterate}{\ensuremath{\mathcal{K}}}

\newcommand{\bra}[1]{\ensuremath{\langle#1|}}
\newcommand{\ket}[1]{\ensuremath{|#1\rangle}}

\newcommand{\fed}{\mbox{\tt EDP}\xspace}
\newcommand{\gfed}{\mbox{\tt GEDP}\xspace}
\newcommand{\gdp}{\mbox{\tt GDP}\xspace}
\newcommand{\ggdp}{\mbox{\tt GGDP}\xspace}
\newcommand{\graphdp}{\mbox{\tt GSDP}\xspace}
\newcommand{\graphlp}{\mbox{\tt GGSDP}\xspace}
\newcommand{\LP}{\mbox{\tt LP}\xspace}
\newcommand{\LPNaive}{\mbox{\tt LP-Naive}\xspace}
\newcommand{\LPin}{\mbox{\tt LP-Int}\xspace}
\newcommand{\LPts}{\mbox{\tt LP-2S}\xspace}
\newcommand{\LPms}{\mbox{\tt LP-MS}\xspace}
\newcommand{\Greedy}{\mbox{\tt Greedy}\xspace}
\newcommand{\ee}{\mbox{\tt E2E}\xspace}
\newcommand{\DP}{\mbox{\tt DP}\xspace}
\newcommand{\DPIterative}{\mbox{\tt DP-Iterative}\xspace}
\newcommand{\DPss}{\mbox{\tt DP-R-SP}\xspace}
\newcommand{\DPOT}{\mbox{\tt DP-R-BP}\xspace}
\newcommand{\DPtwoG}{\mbox{\tt DP-R-EP}\xspace}
\newcommand{\starset}{\ensuremath{S}\xspace}
\newcommand{\stargraph}{\ensuremath{\mathcal{R}}\xspace}
\newcommand{\spvalue}{\ensuremath{K_{\infty}}\xspace}

\newcommand{\GHZDPFSS}{\mbox{\tt DP-F-SP}\xspace}
\newcommand{\GHZDPFBP}{\mbox{\tt DP-F-BP}\xspace}
\newcommand{\GHZDPFtwoG}{\mbox{\tt DP-F-EP}\xspace}

\newcommand{\GHZLPOpt}{\ensuremath{\mathtt{LP}^{*}}\xspace}

\newcommand{\GHZLPVIn}{\mbox{\tt LP-V-Int}\xspace}
\newcommand{\GHZLPVTs}{\mbox{\tt LP-V-2S}\xspace}
\newcommand{\GHZLPVMs}{\mbox{\tt LP-V-MS}\xspace}

\newcommand{\GHZLPTwoF}{\mbox{\tt LP-2F}\xspace}
\newcommand{\GHZLPTwoFIn}{\mbox{\tt LP-2F-Int}\xspace}
\newcommand{\GHZLPTwoFTs}{\mbox{\tt LP-2F-2S}\xspace}
\newcommand{\GHZLPTwoFMs}{\mbox{\tt LP-2F-MS}\xspace}

\newcommand{\GraphDPV}{\mbox{\tt DP-R}\xspace}
\newcommand{\GraphDPF}{\mbox{\tt DP-F}\xspace}

\newcommand{\GraphLPOpt}{\ensuremath{\mathtt{LP}^{*}}\xspace}
\newcommand{\GraphLPV}{\mbox{\tt LP-V}\xspace}
\newcommand{\GraphLPF}{\mbox{\tt LP-F}\xspace}

\newcommand{\mpfont}{\scriptsize}

\ifx\noeditingmarks\undefined
    \newcommand{\MPworker}[2]{{\color{#1}\vrule\vrule}{\marginpar{\color{#1}\mpfont #2}}}
\else
    \newcommand{\MPworker}[2]{}
\fi

\def\BibTeX{{\rm B\kern-.05em{\sc i\kern-.025em b}\kern-.08em
    T\kern-.1667em\lower.7ex\hbox{E}\kern-.125emX}}

\begin{document}

\title{Near-Optimal Techniques for Distributed Entanglement Generation and Purification \vspace{-0.1in}}

\author{
\IEEEauthorblockN{Yukun Yang, Xiaojie Fan, C.R. Ramakrishnan, Himanshu Gupta}
\IEEEauthorblockA{
\textit{Department of Computer Science, Stony Brook University, Stony Brook, NY 11790, USA}\\
Corresponding author: Himanshu Gupta (\texttt{hgupta@cs.stonybrook.edu})
}
}

\tableofcontents
\clearpage
\maketitle
\thispagestyle{plain}
\pagestyle{plain}
\begin{abstract}
We address the problem of efficiently distributing high-fidelity entangled states across quantum-network nodes, a fundamental requirement for quantum communication, distributed quantum computing, and emerging quantum internet applications. We consider repeater-based architectures that combine entanglement swapping, fusion, or graph-state growth with purification operations, which improve fidelity by consuming additional entangled resources. 
Prior work has studied routing and purification strategies, often with restrictions on purification placement or the generation structure. We develop a purification-aware optimization framework that jointly selects construction and purification operations to maximize generation rate under a fidelity threshold. The framework can incorporate different purification models through their fidelity-update, success-probability, and latency rules.

Our framework introduces purification-augmented generation trees and develops dynamic programming (DP) algorithms for tree-based entangled-pair (EP) distribution under fidelity constraints. We further develop hypergraph-flow linear programming (LP) formulations for level-based structures, enabling concurrent generation across multiple trees and demands. 
For multipartite entanglement, we extend the framework to GHZ states and arbitrary graph states, constructing them from remote EP resources.
We introduce graph-state construction primitives and derive closed-form fidelity-vector updates for GHZ fusion and purification, as well as for graph-state growth and purification.
We provide optimal formulations under the adopted models, discretization, and imposed purification bounds, together with computationally efficient DP and LP heuristics. We also bound fidelity- and rate-discretization errors and establish convergence for fixed constructions under stated regularity conditions.

NetSquid simulations for EP generation (using our LP approach) show improvements over the compared prior approaches, with an average 160\% increase in generation rate and gains of up to 330\% in some scenarios.  For GHZ and graph states, the simulations demonstrate the effectiveness of our heuristics in balancing generation rate, output fidelity, and computational overhead.

\end{abstract}

\section{\bf Introduction}
\label{sec:intro}

Quantum networks are a foundational component of the envisioned quantum internet, enabling long-distance quantum communication, distributed quantum computation, and secure multi-party protocols. At the core of these capabilities is reliably distributing high-fidelity entangled states between distant nodes. Applications such as quantum key distribution (QKD)~\cite{qkd,qkd-2,qkd_pirandola2020}, entanglement-assisted distributed computation~\cite{caleffi2024distributed,sundaram2024distributing,cirac1999distributed,sundaram2023distributing,sundaram2022distribution,g2021efficient,main2025distributed}, and distributed quantum sensing~\cite{zhan2024optimizing,eldredge2018optimal,hillery2023discrete} all rely on the ability to generate, distribute, and manage entangled resources under realistic noise and decoherence conditions.

As quantum repeaters and entanglement-swapping architectures evolve from theoretical proposals to prototype deployments, a central challenge is to efficiently distribute entanglement while maintaining fidelity above application-specific thresholds. Low-fidelity entangled pairs can introduce errors in teleportation-based communication and computation, and can reduce the security of QKD protocols. Thus, high-fidelity entanglement is not merely a quality metric, but an operational requirement for useful quantum-network services.

Because purification improves fidelity by consuming additional entangled resources, efficient distribution requires jointly choosing routes, swapping or fusion structures, and purification steps to maximize the generation rate under a fidelity threshold~\cite{swapping-tqe-22}. We therefore develop optimal\footnote{Optimality is relative to the stated operational and analytical models, discretization, and any imposed purification-iteration bound; see \S\ref{sec:optimality-scope} and Appendix~\ref{app:disc-gap}.} dynamic programming (DP) and linear programming (LP) formulations, together with computationally efficient heuristics, for fidelity-constrained distribution of EPs, GHZ states, and general graph states.

\para{Prior Work and Our Approach.}
Prior work studies routing and operation ordering~\cite{zhao2023segmented,swapping-tqe-22}, or incorporates purification under restrictions on its placement or the generation structure~\cite{zhao2022e2e,nickerson2013topological,de2020protocols}; \S\ref{sec:related} provides a detailed comparison.
In contrast, our framework incorporates purification directly into the entanglement-generation structure. We develop DP algorithms for tree-based generation and hypergraph-flow LP formulations for concurrent generation. For GHZ and graph states, we use the corresponding fidelity vectors to track the error distribution through construction and purification.

\begin{table*}[t]
\centering
\scriptsize
\setlength{\tabcolsep}{5pt}
\begin{tabular}{|l|l|l|l|}
\hline
\textbf{Target state} & \textbf{Methods} & \textbf{Main contribution / guarantee} & \textbf{Empirical Takeaway (see \S\ref{sec:eval})} \\
\hline

\makecell[cl]{EPs} &
\makecell[cl]{\textbullet\enspace Tree-based DP;\\
\textbullet\enspace Level-based Hypergraph-flow LP} &
\makecell[cl]{\textbullet\enspace Optimal DP for purification-augmented swapping trees;\\
\textbullet\enspace Optimal LP for level-based swapping structures;\\
\textbullet\enspace Supports different purification models.}
&
\makecell[cl]{\textbullet\enspace LP generally achieves the highest generation rate;\\
\textbullet\enspace DP gives a better time/memory trade-off.} \\
\hline

\makecell[cl]{GHZ states} &
\makecell[cl]{\textbullet\enspace Full second-stage DP and LP;\\
\textbullet\enspace Rate- and fidelity-discretized\\
\phantom{\textbullet}\enspace DP heuristics;\\
\textbullet\enspace LP heuristics with vector or\\
\phantom{\textbullet}\enspace two-scalar fidelity labels;\\
\textbullet\enspace Greedy baseline} &
\makecell[cl]{\textbullet\enspace Closed-form GHZ fusion/purification fidelity models;\\
\textbullet\enspace Optimal second-stage DP for tree-based structures;\\
\textbullet\enspace Optimal second-stage LP for level-based structures;\\
\textbullet\enspace Scalable reduced/restricted variants.}
&
\makecell[cl]{\textbullet\enspace Rate-discretized DP offers the best\\
\phantom{\textbullet}\enspace overall rate--fidelity trade-off;\\
\textbullet\enspace LP with vector fidelity labels outperforms\\
\phantom{\textbullet}\enspace LP with two-scalar fidelity labels and\\
\phantom{\textbullet}\enspace generally plans faster than rate-discretized DP,\\
\phantom{\textbullet}\enspace but meets the threshold less consistently.} \\
\hline

\makecell[cl]{General\\
graph states} &
\makecell[cl]{\textbullet\enspace Full second-stage DP and LP;\\
\textbullet\enspace Rate- and fidelity-discretized\\
\phantom{\textbullet}\enspace DP heuristics;\\
\textbullet\enspace LP heuristics with vector or\\
\phantom{\textbullet}\enspace scalar fidelity labels;\\
\textbullet\enspace Greedy baseline} &
\makecell[cl]{\textbullet\enspace Target-edge construction framework using\\
\phantom{\textbullet}\enspace graph-growth primitives and purification;\\
\textbullet\enspace Optimal DP and LP formulations for tree-based\\
\phantom{\textbullet}\enspace and level-based graph-state construction;\\
\textbullet\enspace Scalable reduced/restricted variants.} &
\makecell[cl]{\textbullet\enspace Fidelity-discretized DP offers the best\\
\phantom{\textbullet}\enspace overall rate--fidelity trade-off;\\
\textbullet\enspace LP with vector fidelity labels outperforms\\
\phantom{\textbullet}\enspace LP with scalar fidelity labels, but its fidelity\\
\phantom{\textbullet}\enspace can fall below the threshold.}\\
\hline
\end{tabular}
\caption{Summary of the main results for fidelity-constrained distribution of bipartite and multipartite entangled states. Optimality statements are in the sense described in \S\ref{sec:optimality-scope}.}
\label{tab:summary-results}
\end{table*}

\para{Our Contributions.}
Our main contributions are as follows.
\begin{enumerate}
\item
We develop optimal DP and LP formulations that jointly optimize construction and purification for EPs and second-stage GHZ and graph-state generation. They optimize purification-augmented trees and level-based structures, respectively, under the stated models (\S\ref{sec:ep}, \S\ref{sec:ghz}, and \S\ref{sec:gen-graphs}).

\item
We derive and incorporate full-vector fidelity updates for GHZ fusion and purification and for graph-state growth and purification, tracking the error distribution needed to compute purification success probabilities and output fidelities (\S\ref{sec:model} and \S\ref{sec:gen-graphs}).

\item
For GHZ and graph states, we develop computationally efficient DP heuristics using rate or fidelity discretization and LP heuristics with restricted search spaces. We also develop LP variants with scalar fidelity labels for comparison (\S\ref{sec:ghz} and \S\ref{sec:gen-graphs}).

\item
We bound fidelity and rate errors and establish convergence for fixed constructions under either rate or fidelity discretization, subject to the stated regularity conditions. We also analyze the effect of bounded purification iterations (\S\ref{sec:optimality-scope} and Appendix~\ref{app:disc-gap}).

\item
Our NetSquid experiments~\cite{netsquid2020} favor LP for EP generation rate and DP for scalability. For GHZ states, results favor rate-discretized DP; for graph states, they favor fidelity-discretized DP. LP with vector fidelity labels generally plans faster than rate-discretized DP but meets the fidelity threshold less consistently (\S\ref{sec:eval}).
\end{enumerate}

Table~\ref{tab:summary-results} summarizes the paper's main technical results.

\section{\bf Background and Models}
\label{sec:background}

\para{Quantum Network (QN).} 
A quantum network is a connected undirected graph with vertices representing  
quantum computers and edges representing the (quantum and classical) direct 
communication links. For a quantum network $Q$, we use \emph{network nodes} (denoted by $V(Q)$) and \emph{links} (denoted by $E(Q)$ to refer
to its vertices and edges. Distribution of \epss and multipartite entangled states such as GHZ and graph states is generally seen as a basic \emph{network service} based on which other quantum operations can be performed over the network~\cite{manzalini2020quantum,sundaram2023distributing,mazza2024simulation,sundaram2022distribution,hillery2023discrete}. For instance, \epss can be used to transmit arbitrary quantum data via teleportation, and implement \emph{remote gates} (operating on qubits from different quantum computers in the network) for executing distributed quantum programs.

\para{Quantum Network Model.}  Following past work on entanglement distribution such as~\cite{caleffi,swapping-tqe-22}, we assume each node in the network has quantum memories (possibly implemented as atomic qubits) partitioned into a collection of non-empty subsets. Each edge incident on a node is assigned a unique subset of such memories, and represents a quantum link capable of establishing an \eps across the two sets of memories connected by that link.  An \eps established across a network link is called a ``\emph{link \eps}''. The technical development of this paper is agnostic to the physical processes used to generate link \epss. We consider the link \eps generation to be a Poisson process. The network model provides two functions: $\id{rate}: E(Q)\rightarrow R^+$, which gives the mean generation rate, and $\id{fid}: E(Q) \rightarrow R^+$, which gives the fidelity of link \epss.  Finally, we assume that pairs of network nodes can directly communicate classical values. 

\begin{figure}
    \centering
    \includegraphics[width=0.5\textwidth]{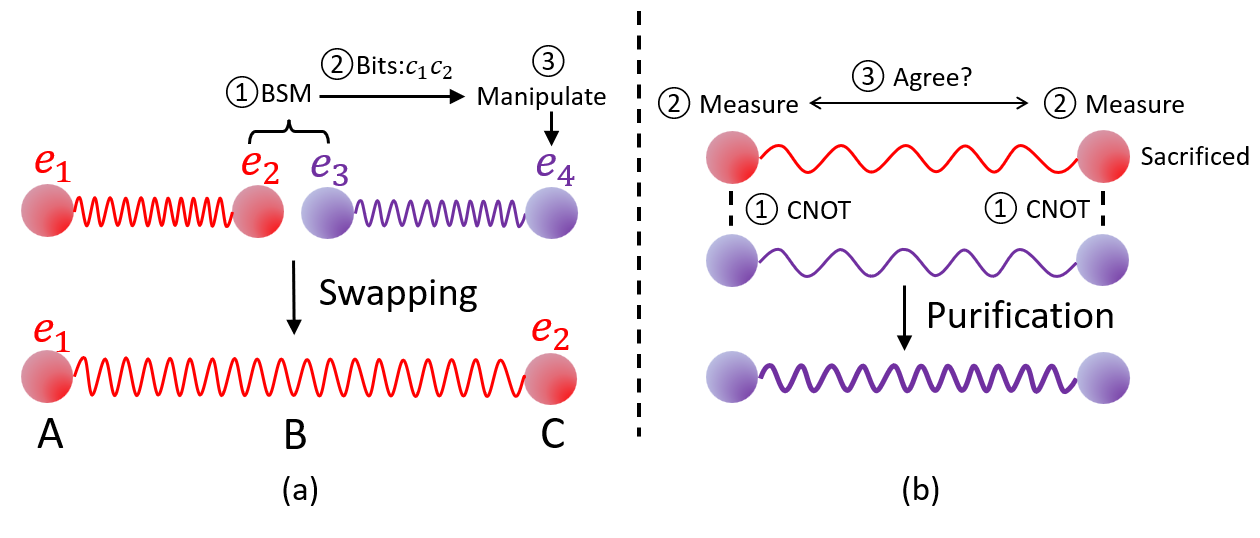}
    \caption{(a) Entanglement swapping over the triplet of nodes $(A, B, C)$, which results in $A$'s qubit entangled with $C$'s qubit. (b) Purification of entanglement (blue, middle) using sacrificial EP (red, top), resulting in a higher-fidelity EP (blue, bottom) if the operation is successful.}
  \vspace{-0.2in}
  \label{fig:teleport_swap}
\end{figure}

\para{Fidelity.}
Although the target \epss{} and GHZ states are pure, imperfections in their generation and distribution can produce a noisy, generally mixed state. Throughout this work, we quantify the fidelity of a realized mixed state $\xi$ with respect to a target pure state $\ket{\psi}$ by its overlap with the target state~\cite{nielsen2001quantum,bennett-96},
$F(\xi) = \bra{\psi}\xi\ket{\psi}$,
where the target state $\ket{\psi}$ is clear from context.

\subsection{\bf Generation of 
Remote \epss}
\label{sec:remote-ep-generation}
Given two link \epss shared between nodes $(x_i, x_j)$ and $(x_j, x_k)$, we can create a remote \eps between $(x_i, x_k)$ by performing an \emph{entanglement swapping} operation (\es) between the two qubits at $x_j$.  The \es operation involves a Bell State Measurement (BSM) at $x_j$ followed by a local operation at $x_i$ or $x_k$ (see Fig.~\ref{fig:teleport_swap}(a)). BSM is generally modeled as a \emph{probabilistic} operation, with a success probability of $\fp$.

\para{Entanglement Swapping Trees.} 
 An efficient way to generate an \eps over a pair of remote network nodes $(s, d)$ using \epss over network links (i.e., edges) is to: 
(i) create a path $P$ in the network graph from $s$ to $d$  with \epss over each of the path's edges, and 
(ii) perform a series of ES over these \epss. 
The \es operations over $P$ can be performed in any order. 
The order can be represented by an 
\emph{entanglement-swapping tree}~\cite{swapping-tqe-22}.  Formally, an entanglement-swapping tree is a complete binary tree consisting of two kinds of nodes:
\begin{itemize}
    \item leaves of the form $\id{link}(x_i,x_j)$ representing link \epss, 
    ($(x_i, x_j) \in E(Q)$ ) and
    \item internal nodes of the form $\id{swap}(x_i, x_j, x_k)$ with $x_i, x_j, x_k \in V(Q)$ representing \epss $(x_i, x_j)$ generated by \es between $(x_i, x_k)$ and  $(x_k, x_j)$ at network node $x_k$. The two children of such a node will be entanglement swapping trees, each rooted at $\id{link}(x_i, x_k)$ or $\id{swap}(x_i, x_k, \_)$, and
    $\id{link}(x_k, x_j)$ or $\id{swap}(x_k, x_j, \_)$. 
\end{itemize} 

\para{Generation Latency and Generation Rate.}
We can compute the latency of generating \epss{} from an entanglement-swapping tree recursively, following~\cite{swapping-tqe-22}. Let $t$ be an entanglement-swapping tree with a $\id{swap}$ node at its root, and let its two children be $l$ and $r$. If the mean generation latencies of $l$ and $r$ are denoted by $l_l$ and $l_r$, respectively, then the generation latency of $t$ is estimated as $l_t = \swaplatency(l_l,l_r)$, where
\begin{equation}
\swaplatency(l_l, l_r) = (\tfrac{3}{2} \max(l_l, l_r) + \ft + \ct)/\fp, \label{eqn:EP_swaplatency}
\end{equation}
where \ft is the latency of \es, \fp is the success probability of \es, and \ct is the latency of classical communication. The factor $\tfrac{3}{2}$ is the steady-state synchronization penalty for waiting until both independently generated operands \epss{} are available: for two equal-rate exponential arrival processes, $\mathbb{E}[\max(X_1, X_2)]=\tfrac{3}{2}\mathbb{E}[X_1]$. Following~\cite{swapping-tqe-22}, we apply this waiting approximation using $\max(l_l,l_r)$ as the dominant operand latency; Appendix~\ref{app:waiting-model} gives the derivation, discusses its scope, and describes the analogous factors used later in the paper.

In the base case $t=\id{link}(x_i,x_j)$, the latency is $l_t = \tfrac{1}{\id{rate}(x_i,x_j)}$.

The simplified multiplicative rate equations displayed for the EP LP and the GHZ and graph-state DP/LP formulations omit additive operations and classical-communication latencies. In particular, omitting $\ft+\ct$ in Eqn.~\ref{eqn:EP_swaplatency} gives an output rate of $\tfrac{2}{3}\fp$ times the smaller input rate. The EP DP retains the full latency recurrence above. These operation and classical-communication latencies are included in the NetSquid evaluation. A controlled comparison of communication delays is reported in \S\ref{sec:eval}.

\para{Fidelity.} 
The fidelity of \epss generated by an entanglement-swapping tree can also be recursively computed as follows.  If $t$ has only a leaf node $\id{link}(x_i, x_j)$, then its fidelity is directly taken from that link's fidelity.  Otherwise, let $t$ have a $\id{swap}$ node at its root with children $l$ and $r$. Let the fidelities of \epss generated by $l$ and $r$ be $f_l$ and $f_r$, respectively.  Following~\cite{dur1999quantum}, the fidelity of \epss generated by $t$ is $\swapfidelity(f_l, f_r)$ where: 
\begin{eqnarray}
\swapfidelity(f_l, f_r)= \tfrac{1}{4} (1+\tfrac{1}{3}(4f_{l}-1)(4f_{r}-1)) \label{eqn:EP_swapfidelity}
\end{eqnarray}
The above assumes noiseless swap operations. We use the same ideal-local-operation convention throughout the analytical development: local gates, Bell-state measurements, and measurement/feed-forward operations affect latency and success probability, but are not assigned an additional fidelity-degradation term. Noisy local operations can be incorporated by replacing the fidelity-update maps, such as $\swapfidelity(\cdot,\cdot)$ and the purification maps below, with hardware-calibrated noisy versions as in~\cite{dur1999quantum,gu2024fendi}. The DP and LP formulations depend only on these update maps, so this substitution does not change the optimization framework. We use the ideal-local-operation form to keep the exposition focused on the routing, swapping, and purification decisions.

\subsection{\bf Incorporating Purification into Generation Schemes}
\label{sec:pur-back}

Ensuring the fidelity of \epss and GHZ states is crucial since they are used as the basis for other operations in distributed computing. \emph{Entanglement Purification}~\cite{bennett-96} is obtaining high-fidelity \epss and GHZ states from multiple copies of lower-fidelity states. 

\para{Purification of \eps.}
Let $t$ (`target'') and $s$ (`sacrificial'') be two pairs of qubits with fidelities $F_t$ and $F_s$ with respect to $\ket{\Phi+}$. A purification operation~\cite{bennett-96,deutsch1996quantum,dur2007entanglement} uses $s$ to improve the quality of $t$: on success, $t$ is retained with an updated fidelity; on failure, $t$ is discarded. The sacrificial pair $s$ is discarded regardless of the outcome.

The fidelity of $t$ after a successful purification step, and the probability of that success, are functions of the input fidelities $F_t$ and $F_s$. We denote these functions by $\purifyfidelity(F_t,F_s)$ and $\purifyprob(F_t,F_s)$, respectively. For example, the Bennett \emph{et al.}\ purification protocol~\cite{bennett-96} gives\footnote{These expressions
assume Bell-diagonal input states in which the three
non-target Bell states have equal probabilities.}

\begin{align}
\purifyprob(F_t, F_s) &= F_{t}F_{s}+F_{t}(\tfrac{1-F_{s}}{3})+(\tfrac{1-F_{t}}{3})F_{s}\nonumber\\
& \hspace*{8ex} 
+ 5(\tfrac{1-F_{t}}{3})(\tfrac{1-F_{s}}{3}) \nonumber \\
\purifyfidelity(F_t, F_s) &= \frac{F_{t}F_{s}+(\tfrac{1-F_{t}}{3})(\tfrac{1-F_{s}}{3})}{\purifyprob(F_t, F_s)}
\label{eqn:EP_One_Purify_Fidelity}
\end{align}
\noindent
In the symmetric case $F_t=F_s=f$, this update increases fidelity whenever $f>1/2$. More generally, our algorithms do not depend on this particular protocol: they only require the corresponding fidelity-update and success-probability maps.
More generally, the optimization framework can accommodate
other purification protocols through their corresponding
state-update and success-probability expressions, together
with the state representation required to evaluate them.
For example, incorporating the Deutsch \emph{et al.}\
protocol~\cite{deutsch1996quantum} generally requires tracking
the individual Bell-state probabilities.

\para{Iterated Purification.}
The above protocol can be used ``iteratively''---to create a higher-fidelity EP from a set of multiple EPs. Note that given a set of EPs, there are many
different schemes by which a set of EPs can be purified to generate a single high-fidelity EP; in essence, each scheme can be represented as a binary tree over the given EPs, with each interior node being the result of purification from its two children (a target and a sacrificial EP). 
In particular, when used for iterative purification, both~\cite{bennett-96,deutsch1996quantum} assume the fidelities of the given EPs to be uniform
and perform a {\em balanced} tree of purifications over the given EPs.
The ``entanglement pumping'' method of~\cite{dur2007entanglement} instead builds a ``left-deep'' purification tree, by iteratively purifying a single target EP $t$ from several given identical copies of $s$. 
In our scalar EP model, we retain only the output fidelity
after each successful purification step and approximate the
remaining probability mass as equally distributed among the
three non-target Bell states; this approximation keeps the
EP state representation compact.
In particular, let  $\iterpurifyfidelity{i}$ denote the fidelity of $t$ 
at the end of the $i$-th step.
Then:
\begin{align}
\iterpurifyfidelity{i}(f_s) &=\left\{
\begin{array}{ll}
f_s & i=0\\
\purifyfidelity(\iterpurifyfidelity{i-1}(f_s), f_s) & i>0
\end{array} \right.
\label{eqn:EP_Purify_Fidelity}
\end{align}
Let $\iterpurifyprob{i}$ denote the probability of success of the $i$-th step.
\begin{align}
\iterpurifyprob{i}(f_s) =\left\{
\begin{array}{ll}
1 & i=0\\
\purifyprob(\iterpurifyfidelity{i-1}(f_s), f_s) & i>0
\end{array} \right.
\label{eqn:EP_Purify_Prob}
\end{align}
In addition to the above iterative schemes (i.e., balanced and left-deep), there have been other schemes proposed in the prior work, including greedy and LP-based approaches. The greedy strategy purifies EPs with the highest or lowest fidelities first~\cite{ladd2006hybrid, van2008system}, while an LP-based scheme uses an LP 
to determine the optimal purification scheduling\cite{chen2024optimum}.

\softpara{Purification Schemes in Our Work.} We note that the algorithms developed in our work (to determine efficient purification-augmented swapping trees) can incorporate any iterative-purification scheme, since our techniques only depend on the function that 
determines the fidelity improvement due to iterated purification; such a function 
can be easily derived for any iterative-purification scheme as done above (Eqns.~\ref{eqn:EP_Purify_Fidelity}-\ref{eqn:EP_Purify_Prob}) for a left-deep scheme, from the expression
for a single-step purification (i.e., Eqn.~\ref{eqn:EP_One_Purify_Fidelity}).
For concreteness, we implicitly assume a left-deep iterated-purification scheme given by Eqn.~\ref{eqn:EP_Purify_Fidelity}, throughout the rest of the paper.

\eat{use another entanglement state to test if the first entanglement state is the expected state, then discard the first state if errors are detected by measuring the second state.

For \epss, the basic purification model is the recurrence model, during each iteration an \eps is sacrificial to detect if errors exist in another \eps between the same pair of nodes. For example, we have two \epss between node $A$ and node $B$, and we use the second \eps as the sacrificial \eps. First, node $A$ and node $B$ implement a controlled-not (CNOT) gate on the two qubits from different \epss, both use the qubit from the first \eps as the control qubit and the qubit from the second \eps as the target qubit. Next, both node $A$ and node $B$ measure the qubits from the second \eps, and transport the measurement results to each other, if their results are the same, the first \eps will be kept, otherwise, it will be discarded. If $t$ sacrificial pairs purify a Bell pair, the fidelity after purification by each sacrificial pair $F^{'}$, can be iteratively calculated via~\cite{dur2007entanglement}
}

\section{\bf Problem Definition}
\label{sec:problem}

In this section, we formally define the problems of optimal generation of \epss and GHZ states using entanglement swapping and purification operations.

\subsection{\bf EP Distribution with Purification}
\label{sec:problem-a}
\noindent

\begin{figure}
    \centering
    \includegraphics[width=0.5\textwidth]{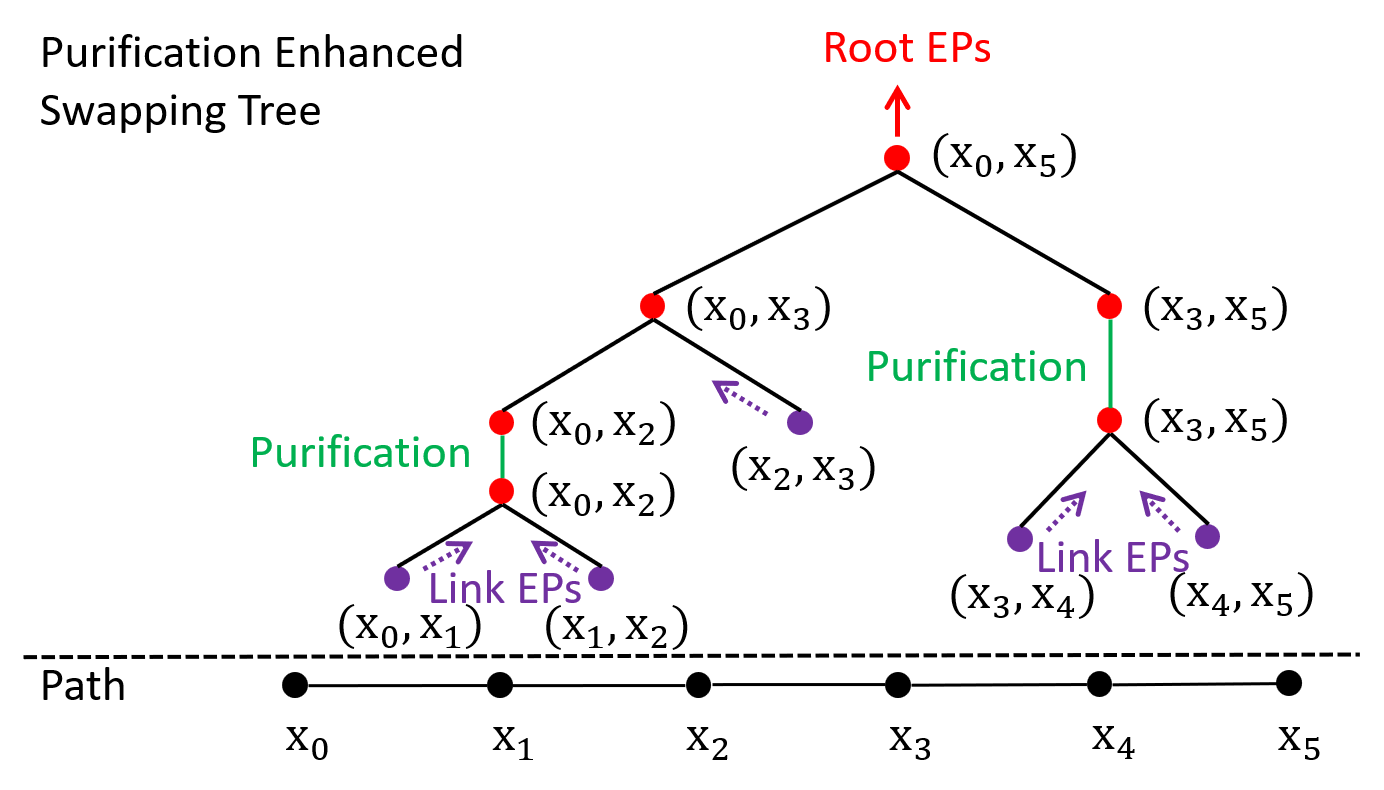}
    \caption{A purification-augmented swapping tree over a path. Binary internal nodes perform entanglement swapping, and unary nodes do purification. The leaves of the tree generate link \epss and internal nodes generate remote \epss.}
\vspace{-0.2in}
\label{fig:tree}
\end{figure}

\para{Purification-Augmented Swapping Tree.} In the presence of purification operations and the goal of establishing \epss of a given fidelity, we generalize entanglement-swapping trees of \S\ref{sec:remote-ep-generation} to purification-augmented swapping trees. In addition to \id{link} and \id{swap} nodes, a purification-augmented swapping tree may contain (unary) nodes of the form $\id{purify}(x_i,x_j,k)$ denoting a purification operation to generate an \eps $(x_i, x_j)$ using $k$ additional sacrificial copies.  Since the purification operation expects \eps $(x_i, x_j)$, the child of a $\id{purify}(x_i, x_j, k)$ node may be any node that produces \eps $(x_i, x_j)$; and a $\id{purify}(x_i, x_j, k)$ may occur wherever \eps $(x_i, x_j)$ may be an operand. See Fig.~\ref{fig:tree} is an example.

Generation latency is calculated from entanglement swapping trees, with the following additional definition that captures purification's impact on latency. Let $t$ be a tree with $\id{purify}(x_i, x_j, k)$ as its root, with $t'$ as its child. Let $l_{t'}$ be the generation latency of $t'$, and $f_{t'}$ be the fidelity of \epss generated by $t'$. Based on iterated purification, the latency of $t$, $l_t = L_k(l_{t'}, f_{t'})$ where $L_i$ is defined as: 
\begin{align}
L_i(l,f) &= \left\{
\begin{array}{ll}
l & i=0\\
\frac{(L_{i-1}(l, f) + l + \pt + \ct)}{\iterpurifyprob{i}(f)} & i>0
\end{array}
\right.
\label{eqn:iterated-purify-latency}
\end{align}
In the above, 
$\pt$ and $\ct$ are the latency of purification and classical communication, and $\iterpurifyprob{i}$ is defined in Eqn.~\ref{eqn:EP_Purify_Prob}.
As with swapping, the simplified multiplicative purification-rate equations displayed for the EP LP and the GHZ and graph-state DP/LP formulations omit $\pt+\ct$; the EP DP retains these terms through Eqn.~\ref{eqn:iterated-purify-latency}.

\para{EP Distribution\footnote{We use the terms {\em generation} and {\em distribution} interchangeably throughout this paper; specifically, each term signifies both processes---in-network generation as well as distribution to the desired nodes in the network.} with Purification (\fed) Problem.}
Given a quantum network $Q$, a pair of nodes $(s, d)$, and a required fidelity threshold $\fth$, the \fed problem is to determine a single purification-augmented swapping tree that maximizes the generation rate (i.e., minimizes the expected generation latency) of EPs $(s,d)$ with fidelity no less than $\fth$.  \hfill $\Box$

We now generalize the \fed problem to consider concurrent swapping trees.

\begin{figure}
    \centering
    \vspace*{0.15in}
   \includegraphics[width=0.35\textwidth]{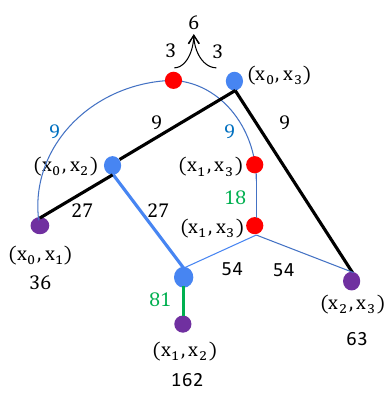}
    \caption{Example level-based swapping structure. This ``aggregate'' of two purification-augmented swapping trees. For illustration, the figure assumes a parent's generation rate is 1/3 of its children's swapping rate and 1/2 of its purification rate. The leaf node $(x_0, x_1)$'s generation rate of 36 units is ``split'' into 9 and 27 for the two different (red and blue) swapping operations.  The two root nodes represent the final/target EP formed in two different ways---for a total generation rate of 6 (3 from each swapping operation). Unary nodes represent the results of purification operations.}
\vspace{-0.2in}
\label{fig:ep-level}
\end{figure}

\para{Level-Based Swapping Structure.}
To maximize the generation rate of EPs,
{\em multiple concurrent} purification-augmented swapping trees may be required to use 
all available network resources.
Since the number of such trees can be exponential, 
we use an ``aggregated'' structure that aggregates multiple 
(purification-augmented swapping) trees into one structure; we refer to this 
as a {\em level-based} swapping structure, as it is composed of multiple 
levels---with each level consisting of EPs (as vertices) 
created by swapping operations and purification operations from the previous levels. 
See Fig.~\ref{fig:ep-level}. 
The bottom level consists of link EPs, and each non-leaf EP 
$S$ is formed by a swapping operation or a purification operation 
over of 
pairs of EPs (or one EP) in the previous layers; however, there may be 
several such operations that create $S$ (in different ways). 
Each EP $S$ may also have multiple ``parents'' (unlike in a tree), i.e., 
an EP $S$ may be used to create several EPs in the next layer; in such a case, 
the generation rate of $S$ is ``split'' across these operations.
Due to the operations from previous layers, each vertex/EP has a resulting generation rate, estimated as discussed in \S\ref{sec:problem}.
Accordingly, we refer to distribution problems whose solutions are level-based structures---aggregating multiple concurrent generation trees rather than selecting a single tree---as \emph{generalized} distribution problems.

\para{Generalized \fed{} (\gfed) Problem.}
Given a quantum network $Q$ and a demand set $\Delta={((s_i,d_i),\fthi)}$ of source-destination pairs with corresponding fidelity thresholds, the \gfed{} problem is to determine a level-based swapping structure that maximizes the aggregate generation rate of EPs satisfying the required fidelity threshold for each requested pair. \hfill$\Box$

\subsection{\bf GHZ and Graph-State Distribution with Purification}
GHZ states can be generated across network nodes using \emph{fusion} operations that combine smaller GHZ states into larger ones. The fusion operations considered in our work are the same as those used in~\cite{ghaderibaneh2023generation}. The fidelity of GHZ states can also be improved using purification operations analogous in spirit to those used for \epss. This leads to the notion of a \emph{purification-augmented fusion tree}, which specifies the sequence of fusion and purification operations used to generate a distributed GHZ state.

Although the resulting fusion trees are structurally analogous to the purification-augmented swapping trees used for \epss, the GHZ setting requires additional ingredients, including GHZ-specific fidelity models for fusion and purification, and generation schemes that build a target GHZ state from intermediate GHZ states generated over subsets of terminals.

The \emph{GHZ Distribution with Purification (\gdp) Problem} is the problem of generating a target GHZ state with fidelity at least a given threshold and a maximum generation rate. The \emph{Generalized GHZ Distribution with Purification (\ggdp) Problem} is the corresponding generalized version for GHZ states. We address these GHZ-distribution problems in \S\ref{sec:ghz}.

We extend the same tree-based and level-based viewpoints to general graph states, using graph-state growth and purification operations. The corresponding \graphdp{} and \graphlp{} problems maximize the target graph-state generation rate subject to a fidelity threshold; we address them in \S\ref{sec:gen-graphs}.

\subsection{\bf Scope of Optimality Claims}
\label{sec:optimality-scope}

Throughout the paper, our optimality claims are relative to two main
aspects of the formulation: (i) the operational and analytical model,
and (ii) the adopted discretization. In some formulations, these claims are also subject to an
explicitly imposed purification-iteration bound.

\softpara{(i) Operational and Analytical Model.}
The operational model specifies the allowed primitives---such as swapping, 
fusion, purification, and graph-state growth operations---and the class of 
generation structures over which the algorithm optimizes. 
The analytical model specifies how these operations are evaluated, including their 
latency rules, success probabilities, and fidelity-update expressions, 
as well as any decoherence model if included. 
Our algorithms are modular with respect to these analytical
expressions: hardware-calibrated fidelity-update and
success-probability expressions can be substituted, and the
same DP or LP formulations optimize with respect to the
substituted model.
Thus, the optimality guarantee is model-specific rather than a claim of model-independent physical optimality.

\softpara{(ii) Discretization.}
Our LP formulations discretize fidelity labels, while our DP
formulations use either fidelity or rate discretization.
The corresponding optimality claims are with respect to the
resulting discretized formulation.
In Appendix~\ref{app:disc-gap}, we bound the resulting fidelity and
rate errors and show that, under its stated regularity
conditions, the fidelity and generation rate of a fixed finite construction converge to their continuous counterparts as the discretization is refined and source-rounding errors vanish.

\softpara{Purification-iteration bounds.}
Where a formulation explicitly bounds the number of purification iterations, we assume that restricting purification trees to at most \((m+1)\) leaves (equivalently, \(m\) sacrificial copies) incurs little performance loss. For our optimality results, we essentially assume that this restriction does not exclude an optimal solution. This assumption is motivated by the diminishing fidelity gains from additional purification rounds, as discussed in Appendix~\ref{app:disc-gap}.

\section{\bf Related Work}
\label{sec:related}

\para{Distribution of \epss{} in quantum networks.}
There has been substantial recent interest in efficient entanglement-distribution protocols for quantum networks; see~\cite{abane2025survey} for a recent survey. Many of these works focus on selecting entanglement routes without explicitly incorporating purification into the routing and swapping decisions~\cite{caleffi,sigcomm20,swapping-tqe-22,zhao2023segmented,sundaram2024optimized,yang2024multitree}.

The seminal work of D"ur \emph{et al.}~\cite{dur1999quantum} showed that interleaving entanglement swapping with purification can create high-fidelity EPs over noisy channels, with only logarithmic growth in physical resources as a function of channel length. Since then, several works have studied EP distribution protocols that combine swapping and purification, but most do so under restrictions that differ substantially from our formulation.

\softpara{Purification at the link or target level.}
A large class of purification-enabled EP-distribution protocols considers purification only at the link level or at the final target level to satisfy an end-to-end fidelity threshold~\cite{li2021effective,panigrahy2023capacity,pouryousef2023quantum,zhan2025design,jia2024entanglement,chen2024optimum,li2022fidelity}. In particular,~\cite{jia2024entanglement,chen2024optimum} develop optimal purification schedules over a \emph{given} entanglement path, while restricting purification to link-level EPs. They also develop approximation schemes for selecting a path based on the resulting purification schedule. These works, as well as~\cite{li2022fidelity}, use a synchronized swapping model, which differs from the memory-assisted waiting model adopted in our work~\cite{swapping-tqe-22}. In the waiting model, the order of swapping operations matters, and the optimization problem is naturally over swapping trees rather than only over paths.

\softpara{Purification at intermediate levels.}
Only a few works~\cite{victora2023entanglement,santos2023shortest,hu2024highfidelity,wang2024scheduling} allow purification to be interleaved more generally with entanglement swapping. However, these approaches still consider limited classes of swapping and purification structures. For example,~\cite{victora2023entanglement} considers the purification protocol independently of path selection and limits purification to at most three hops. The works~\cite{santos2023shortest,hu2024highfidelity} develop incremental hop-by-hop path-selection approaches that maintain fidelity requirements, possibly using purification; however, they implicitly restrict the swapping/purification structure, with~\cite{santos2023shortest} considering balanced trees and~\cite{hu2024highfidelity} considering left-deep trees. Similarly,~\cite{wang2024scheduling} studies link-level, target-level, and threshold-based purification over a given path in a {\em synchronized} network model.

The most closely related EP-distribution work is~\cite{zhao2022e2e}, which identifies the top $k$ shortest paths between a source and destination, purifies link EPs along each path to exceed a link-fidelity threshold, iteratively purifies the link most critical to the end-to-end fidelity until the target threshold is met, and then uses linear programming to allocate limited resources across paths. In contrast, our EP formulation directly optimizes purification-augmented swapping trees and level-based swapping structures under a global fidelity threshold, thereby jointly deciding where purification occurs and how swapping operations are arranged. We compare against~\cite{zhao2022e2e} in \S\ref{sec:eval}.

\para{Distribution of GHZ states.}
Most prior work on GHZ states focuses on local generation or purification, where all qubits reside at the same node, rather than on distributing GHZ states over a quantum network. For example,~\cite{wallnofer2019multipartite,kay2007multipartite,qin2023efficient,carle2013purification} study multipartite purification or local GHZ-state preparation but do not address network-level questions such as routing remote EPs, selecting fusion structures, or contending for network resources. Our prior work on GHZ distribution~\cite{ghz-qce-23} develops techniques for selecting efficient fusion trees, but does not incorporate purification.

Some works combine GHZ purification with distribution, but only in restricted settings. In particular,~\cite{nickerson2013topological,de2020protocols} study purification protocols for GHZ states, including purification using GHZ states of the same or smaller size, and combine these protocols with GHZ distribution. However, their distribution schemes are limited to specific, simple network topologies. At a higher architectural level, Pirker and D"ur~\cite{pirker2018modular} propose modular quantum-network architectures based on shared multipartite resource states, including GHZ and decorated graph states, with purification during network-state configuration. Their focus is on flexible, multipartite network-state architectures and request handling, rather than on fidelity-threshold-constrained routing and optimization of end-to-end generation plans over a given physical network.

More broadly, prior GHZ-distribution works do not provide the closed-form fidelity expressions needed for fidelity-aware optimization across varying input sizes and fidelities. In contrast, our work derives analytical fidelity models for GHZ fusion and purification and uses them to optimize purification-augmented fusion trees and level-based fusion structures subject to a fidelity threshold.

\para{Distribution of general graph states.}
Prior work on graph-state distribution has largely focused on objectives and models different from ours. Meignant \textit{et al.}~\cite{Meignant2019GraphStates} propose a distributed scheme that first constructs GHZ states via star expansion and then uses edge-decorated complete-graph constructions together with local measurements to obtain arbitrary graph states; their focus is on minimizing Bell-pair consumption in an idealized setting. Ji \textit{et al.}~\cite{Ji2025GraphTransformation} recently propose a graph-transformation approach that searches for a lower-cost LC-equivalent graph state, distributes that alternative state, and then recovers the target graph state using compressed local Clifford operations; their optimization emphasis is on EP cost and success probability. Huang \textit{et al.}~\cite{Huang2025P2PGSD} formulate graph-state distribution in a shot-based spacetime-network model, where inter-shot links capture memory usage, and develop centralized and peer-to-peer algorithms that primarily optimize shot count, cumulative memory usage, and computation time.

The above works do not incorporate purification into the graph-state distribution. In contrast, our work studies fidelity-aware stochastic graph-state generation under network resource constraints, with the objective of maximizing generation rate while explicitly accounting for purification and fidelity constraints.

\softpara{Relation to our prior work.}
This paper is also closely related to our earlier tree-based line of work, none of which incorporates purification into the generation structure. In particular,~\cite{swapping-tqe-22} developed optimized entanglement-swapping trees for EP distribution,~\cite{ghaderibaneh2023generation} developed fusion-based techniques for GHZ distribution, and~\cite{fan2024optimized} developed techniques for distributing more general graph states, including paths, trees, and grids. The present work builds on the tree-based and LP-based optimization viewpoints developed in these earlier papers, but substantially extends them by integrating purification directly into the generation structures and by optimizing under explicit fidelity constraints.

A preliminary conference version of this work appeared in~\cite{purification-qcnc}, where we introduced purification-augmented swapping and fusion trees for EP and GHZ distribution. The present journal version significantly extends that preliminary version by sharpening the theoretical treatment, formalizing the discretized-fidelity optimality claims, developing and organizing the fidelity-update models more systematically, extending the framework to general graph states, and expanding the evaluation of the resulting DP and LP schemes.

\section{\bf EP Distribution with Purification: Optimal Solutions}
\label{sec:ep}

In this section, we present our solutions to the \fed and \gfed problems based on dynamic programming (DP) and linear programming (LP), respectively.

\subsection{\bf Dynamic Programming (DP) for EP Distribution}

Our algorithm for the \fed problem uses dynamic programming to construct an optimal purification-augmented swapping tree that minimizes generation latency under a target fidelity constraint. We begin by outlining the key challenges and the basic idea.

\para{Challenges and Basic Idea.}
To develop a dynamic programming (DP) algorithm for the \fed problem, we encounter two key challenges:
(i) \eps over {\em any} pair of network nodes can serve 
as an intermediate state towards generating the given target EP, 
regardless of the intermediate state's fidelity (since fidelity 
can be improved via purification at a later stage) 
and generation rate. This leads to an infinite number of intermediate
states/subproblems to be computed. 
(ii) 
Second, the potential subproblems, corresponding to EP-generation over
pairs of network nodes, may depend on each other, leading to a potentially
cyclic dependency among them. 
Below, we propose a dynamic programming (DP) algorithm by defining appropriate 
subproblems and recursive equations to compute the subproblems recursively, wherein we address the above challenges by: (i) discretizing the fidelity values, and 
(ii) adding a parameter (which can be appropriately bounded) to break the 
dependency cycles among the subproblems considered.

\softpara{Discretized Fidelity Values.}
To make the set of intermediate states finite, we restrict fidelity labels to a discrete grid $\discretefid={f_1,f_2,\ldots,f_d}$. For an arbitrary fidelity value $f$, we write $\floor{f}$ for the largest grid value $f_i\leq f$ and $\lceil f \rceil$ for the smallest grid value $f_i\geq f$. In the DP recurrences, output fidelities are rounded down to the grid, yielding a conservative discretized model. 
Appendix~\ref{app:disc-gap} bounds the fidelity and rate errors for a fixed construction and establishes convergence to its continuously evaluated values under the stated regularity conditions as the grid is refined.

\para{DP Algorithm.}
For our DP algorithm, we define
subproblems for each triplet $\langle x, y, f\rangle$ where $(x,y)$ is a pair of nodes in the network and $0 \leq f \leq 1$ is the discretized fidelity threshold $f$. The goal of each subproblem is to determine the minimum generation latency\footnote{Even though the \fed problem is to maximize the generation rate, we consider minimizing generation latency here, for clarity of presentation, since generation rate is just the 
inverse of generation latency (when there is no ''pipelining''; see Theorem~\ref{thm:ep-dp}).} for the given triplet of parameters.  

\softpara{Subproblems $L(x,y,f,h)$.}
In particular, let $T(x,y,f,h)$ be a purification-augmented swapping tree with the least latency among all trees that generate \epss over $(x,y)$  with at least fidelity $f$ and have a height at most $h$; let $L(x,y,f,h)$ be the generation latency of $T(x,y,f,h)$. 
As mentioned above, the parameter $h$, representing the maximum height of the purification-augmented swapping tree, has been added to break the cyclic dependency 
of the subproblems; e.g., \epss over $(x,y)$ may be useful as intermediate states
to generate \epss over $(r, s)$ and {\em vice-versa}. We show later (see Lemma~\ref{lemma:nocycle}) that the maximum height of an optimal purification-augmented swapping tree is $2n$ where $n$ is the number of network nodes.

\softpara{Base Case.}
The base case corresponds to $h = 0$. In this case, the tree $T(x,y,f,0)$ is essentially a single-node\footnote{The term {\em node} may refer to a node within a purification-augmented swapping/fusion tree or a node in the network; generally, this is clear from the context, else we use {\em network nodes} for the latter.} tree corresponding to a pair of network nodes $(x,y)$. An \eps over a pair of network nodes $(x,y)$ can be generated {\em directly} (without swapping) only if $(x,y)$ is a direct link in the network. Thus, for all $x,y,f$ values, we have:
\begin{equation}
L(x,y,f,0)=
\begin{cases}
\tfrac{1}{\id{rate}(x,y)}, & \text{if } (x,y)\in E(Q)\text{ and }\id{fid}(x,y)\ge f,\\[2pt]
\infty, & \text{otherwise.}
\end{cases}
\label{eqn:dp-recursion-link-latency}
\end{equation}
Above, we use the functions {\em rate} and {\em fid} defined in our quantum network model (see \S\ref{sec:background}).

\softpara{Recursive Step.}
There are two cases to consider for the recursive step.
\begin{enumerate}
    \item The last operation in $T(x,y,f,h)$ is a {\em swapping} operation, i.e., 
    the \eps over $(x,y)$ are formed by swapping \eps over $(x,z)$ and $(z,y)$ for
    some network-node $z$. 
    
    \item The last operation in $T(x,y,f,h)$ is a {\em purification} operation, i.e., 
    the \eps over $(x,y)$ are formed by purifying \eps (of lower fidelity) over $(x,y)$. 
\end{enumerate}
In both the above cases, the children of the root in $T(x,y,f,h)$ must correspond to
{\em optimal} subtrees for the desired/appropriate parameter values. 
This means the problem has the \emph{optimal substructure property}, i.e., the optimal solution to a subproblem can be formed from optimal solutions to other subproblems, enabling a dynamic programming solution.  
More formally, we can write the following recurrence for 
$L(x,y,f,h)$ as follows: 
\begin{align}
    L(x,y, f, h) &= \min\{l_s, l_p, L(x,y,f,h-1)\} 
    \label{eqn:dp-recursion-latency}
\end{align}
\noindent
where $l_s$ and $l_p$ are defined as follows.
\begin{align}
l_s &=
    \min_{\substack{z\\f_1,f_2}} \Bigl(\swaplatency(L(x,z,f_1,h-1), L(z,y,f_2,h-1))\Bigr)
    \label{eqn:dp-recursion-swap-latency}\\
&    \quad\mbox{where $f_1$ and $f_2$ are such that}\ f \leq \lfloor \swapfidelity(f_1, f_2) \rfloor; \nonumber\\
l_p &= \min_{\substack{i\leq m\\f_0}}\{L_i(L(x,y,f_0,h-1), f_0) \mid
\iterpurifyfidelity{i}(f_0) \geq f\} %
\label{eqn:dp_purification_latency}
\end{align}
Above, \swaplatency, \swapfidelity, $L_i$, and \iterpurifyfidelity{i} are as defined in Eqns.~\ref{eqn:EP_swaplatency}, \ref{eqn:EP_swapfidelity}, \ref{eqn:iterated-purify-latency}, 
and~\ref{eqn:EP_Purify_Fidelity} respectively, and 
$m$ in the $l_p$ expression is the upper bound on the
number of sacrificial copies used in a purification tree; see
\S\ref{sec:optimality-scope}.

\softpara{DP Solution.}
We compute the $T()$ and $L()$ values in order of increasing $h$, and 
return the $T(s,d,\lceil f \rceil, 2n)$ as the solution.\footnote{Recall that $\lceil f \rceil$ denote the
the smallest discretized fidelity value in $\discretefid$ such that $\lceil f \rceil \geq f$; here, we use $\lceil~\rceil$ instead of $\floor{~}$ to ensure that the EP generated by $T(s,d,\lceil f \rceil, 2n)$ achieves fidelity {\em at least} $f$.} Below, we show that the solution returned is optimal.

\begin{figure*}[t]
    \centering
    \vspace*{0.15in}
    \includegraphics[width=0.7\textwidth]{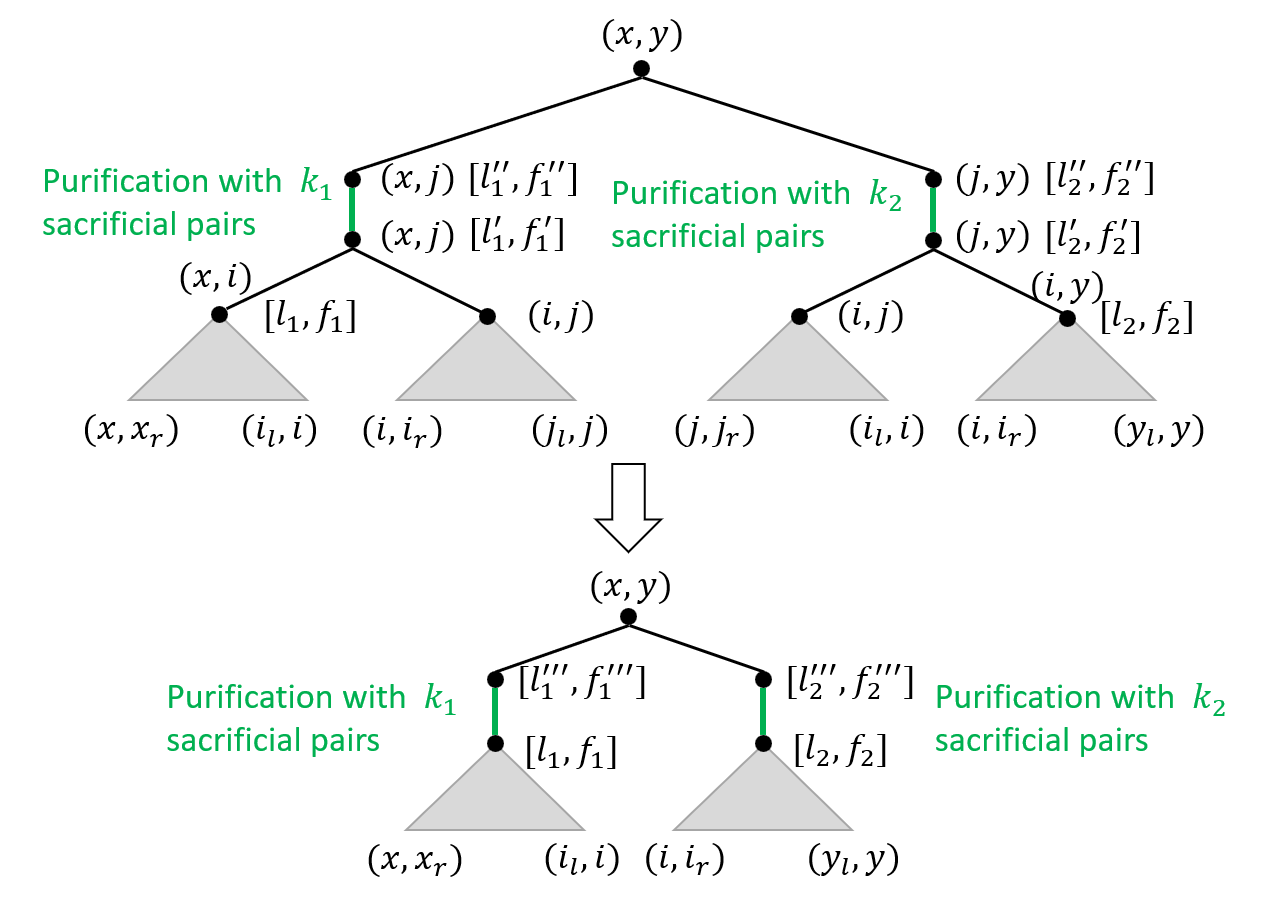}
    \caption{Proving that the network path induced by the leaves of an optimal purification-augmented swapping tree can be chosen to be acyclic. In the top tree, the leaves induce a path from $x$ to $y$ containing the cycle segment $i \to j \to i$. Removing this cycle and appropriately relocating the purification operations yields the bottom tree, whose induced leaf-path is acyclic. Here, $(u,v)$ denotes an \eps over network nodes $u$ and $v$, and $[L,f]$ denotes the corresponding \eps's expected generation latency and fidelity. The resulting tree has no larger generation latency and no smaller fidelity than the original tree ($l_1''' \leq l_1''$, $f_1''' \geq f_1''$, $l_2''' \leq l_2''$, and $f_2''' \geq f_2''$), preserving optimality.}
    \label{fig:no-cycle}
\end{figure*}

\para{DP Algorithm is Optimal.}
We now show that the DP formulation defined by Eqns.~\ref{eqn:dp-recursion-link-latency}--\ref{eqn:dp_purification_latency} returns an optimal solution for the \fed problem. We start by establishing a structural bound on the height of an optimal purification-augmented swapping tree; this bound is used in the proof of optimality.

\begin{lem-prf}
\label{lemma:nocycle}
For any instance $(Q,s,d,f^\tau)$ of the \fed problem, there exists an optimal purification-augmented swapping tree of height at most $2n$, where $n=|V(Q)|$.
\end{lem-prf}
\begin{prf}
Let $T$ be an optimal purification-augmented swapping tree for the given \fed instance. Note that each leaf of $T$ corresponds to a {\em link} in the network, and the sequence of leaves of $T$ forms a network {\em path}\footnote{It is easy to see this for a swapping tree; similarly, this also holds for a purification-augmented swapping tree.} from $s$ to $d$ (not necessarily acyclic).

Below, we show that $T$ can be chosen so that the network path corresponding to its sequence of leaves is acyclic. Hence, the number of leaves of $T$ is at most $n$, which implies that the height of $T$ is at most $2n$, since any root-to-leaf path in $T$ contains at most $n$ swapping operations and at most $n$ (iterated)-purification operations.

\softpara{An Acyclic Optimal Tree.}
Suppose that the network path induced by the leaves of $T$
contains a cycle. Then, as illustrated in Fig.~\ref{fig:no-cycle},
we can remove one such cycle and relocate the purification
operations to obtain another purification-augmented swapping
tree whose generation latency is no larger while its fidelity
is no smaller.\footnote{
Removing a cycle may require combining consecutive purification
operations. Under the bounded-purification assumption in
\S\ref{sec:optimality-scope}, we assume that the combined sequence
can be replaced by an $m$-bounded purification operation
without increasing latency or reducing fidelity.
In Fig.~\ref{fig:no-cycle}, purification is relocated from
the EP over $(x,j)$ to the EP over $(x,i)$. Since the latter
requires fewer swapping operations, it has no greater generation
latency and no lower fidelity. By monotonicity of the
purification updates, the relocated purification is therefore
no worse.}
The resulting tree remains optimal. Repeating this process
removes all cycles and yields an acyclic induced leaf-path
without losing optimality.

\end{prf}

\medskip
\begin{thm-prf}
\label{thm:ep-dp}
For any instance $(Q,s,d,f^\tau)$ of the \fed problem, let $n=|V(Q)|$. Then the purification-augmented swapping tree $T(s,d,\lceil f^\tau\rceil,2n)$ returned by the DP formulation in Eqns.~\ref{eqn:dp-recursion-link-latency}--\ref{eqn:dp_purification_latency} is an optimal solution to the \fed problem under the adopted discretized-fidelity model and the stated purification-iteration bound.
\end{thm-prf}

The proof of Theorem~\ref{thm:ep-dp} is given in Appendix~\ref{app:ep-optimality-proofs}.
\subsection{{\bf Linear Programming (LP) for Generalized EP Distribution}}
\label{sec:ep-lp}
In this section, we consider the \gfed problem and describe a solution based on linear programming (LP) to generate an optimal level-based swapping structure (essentially, a {\em collection} of purification-augmented swapping trees, as described in~\S\ref{sec:problem-a})
that maximizes the generation rate of EPs with fidelity exceeding the required threshold.

\begin{figure}
    \centering
   \includegraphics[width=0.48\textwidth]{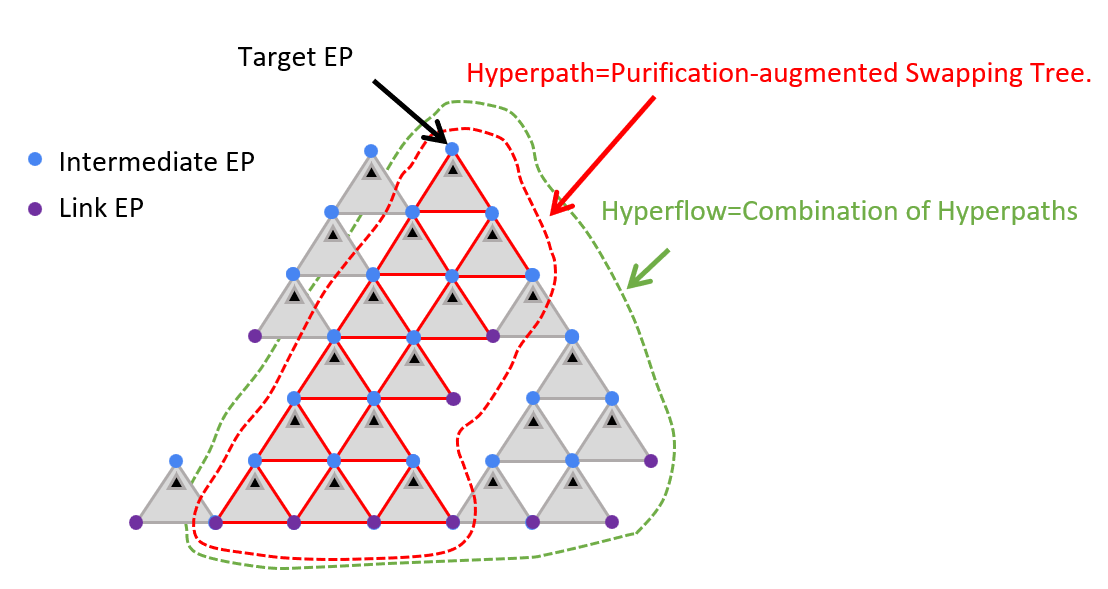}
    \caption{Here, each hyperarc represents a swapping operation, and each hyperpath represents a potential purification-augmented swapping tree (for clarity, the hyperarcs corresponding to purification operations are omitted). A hyperflow incorporates multiple hyperpaths, akin to a network flow in a simple graph, and represents a level-based swapping structure.}
\label{fig:hyperflow}
\end{figure}

\para{Basic Idea.}
We use a directed-hypergraph flow model as a generic optimization template for aggregated level-based structures. In the EP setting, the template is instantiated over remote-EP states indexed by endpoint pair and discretized fidelity, with hyperarcs corresponding to swapping and EP-purification operations under the adopted waiting model. Given a quantum network and a source-destination pair, we start by constructing a hypergraph that incorporates all possible level-based swapping structures 
for generating the target EPs $\{(s_i, d_i)\}$. In this hypergraph:
\begin{enumerate}[(i)]
\item
Each hypervertex represents a possible EP $(u, v, f)$, where $f \in \discretefid$ is a discretized fidelity value;
\item
Each hyperarc (i.e., a directed hyperedge) represents a valid operation that consumes one or more input EPs and produces a new EP with a particular fidelity value; thus, a hyperarc goes from 1-2 hypervertices (consumed EPs) to a hypervertex (resulting EP).
\item
A ``hyperpath'' from the special hypervertex {\tt start} (representing the source for link-EPs) to the hypertext for $(s_i, d_i)$ corresponds
to purification-augmented swapping tree that generates \eps over $(s_i, d_i)$.

\item
A ``hyperflow'' is a level-based swapping structure that combines multiple hyperpaths, analogous to how a network flow in a standard graph represents a combination of simple paths (Fig.~\ref{fig:hyperflow}).

\item
Finally, as in network flows in simple graphs, we assign flow variables to hyperarcs; herein,
the flow variable for a hyperarc represents the generation rate of the EP corresponding to the result of the hyperarc's operation. 
\end{enumerate}
To determine the optimal hyperflow (and thus the optimal level-based swapping structure) in 
the above hypergraph, we formulate a linear program that maximizes the hyperflow from the {\tt start} node to the {\tt term} node (representing the aggregate of generation rates of all the target EPs), subject to the constraints on 
resources and flow conservation. This transforms the \gfed problem into a hypergraph flow optimization problem. We formalize the above ideas below.

\eat{
For example, the swap operation that costs \epss $ep_1$ and $ep_2$ to generate $ep_3$ is represented by a hyperarc with heads $ep_3$ and two tails $(ep_1, ep_2)$, and the purification operation using $ep_2$ as sacrificial \eps to purify $ep_1$ is represented by another kind of hyperarc with heads $ep_3$ and two tails $(ep_1, ep_2)$. We formulate an LP based on the hypergraph we defined and maximize the total in-flow to the hypervertices representing the node pairs in $P$ with fidelity above $F_{t}$.
To limit the size of the hypergraph, we \blue{discretize} %
fidelity into 
$F=\{f_{1}, f_{2}, ..., f_{n}\}$, and \blue{constrain} the fidelity of the \epss represented by hypervertex can only be elements in set $F$.
We first discuss the model using existing resources addressed in \S\ref{sec:problem}.
}

\para{Hypergraph.}
Given a \gfed problem over a quantum network $Q$ and demand is $\Delta$, we construct a hypergraph $H_{Q,\Delta}$ with hypervertices $V(H_{Q,\Delta})$ and hyperarcs $E(H_{Q,\Delta})$ as follows:

\begin{figure}
    \centering
   \includegraphics[width=0.48\textwidth]{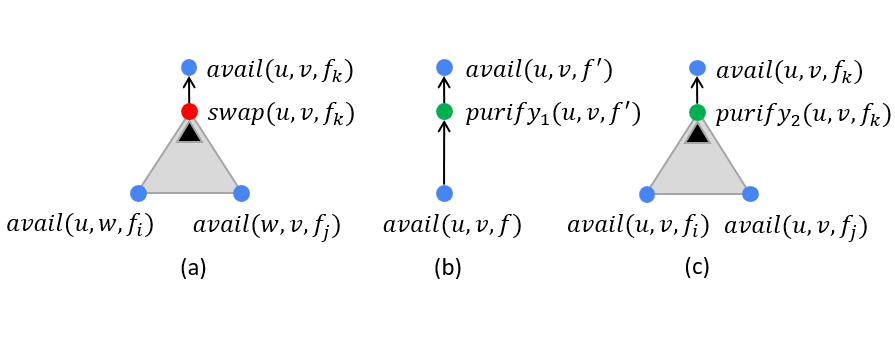}
    \caption{Hyperarc representations of different operations in the hypergraph. (a) Entanglement swapping operation. (b) Purification using two copies of EPs with the same fidelity. (c) Purification using two EPs with different fidelities but the same endpoints.}
    \vspace{-0.2in}
\label{fig:hyperarcs}
\end{figure}

\softpara{$V(H_{Q,\Delta})$ consists of:}
\begin{enumerate}
  \item Two distinguished vertices \emph{start} and \emph{term}
  \item For each $u,v \in V(Q)$ and discretized fidelity value $f \in \discretefid$, there are 4 hypervertices representing \epss $(u,v)$ of fidelity $f$, produced using different operations: 
  \begin{enumerate}
      \item $\id{swap}(u, v, f)$:  \epss produced by an entanglement swap
      \item $\id{purify}_1(u, v, f)$: \epss purified using two EPs of same fidelity
      \item $\id{purify}_2(u, v, f)$:  \epss purified using two EPs of different fidelity.\footnote{Note that we didn't have such a purification operation in the DP algorithm, since such an operation can't arise when generating \eps over a single network path, as is the case for a single purification-augmented swapping tree.}
    \item $\id{avail}(u, v, f)$: \epss available for consumption (for further swap and purify operations). This is the hypervertex where \eps of the same type get aggregated and then split across 
    further operations.
  \end{enumerate}
\end{enumerate}

\begin{figure}[h]
    \includegraphics[width=0.5\textwidth]{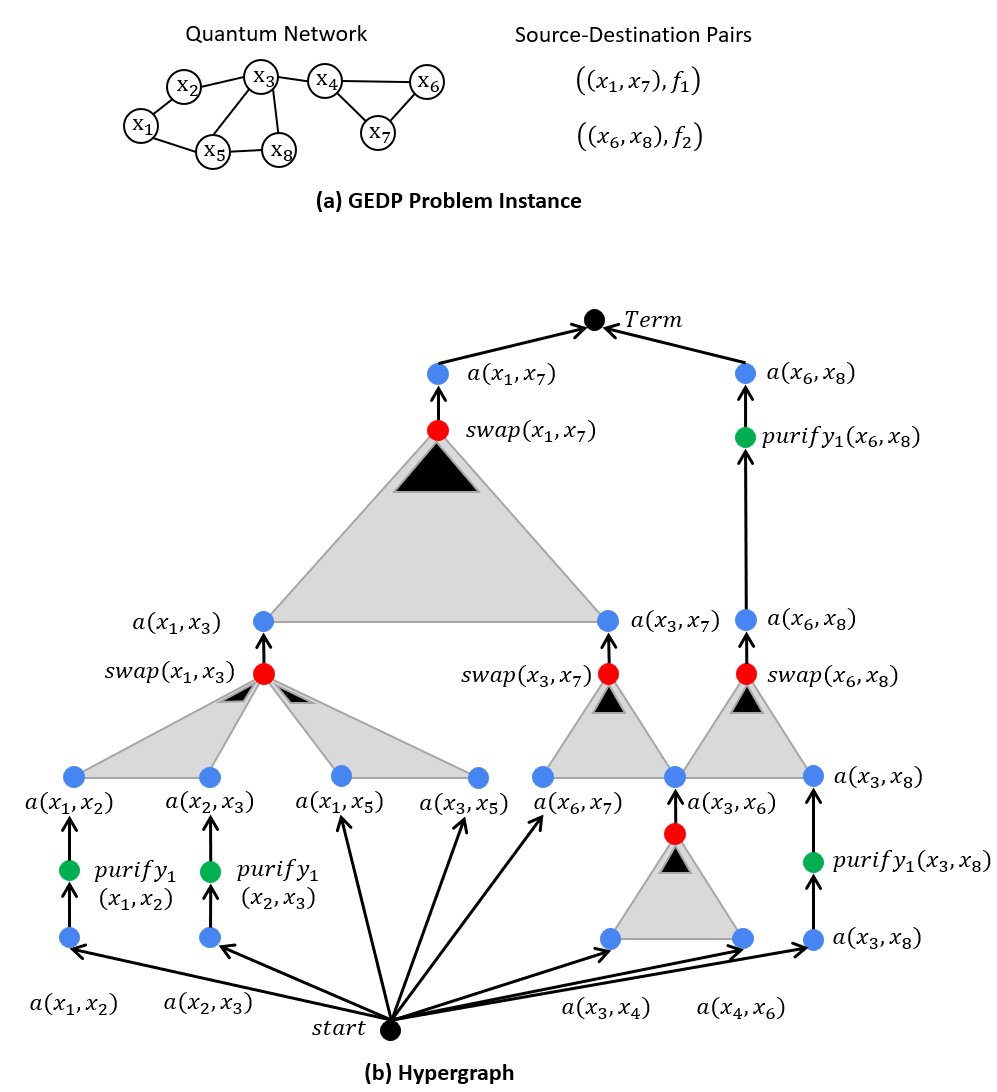}
    \caption{(a) A \gfed problem instance with two source-destination pairs. (b) The corresponding hypergraph representation. For clarity, only a small subset of the relevant hypervertices and hyperarcs is shown, and each hypervertex is labeled using its endpoint pair $(u,v)$ (instead of the full triplet $(u,v,f)$ used in the actual hypergraph, where $f$ denotes the fidelity). Here, $a() = Avail()$.}
    \label{fig:hypergraph}
\end{figure}

\softpara{$E(H_{Q,\Delta})$ consists of:}
We create hyperarcs to represent a swapping or purification operation; these hyperarcs go from one or two $\id{avail}()$ hypervertices to a $\id{swap}()$ or $\id{purify}_*()$ hypervertex.
In addition, we also create ``consolidation'' hyperarcs that go from
each $\id{swap}()$ or $\id{purify}_*()$ to its corresponding 
$\id{avail}()$ to make the \eps available for further consumption. 
Overall, we have the following set of hyperarcs (see Fig.~\ref{fig:hyperarcs}): 
\begin{enumerate}
  \item{} [Start]: representing the generation of link-\epss.
  \begin{small}
      \begin{itemize}
          \item $(\{start\},\{\id{avail}(u,v,\id{fid}(u,v))\}), \quad \forall (u,v) \in E(Q)$
      \end{itemize}
  \end{small} 
  \item{}[Swap]: representing the entanglement swap operation.
  \begin{small}
  \begin{itemize}
  \item  $(\{\id{avail}(u,w,f_{i}),\id{avail}(w,v,f_{j})\}, \{\id{swap}(u,v,f_{k})\})$ \\
  $\ \forall u,v,w \in V$, $f_i, f_j \in \discretefid$, 
  $f_{k}=\lfloor\swapfidelity(f_i, f_j)\rfloor$
  \item $(\{\id{swap}(u,v,f)\},\{\id{avail}(u,v,f)\})$, \quad $\forall u,v \in V$, $f \in \discretefid$
  \end{itemize}
  \end{small}
  \item{[SelfPurify] representing purification using two copies of EPs with the same fidelity.
  \begin{itemize}
  \begin{small}
  \item  $(\{\id{avail}(u,v,f)\}, \{\id{purify}_1(u,v,f')\})$ \\
  $\forall u, v \in V, f\in \discretefid$, 
  $f' = \lfloor F_{\mathrm{pur}}(f,f) \rfloor$

  \item $(\{\id{purify}_1(u,v,f)\}, \{\id{avail}(u,v,f)\}), \quad \forall u,v \in V, f\in \discretefid$
  \end{small}
  \end{itemize}}
  \item{[PurifyDistinct] representing purification using two EPs with different fidelity.
  \begin{itemize}
  \begin{small}
  \item  $(\{\id{avail}(u,v,f_{i}), \id{avail}(u,v,f_{j})\}, \{\id{purify}_2(u,v,f_{k})\})$ \\
  $\forall u, v \in V, f_i, f_j \in \discretefid$, 
  $f_i \neq f_j$,
  $f_{k}=\lfloor \purifyfidelity(f_i, f_j) \rfloor$ 
  \item $(\{\id{purify}_2(u,v,f)\}, \{\id{avail}(u,v,f)\}), \quad
  \forall u,v \in V$, $f \in \discretefid$
  \end{small}
  \end{itemize}}
  \item{[Term] representing the generation of target \epss above the given fidelity threshold. 
  \begin{itemize}
  \begin{small}
  \item $(\{\id{avail}(s,d,f')\},\{\id{term}\})$, $\forall (s,d,f) \in \Delta,$\\
  $f' \in \discretefid$ s.t. $f'\geq f$
  \end{small}
  \end{itemize}}
\end{enumerate}
See Fig.~\ref{fig:hypergraph} for the hypergraph corresponding to a \gfed problem instance. Note that the non-linearity of the functions 
\swapfidelity() (Eqn.~\ref{eqn:EP_Purify_Fidelity}) and 
\iterpurifyfidelity{i} (Eqn.~\ref{eqn:EP_Purify_Prob}) 
doesn't affect the linearity of the constraint equations (and thus, the overall program is still an LP); in particular, these non-linear functions only
determine the hyperarcs in the hypergraph.

\para{LP Variables, Constraints \& Optimization Objective.}
\begin{itemize}    
    \item \emph{Variables:} $z_e$, for each hyperarc $e$ in $H$, represents the EP generation rate at $e$'s tail.  %

    \item \emph{Flow Constraints,} which vary with vertex types. Below, we use  $out(v)$ and $in(v)$ to denote the outgoing and incoming hyperarcs at $v$, respectively. %

    \begin{itemize}
        \item For each vertex $v$ s.t.
        $v=\id{avail}(\cdot)$:
              \[ \sum_{e \in in(v)} z_e =\sum_{e' \in out(v)} z_{e'} \]
        This equation represents aggregation and then splitting of rates at an $id{avail}(\cdot)$ hypervertex.
        \item For each vertex $v$ s.t. $v = \id{swap}(\cdot)$:
              \[ \sum_{e \in in(v)} \tfrac{2}{3} z_e \fp  = \sum_{e' \in out(v)} z_{e'} \]
              Here, the factor $\fp$ accounts for the loss in generation rate due to the swapping operation failure, and the $\tfrac{2}{3}$ factor accounts for the loss in generation rate due to one of the EPs waiting for the other EP to arrive before performing the operation (see Eqn.~\ref{eqn:EP_swaplatency}).

        \item For each vertex $v$ s.t. $v=\id{purify}_1(\cdot)$:
              $$ \sum_{e \in in(v)}  \tfrac{1}{2}  z_e  p_p(e)  = \sum_{e' \in out(v)} z_{e'} $$
              
                Since the self-purification operation consumes two EPs with the same fidelity to produce one, half of the EPs are sacrificed; thus, the $\tfrac{1}{2}$ factor above. The rate loss due to the purification operation is accounted for by $p_p$, the success probability of the purification operation.\footnote{\label{fn:pp}We note that $p_p$ depends on the {\em fidelity} of the ``operand'' in $e$, and thus should be a ``function'' of $e$. Thus, each equation may use a different $p_p$ value. This doesn't create any complications, as the $p_p$ value is still a constant for a given $e$ (and thus, for a given equation).}
                
        \item For each vertex $v$ s.t. $v = \id{purify}_2(\cdot)$:
              $$ \sum_{e \in in(v)}  \tfrac{2}{3}z_e p_p(e)  = \sum_{e' \in out(v)} z_{e'} $$
              Here, $p_p$ models the rate loss due to purification, and the $\tfrac{2}{3}$ factor arises from waiting for \emph{both} \epss to arrive before the operation.\footnotemark[\getrefnumber{fn:pp}]

    \end{itemize}

    \item \emph{Capacity Constraints:} 
    For all hyperarcs $e$ from \id{start}
 to $\id{avail}(u,v,f)$: \\
 $z_e \leq \id{rate}(u,v)$, where $\id{rate}$ is specified by the quantum network graph $Q$.
    
    \item \emph{Objective}: Maximize $\sum_{e \in in(term)} z_e$    
\end{itemize}

\para{LP Size and Time Complexity.}
For a network with $n$ nodes, the number of hypervertices in the hypergraph is $O(n^2|\discretefid|)$, and the number of hyperarcs is $O(n^3|\discretefid|^2)$, resulting from combinations of swapping and purification operations. The LP introduces one variable per hyperarc and one constraint per hypervertex. Thus, the LP contains:
\begin{itemize}
    \item $O(n^3|\discretefid|^2)$ variables,
    \item $O(n^2|\discretefid|)$ equality constraints.
\end{itemize}
Using interior-point methods for LP solving\cite{lee2015efficient}, which run in time polynomial in the number of variables and constraints, the overall time complexity is:

\[
\tilde{O}(n^5|\discretefid|^{5/2}).
\]

\para{Performance Guarantee.}
The following theorem states that the LP formulation over the hypergraph $H_{Q,\Delta}$ exactly captures the \gfed{} problem under the adopted discretized-fidelity model, i.e., optimizing the LP is equivalent to optimizing over all feasible level-based swapping structures.

\begin{thm-prf}
\label{thm:ep-lp}
For any instance $(Q,\Delta)$ of the \gfed{} problem, under the adopted discretized-fidelity model, the LP formulation over the hypergraph $H_{Q,\Delta}$ has no relaxation gap with respect to the level-based formulation. In particular, its optimum objective value equals the maximum achievable generation rate over all feasible level-based swapping structures for the same instance. Moreover, any optimal LP solution induces an optimal level-based swapping structure.
\end{thm-prf}

The proof is given in Appendix~\ref{app:ep-optimality-proofs}.

\section{\bf GHZ Distribution with Purification}
\label{sec:ghz}

In this section, we develop efficient generation schemes for GHZ states. Although the resulting fusion trees are structurally analogous to the purification-augmented swapping trees used for \eps, the GHZ setting requires several additional ingredients: a fidelity model for fusion and GHZ purification, a treatment of the exponentially many intermediate GHZ states created by fusion, and generation schemes that combine first-stage remote-\eps generation with second-stage GHZ construction. We begin by discussing the challenges of extending our algorithms from the previous section to the GHZ states, and then, in the following subsections, describe the fusion and purification models 
and the generation schemes based on dynamic programming and linear programming. 

\para{GHZ States as Star Graph States.}
The GHZ state is a type of multipartite entanglement over some qubits. In particular, we will use the term $r$-GHZ to refer to the GHZ states with $r$ qubits, i.e, $\ket{\text{$r$-GHZ}}=\frac{1}{\sqrt{2}} \left( \ket{0}^{\otimes r} + \ket{1}^{\otimes r} \right)$.
Note that a GHZ state is locally ``equivalent'' to a {\em star} graph state: $\frac{1}{\sqrt{2}} \left( \ket{0}\ket{+}^{\otimes r-1} + \ket{1}\ket{-}^{\otimes r-1} \right)$, the star graph with an edge linking the first qubit (the {\em center} qubit) to each of the other qubits. The choice of the center qubit is arbitrary and can be changed by local operations. {\em To avoid confusion, we generally use the term {\em links} for the quantum network edges, and {\em edges} for the graph state edges.}

\para{Challenges in Generating GHZ States.}
Three key challenges arise when extending the EP-generation schemes from the previous section to GHZ-state generation.

\begin{enumerate}
\item
First, an $r$-GHZ state can be constructed by fusing two smaller GHZ states (in particular, an $r_1$-GHZ and an $r_2$-GHZ state where $r=r_1+r_2-1$); this leads to an exponential number of intermediate GHZ states that can help generate the target GHZ state. 

\item
Second, generating the 2-GHZ states (essentially, \eps) corresponding to the GHZ-edges in the target GHZ state entails generating EPs via swapping and purification of smaller \eps, as in 
the previous section. But, more importantly, for an optimal overall GHZ-generation scheme, we need to ``intertwine'' generation of 2-GHZs (using swapping and purification of smaller \eps) with fusion of smaller GHZs into bigger GHZs. 
The straightforward approach to address the above would be to add the {\em fusion} operation to purification-augmented swapping trees, {\em and} allow for an arbitrary set of leaves (rather than just the links over a network path). However, this approach precludes a direct generalization of our techniques
from the prior section. 

\item
Another obstacle is that a single scalar fidelity does not determine the outcome of a later GHZ purification operation. Two states with the same scalar fidelity can have different weights on the GHZ-basis components, including different phase labels and different bit-error strings, and can consequently have different purification success probabilities and output fidelities.

\end{enumerate}

We address the above challenges with the following strategy. 
We consider {\bf two-stage approaches} wherein the first stage generates the remote EPs (i.e., 2-GHZs) for each GHZ-edge of the target GHZ state, and the second stage generates the target GHZ from the 2-GHZs.
For the {\bf first stage}, we use the techniques from the previous section, while for the {\bf second stage}, we design exponential-time optimal schemes and then reduce their computation time by appropriate optimizations (see \S\ref{sec:ghz-dp}-\ref{sec:ghz-lp}).
To address the third challenge, we represent each intermediate GHZ resource with its GHZ-basis fidelity vector and define fusion and purification as vector-update maps.
We start by describing these fusion and purification models for the GHZ states in the following subsection.

\subsection{\bf GHZ Fusion and Purification Model}
\label{sec:model}

Similar to the swapping and purification operations used for generating \eps, we use fusion and purification operations to generate GHZ states. The objective of this subsection is to give the state representation and operation maps for the fusion and purification operations used by the second-stage GHZ algorithms;
tedious details (e.g., the circuit-level derivations) are deferred to Appendix~\ref{app:purification-detail}.

For GHZ generation, we represent each intermediate state
by its full GHZ-basis fidelity vector, which tracks the error
distribution through fusion and purification. A scalar fidelity
alone does not determine subsequent purification outcomes: two GHZ states with the same fidelity may have different weights
on the GHZ-basis components, including different phase labels
and bit-error strings, and can therefore have different
purification success probabilities and output fidelities.
The vector representation retains this information when
evaluating successive operations.
The same approach could also be used for EPs by tracking
their individual Bell-state components. In the EP formulations
in \S\ref{sec:ep}, we instead retain a scalar representation
to keep the planning model compact.

\para{Vector Representation of the GHZ Fidelity.}
For an $n$-qubit GHZ state, we use the orthonormal basis
\[
    \left|\psi^{(n)}_{p,\mathbf{b}}\right\rangle
    =
    \frac{
        |0,\mathbf{b}\rangle
        +
        (-1)^p|1,\overline{\mathbf{b}}\rangle
    }{\sqrt{2}},
    \qquad
    p\in\{0,1\},\quad
    \mathbf{b}\in\{0,1\}^{n-1},
\]
where $p$ is the phase-error bit and
$\mathbf{b}=(b_1,\ldots,b_{n-1})$ is the ordered bit-error string relative to the first qubit in the GHZ-basis ordering. We represent the state by the GHZ-basis fidelity vector
\[
    \mathbf{P}^{(n)}
    =
    \left\{
        P^{(n)}[p,\mathbf{b}]
    \right\}_{p,\mathbf{b}},
    \qquad
    \sum_{p,\mathbf{b}}P^{(n)}[p,\mathbf{b}]=1.
\]
The fidelity with respect to the ideal GHZ state is the distinguished component
\[
    F^+\!\left(\mathbf{P}^{(n)}\right)
    =
    P^{(n)}[0,\mathbf{0}].
\]
Thus $F^+$ remains the scalar quantity compared with the final fidelity threshold, but every intermediate fusion and purification update is computed from the complete vector $\mathbf{P}$.

In particular, for a Bell-pair source with fidelity $F$, the depolarizing source model used in our NetSquid implementation gives
\[
    \mathbf{P}^{(2)}
    =
    \left(
        F,\frac{1-F}{3},
        \frac{1-F}{3},\frac{1-F}{3}
    \right),
\]
where the entries are ordered as
$(p,b)=(0,0),(0,1),(1,0),(1,1)$. The algorithms do not require the source to be isotropic; an arbitrary measured or simulated Bell-diagonal source vector can be supplied.

\para{Fusion Operations; Fidelity Updates.} 
There are two main fusion operations~\cite{ghaderibaneh2023generation} for generating GHZ states from smaller GHZ states: 
\emph{fusion retain} and \emph{fusion discard}.
In essence, a fusion operation involves two qubits, say $q_i$ and $q_j$, one from each of the given two GHZ states, and yields a GHZ state over all the qubits except for $q_i$ and/or $q_j$; in the case of {\em fusion-retain} operation, only one of the qubits is discarded, while in the case of the {\em fusion-discard} operation, both $q_i$ and $q_j$ are discarded. 
If the qubits $q_i$ and $q_j$ reside in a single network node, the fusion operations require only local (quantum) operations and classical communication~\cite{ghaderibaneh2023generation}. This paper uses only the {\tt fusion-retain} operation.\footnote{In general, both fusion operations can help generate GHZ states as in~\cite{ghaderibaneh2023generation}, but, in our context, the {\em second} stage of generating GHZ states from EPs over GHZ-edges doesn't involve any ``non-terminal'' network nodes. Hence, there is no need to discard any qubits that are involved.}

Below, we describe the sequence of operations in fusion retain and derive the vector update used to compute the fidelity of the resulting GHZ states. The fusion-discard operation and its fidelity update are given in Appendix~\ref{app:purification-detail}.
While prior works~\cite{zwerger2012measurement,pirker2018modular,ghaderibaneh2023generation} discuss fusion operations for creating and distributing GHZ states, they do not provide expressions for the fidelity of the resulting states. 
In contrast, our work provides analytical vector updates for the resulting GHZ state under fusion operations. These expressions are essential for developing and analyzing our fidelity-aware GHZ generation schemes.

\softpara{Fusion Retain.} Consider an $n_1$-GHZ state $\ket{\psi_1}$ and an $n_2$-GHZ state $\ket{\psi_2}$.
The {\tt fusion-retain} operation consists of the following:
\begin{enumerate}
    \item Apply a CNOT from a selected qubit $q_1$ in $\ket{\psi_1}$ to a selected qubit $q_2$ in $\ket{\psi_2}$. 
    \item Measure and discard $q_2$ in the $Z$-basis.
    \item If the measurement outcome is $1$, apply Pauli-$X$ to every retained qubit originating from $\ket{\psi_2}$.
\end{enumerate}
The resulting state contains $n_1+n_2-1$ qubits. The selected qubit $q_1$ from $\ket{\psi_1}$ is retained and is used as the first qubit in the output GHZ-basis ordering, while $q_2$ is measured and discarded. The remaining bit-error labels are ordered by the retained non-reference qubits. For $i\in\{1,2\}$, let $\mathcal{L}_i$ denote the ordered list of such retained non-reference qubits originating from $\ket{\psi_i}$, and let $\mathcal{L}_{\mathrm{out}}$ be the canonical ordered list of the output non-reference qubits. The fusion circuit implements
\[
    p_{\mathrm{out}}=p_1\oplus p_2,
    \qquad
    \mathbf{b}_{\mathrm{out}}
    =
    \Pi_{\mathcal{L}_1,\mathcal{L}_2
    \rightarrow\mathcal{L}_{\mathrm{out}}}
    (\mathbf{b}_1,\mathbf{b}_2),
\]
where $\Pi$ places each bit-error entry from $\mathbf{b}_1$ and $\mathbf{b}_2$ at the position of its corresponding retained non-reference output qubit.

For fixed input strings $\mathbf{b}_1$ and $\mathbf{b}_2$, define
\[
\mathbf{c}
=
\Pi_{\mathcal{L}_1,\mathcal{L}_2
\rightarrow\mathcal{L}_{\mathrm{out}}}
(\mathbf{b}_1,\mathbf{b}_2).
\]
For independent input distributions $\mathbf{P}_1$ and $\mathbf{P}_2$, the output probabilities are
\begin{align*}
P_{\mathrm{out}}[0,\mathbf{c}]
&=
P_1[0,\mathbf{b}_1]P_2[0,\mathbf{b}_2]
+
P_1[1,\mathbf{b}_1]P_2[1,\mathbf{b}_2],
\\
P_{\mathrm{out}}[1,\mathbf{c}]
&=
P_1[0,\mathbf{b}_1]P_2[1,\mathbf{b}_2]
+
P_1[1,\mathbf{b}_1]P_2[0,\mathbf{b}_2].
\end{align*}
For isotropic inputs, this equation reduces to
\[
    \fusionfidelity(F_1,F_2)
    =
    F_1F_2+
    \frac{(1-F_1)(1-F_2)}
    {(2^{n_1}-1)(2^{n_2}-1)}.
\]
This scalar expression is shown as an isotropic special case.

\para{Purification Operations; Fidelity Updates.}
Purification of arbitrary two-colorable graph states, including GHZ states as a special case, has been studied in~\cite{dur2007entanglement, aschauer2005multiparticle}. In particular,\cite{aschauer2005multiparticle} derives the fidelity of GHZ states after purification under imperfect local operations (e.g., noisy CNOT gates). However, these works only consider 'homogeneous' cases in terms of size and fidelity, i.e., they only consider the purification of an $n$-GHZ state using another $n$-GHZ state of the same fidelity. In addition,~\cite{dp_ghz_from_bell20} explores stabilizer-based protocols for distilling multipartite GHZ states, allowing purification using an $m$-GHZ state to improve an $n$-GHZ state (with $m = n$ or $m < n$, where $m$ is even), without requiring identical fidelities. However, this work does not provide an explicit vector formula for the purified GHZ state. In the following, we present vector update rules that characterize the fidelity and success probability of GHZ purification under arbitrary input vectors.

Specifically, we wish to consider more general cases wherein an $n$-GHZ state may be purified using either another $n$-GHZ state or an $n_1$-GHZ ($n_1 \leq n$) state, and the states may have different error distributions. Consequently, our procedure can then be used in a way analogous to ``entanglement pumping''~\cite{dur1999quantum}.
Considering such general purification of GHZ states broadens the search space for our purification-augmented fusion tree (formally defined later), enabling more efficient GHZ-generation strategies. Moreover, by allowing smaller-sized and/or lower-fidelity GHZ states to purify larger-sized and/or higher-fidelity ones, our approach can better leverage available resources in the network and improve overall throughput. Such flexibility makes our protocols more adaptable to realistic and resource-constrained quantum networks.

In the following paragraphs, we develop a fidelity model for a general purification procedure for GHZ states. Note that the purification of EPs can be viewed as a special case of the purification of $n$-qubit GHZ states, where $n=2$. We describe three operations used by the DP: full bit-pattern purification, subset bit-pattern purification, and full phase purification. Suppose we have a set (denoted by $t$, for ``target'') of $n \geq 2$ qubits representing an $n$-GHZ state with vector $\mathbf{P}_{\rm t}$. Let $s$ be a disjoint set of $m$ qubits, where $2\leq m\leq n$, representing a sacrificial $m$-GHZ state with vector $\mathbf{P}_{\rm s}$. We want to purify $t$ to increase its overlap with the ideal GHZ state, $P_{\rm t}[0,\mathbf{0}]$.

\begin{figure}
    \centering
    \includegraphics[width=0.4\textwidth]{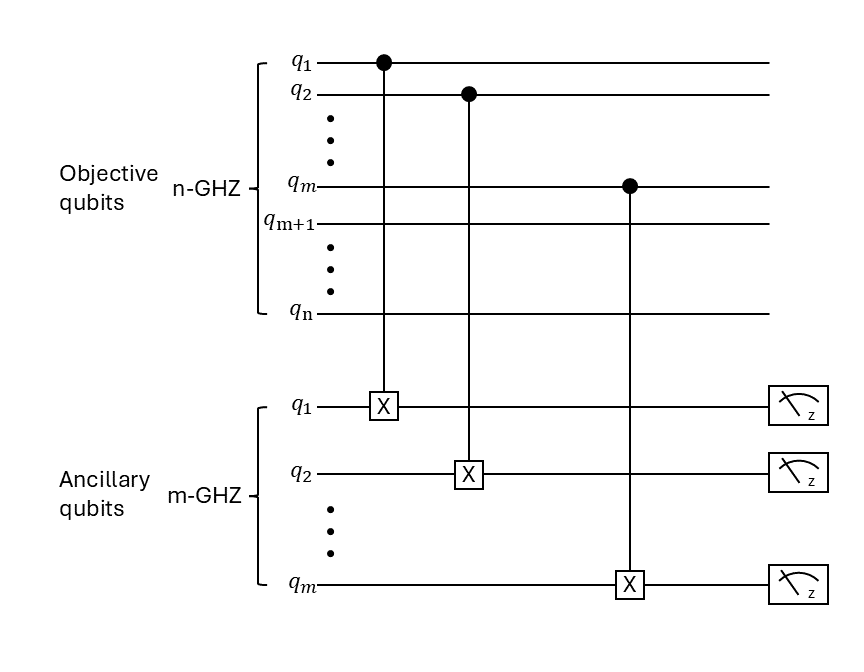}
    \caption{Bit-pattern check purification for an $n$-GHZ target and an $m$-GHZ sacrifice. The checked target qubits control the CNOT gates, the sacrifice qubits are measured in the $Z$ basis, and the target is retained when all sacrifice outcomes are equal.}
    \label{fig:bit_check}
    \vspace*{-0.2in}
\end{figure}

\softpara{Full Bit-Pattern Purification.}
For full bit-pattern purification, the sacrifice has the same size as the target ($m=n$). The CNOTs go from the target qubits to the corresponding sacrifice qubits, and the sacrifice is measured in the $Z$ basis as shown in Fig.~\ref{fig:bit_check}. We deem the bit-pattern purification procedure to have succeeded and retain $t$ if the outcome of \emph{all} the measurements is the same, i.e., all outcomes are either $0$ or all are $1$. If any outcome differs, we deem the procedure failed and discard $t$.

For input labels
$(p_{\rm t},\mathbf{b}_{\rm t})$ and
$(p_{\rm s},\mathbf{b}_{\rm s})$, where the subscripts ${\rm t}$ and ${\rm s}$ denote the target and sacrifice states, respectively, the full bit-pattern purification circuit accepts exactly when their bit-error strings agree, namely
$\mathbf{b}_{\rm t}=\mathbf{b}_{\rm s}$. Under this acceptance condition, the accepted output label is
\[
    p_{\rm out}=p_{\rm t}\oplus p_{\rm s},
    \qquad
    \mathbf{b}_{\rm out}=\mathbf{b}_{\rm t}.
\]
Thus, for two independent equal-size input vectors, the unnormalized accepted output vector, success probability, and normalized output vector are
\begin{align*}
    \widetilde{P}_{\rm B}[p,\mathbf{b}]
    &=
    \sum_{p_{\rm t}\oplus p_{\rm s}=p}
        P_{\rm t}[p_{\rm t},\mathbf{b}]
        P_{\rm s}[p_{\rm s},\mathbf{b}],
    \\
    q_{\rm B}
    &=
    \sum_{p,\mathbf{b}}
        \widetilde{P}_{\rm B}[p,\mathbf{b}],
    \\
    P_{\rm B}[p,\mathbf{b}]
    &=
    \frac{\widetilde{P}_{\rm B}[p,\mathbf{b}]}{q_{\rm B}}.
\end{align*}

Note that for the EPs ($n=m=2$) and in the homogeneous isotropic case ($F_t=F_s=F$), the full bit-pattern purification vector update gives
\[
F_{out}=\frac{F^2 + \frac{(1-F)^2}{9}}
{F^2 + 2\frac{F(1-F)}{3} + 5\frac{(1-F)^2}{9}}
\]
which matches with the Bennett et al.'s formula~\cite{bennett-95}.

\softpara{Subset Bit-Pattern Purification.}
For subset bit-pattern purification, the sacrifice size is smaller than the target size. We write the sacrifice size as $k+1<n$ and, for clarity of notation, write the checked target qubits as $q_0,q_1,\ldots,q_k$, where $q_0$ is the first qubit in the GHZ-basis ordering. The implementation uses the actual selected physical qubits and does not assume that all qubits have identical noise.

For input labels
$(p_{\rm t},\mathbf{b}_{\rm t})$ for the $n$-qubit target and
$(p_{\rm s},\mathbf{b}_{\rm s})$ for the $(k+1)$-qubit sacrifice, write
$\mathbf{b}_{\rm t}=(b_{{\rm t},1},\ldots,b_{{\rm t},n-1})$ and
$\mathbf{b}_{\rm s}=(b_{{\rm s},1},\ldots,b_{{\rm s},k})$. The subset bit-pattern check accepts when the sacrifice bit-error string equals the checked target substring,
\[
    (b_{{\rm t},1},\ldots,b_{{\rm t},k})
    =
    \mathbf{b}_{\rm s}.
\]
The target bit-error string is retained, while the phase label is updated:
\[
    p_{\rm out}=p_{\rm t}\oplus p_{\rm s},
    \qquad
    \mathbf{b}_{\rm out}=\mathbf{b}_{\rm t}.
\]
Hence, writing the output bit-error string as
$\mathbf{b}=(b_1,\ldots,b_{n-1})$,
\begin{align*}
    \widetilde{P}_{{\rm B},k}[p,\mathbf{b}]
    &=
    \sum_{p_{\rm t}\oplus p_{\rm s}=p}
        P_{\rm t}[p_{\rm t},\mathbf{b}]
        P_{\rm s}[p_{\rm s},(b_1,\ldots,b_k)],
    \\
    q_{{\rm B},k}
    &=
    \sum_{p,\mathbf{b}}
        \widetilde{P}_{{\rm B},k}[p,\mathbf{b}],
    \\
    P_{{\rm B},k}[p,\mathbf{b}]
    &=
    \frac{\widetilde{P}_{{\rm B},k}[p,\mathbf{b}]}
    {q_{{\rm B},k}}.
\end{align*}

\begin{figure}
    \centering
    \includegraphics[width=0.4\textwidth]{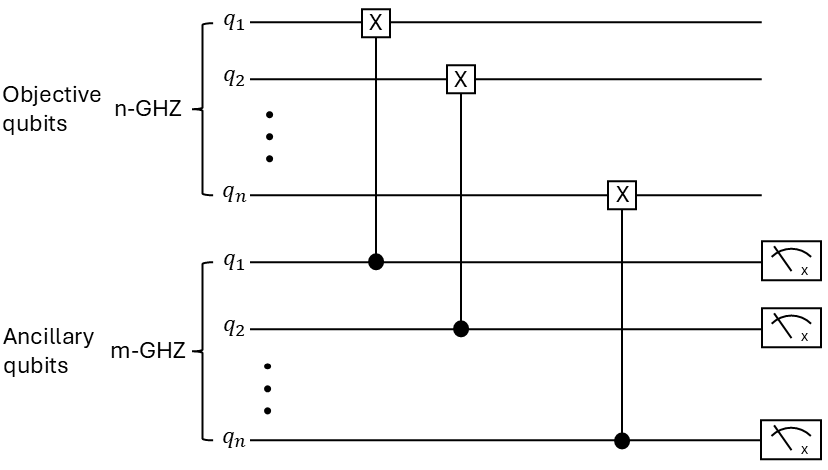}
    \caption{Full phase-check purification for two equal-size GHZ states. The sacrifice qubits control the CNOT gates, are then measured in the $X$ basis, and the target is retained when the measurement parity is even.}
    \label{fig:phase_check}
\end{figure}

\softpara{Full Phase Purification.}
Full phase purification uses two equal-size GHZ states ($m=n$). The CNOT direction is reversed relative to the bit-pattern check: every sacrifice qubit controls the corresponding target qubit. The sacrifice qubits are measured in the $X$ basis, and the target is retained when the measurement parity is
even; see Fig.~\ref{fig:phase_check}.

For input labels
$(p_{\rm t},\mathbf{b}_{\rm t})$ and
$(p_{\rm s},\mathbf{b}_{\rm s})$, the full phase purification circuit accepts exactly when their phase labels agree, namely $p_{\rm t}=p_{\rm s}$. Under this acceptance condition, the accepted output label is
\[
    p_{\rm out}=p_{\rm t},
    \qquad
    \mathbf{b}_{\rm out}
    =
    \mathbf{b}_{\rm t}\oplus\mathbf{b}_{\rm s}.
\]
Thus
\begin{align*}
    \widetilde{P}_{\rm H}[p,\mathbf{b}]
    &=
    \sum_{\mathbf{b}_{\rm t}\oplus\mathbf{b}_{\rm s}=\mathbf{b}}
        P_{\rm t}[p,\mathbf{b}_{\rm t}]
        P_{\rm s}[p,\mathbf{b}_{\rm s}],
    \\
    q_{\rm H}
    &=
    \sum_{p,\mathbf{b}}
        \widetilde{P}_{\rm H}[p,\mathbf{b}],
    \\
    P_{\rm H}[p,\mathbf{b}]
    &=
    \frac{\widetilde{P}_{\rm H}[p,\mathbf{b}]}{q_{\rm H}}.
\end{align*}

\begin{figure}[t]
    \centering
    \vspace*{-0.1in}
    \includegraphics[width=0.45\textwidth]{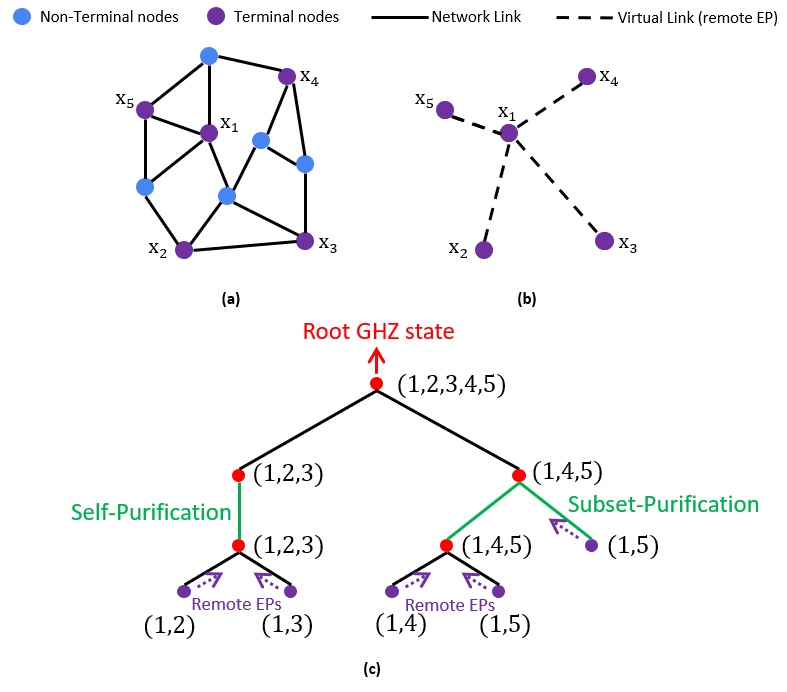}
    \caption{Concepts used in our GHZ generation scheme. (a) The quantum network with terminal nodes in purple; these are the nodes in the network over which the desired/target 4-qubit GHZ state is distributed. (b) Star graph \stargraph for the 4-qubit/node GHZ state with vertices numbered as $\{v_1, v_2, v_3, v_4\}$; here, the choice of central node $v_1$ and the numbering of the remaining vertices may be done strategically (see \S\ref{sec:ghz-dp}). (c) A purification-augmented fusion tree over the star graph edges; here, all the binary operations are {\tt fusion-retain} except for the one marked in green as {\tt subset-purification}. For clarity, we only use the subscript number $i$ to denote the node $v_i$.}
    \vspace*{-0.2in}
    \label{fig:2-stage-ghz}
\end{figure}

\subsection{\textbf{Dynamic Programming for GHZ Distribution}}
\label{sec:ghz-dp}

In this subsection, we formally define the problem representing
the second stage of our GHZ generation approach. Then, we describe our optimal dynamic programming algorithm and several computationally 
efficient strategies. First, we introduce some notations, describe our high-level two-stage approach to generate GHZ states, and develop the notion of  
purification-augmented fusion trees.

\para{Star Graph $\stargraph$ for the Target GHZ State.} 
Recall that the quantum network is $G(V, E)$. Let $\starset \subseteq V(G)$ be a set of network nodes (referred to as {\em terminal nodes}) over which we wish to distribute a GHZ state (one qubit over each node in \starset).
As mentioned before, a (distributed) GHZ state can be looked upon as a (distributed) 
star graph state. For clarity of presentation, let us assume $\starset = \{v_1, v_2, \ldots, v_m\}$, where each $v_i$ is a network node, and let us denote the (distributed) {\em star graph state} over 
$\starset$ be the ``overlay'' graph $\stargraph(V=\starset, E = \{(v_1, v_i) | 2 \leq i \leq m\})$. See Fig.~\ref{fig:2-stage-ghz}.

\para{Two-Stage Strategy.} To generate high-fidelity GHZ states, we apply a simplified strategy inspired by the fact that a GHZ state is locally equivalent to a star graph state. In particular, as mentioned earlier in this section, we will use the following two-stage generation scheme~\cite{fan2024optimized}. See Fig.~\ref{fig:2-stage-ghz}.
\begin{enumerate}
\item
In the first stage (Figs.~\ref{fig:2-stage-ghz}(a) and~(b)), we generate (remote) EPs over the edges in the star graph state \stargraph\ (corresponding to the target GHZ state). Thus, this first stage can be treated as an instance of the \gfed problem (\S\ref{sec:ep}) of generating EPs over the edges in \stargraph.\footnote{The first
stage can also be solved using an iterated version of the \fed problem, wherein we iteratively determine the optimal purification-augmented swapping tree in the residual network graph for each of the edges in \stargraph.}

\item
In the second stage (Fig.~\ref{fig:2-stage-ghz}(c)), we perform fusion and purification operations on the \epss being generated in the first stage to generate the target GHZ states. This is done using an appropriate purification-augmented fusion tree (defined below) over the edges of \stargraph, the star graph state. In this subsection (and the next), our focus is solely on designing efficient protocols for this second stage---i.e., of generating the target GHZ states from the EPs over the star graph edges.
\end{enumerate}

Before formally defining our second-stage GHZ generation problem, we define purification-augmented fusion trees, a variation of the purification-augmented swapping tree defined in \S\ref{sec:problem-a}.

\para{Purification-Augmented Fusion Tree.} Informally, a purification-augmented fusion tree $T$ over a given star graph \stargraph is a tree with edges of \stargraph as the leaves (representing the EPs or 2-GHZs over them), and 
the root of $T$ corresponds to the target GHZ state. The tree represents a sequence of {\tt fusion-retain}, {\tt full bit-pattern purification}, {\tt subset bit-pattern purification}, and {\tt full phase purification} operations over the leaves. Each intermediate node is associated with a GHZ-basis fidelity vector $\mathbf{P}$, and its scalar fidelity is evaluated only as the ideal component $F^+(\mathbf{P})=P[0,\mathbf{0}]$.
More formally, a purification-augmented fusion tree over a given star graph \stargraph has four structural types of nodes:
\begin{itemize}
    \item {\em Leaves for \stargraph\ Edges.} Leaves of the form $\id{EP}(v_1,v_i)$ representing \epss (2-GHZ states) over the edge $(v_1,v_i)$ in the star graph \stargraph. The corresponding source vector is the Bell/GHZ-basis vector $\mathbf{P}^{(2)}$ defined in \S\ref{sec:model}.
    
    \item {\em Fused Nodes.} Internal nodes of the form $\id{fuse}(S_i,S_j)$ representing a binary {\tt fusion-retain} operation over GHZ states on $(S_i\cup\{v_1\})$ and $(S_j\cup\{v_1\})$, where $S_i$ and $S_j$ are disjoint subsets of $\{v_2,\ldots,v_m\}$. Such a node has two children, and its output vector is given by the fusion vector update in \S\ref{sec:model}, after permuting the output bit-string order to the canonical order of $S_i\cup S_j$.
    
    \item {\em Full-Purify Nodes.} Internal nodes of the form
    $\id{full\_purify}(S,\mathsf{op})$, where $\mathsf{op}\in\{\mathsf{Bit},\mathsf{Phase}\}$, denoting purification of a GHZ state over $S\cup\{v_1\}$ using a sacrificial GHZ state over the same qubit set. This is one structural node type; the operation label $\mathsf{op}$ determines whether the node applies the full bit-pattern check update or the full phase-check update of \S\ref{sec:model}.
    
    \item {\em Subset Bit-Pattern Purify Nodes.} Internal nodes of the form
    $\id{subset\_purify}(S,S')$ with $S'\subset S$, denoting subset bit-pattern purification of a target GHZ state over $S\cup\{v_1\}$ using a sacrificial GHZ state over $S'\cup\{v_1\}$. The operation uses the subset bit-pattern check vector update defined in \S\ref{sec:model}.
\end{itemize} 

\para{(Second-Stage) GHZ Distribution with Purification (\gdp) Problem.} 
Consider a star graph $\stargraph(V= \{v_1, v_2, \ldots, v_m\}, E = \{(v_1, v_i)\ |\ 2 \leq i \leq m\})$, with each edge $(v_1, v_i)$ associated with an \eps generation rate $r_i$ and a source vector $\mathbf{P}^{(2)}_i$, for $i = 2, 3, \dots, m$. Given
a fidelity threshold $\fth$, the \gdp problem is to determine a (single) purification-augmented fusion tree that generates the $m$-GHZ state over $V(\stargraph) = \{v_1, v_2, \ldots, v_m\}$ with the maximum generation 
rate and final ideal-component fidelity $F^+(\mathbf{P}_{\rm out})\geq \fth$. \hfill $\Box$

\para{Discretizing Rate vs.\ Fidelity}.
We consider two discretization choices: fidelity and
generation rate. 
Our hypergraph-flow LP formulations require fidelity
discretization, while DP formulations can use either fidelity
or rate discretization.
To illustrate both choices, we presented the
EP DP in \S\ref{sec:ep} using fidelity discretization and
present the GHZ DP below using rate discretization.
Fidelity-discretized GHZ DP variants are briefly discussed
at the end of this subsection.

The restriction of our LP formulations to fidelity discretization
follows from their additive flow structure: generation rates
can be aggregated and split across operations, whereas fidelities cannot.
Consequently, discretizing rates and using fidelities as the
flow variables would not yield a linear formulation, since
fidelity evolves through nonlinear operation-specific updates.
We therefore discretize fidelity labels and retain generation
rates as continuous flow variables. The fidelity updates and
associated success probabilities are evaluated when constructing
the hypergraph, leaving the LP linear in the rate variables.

\para{Optimal DP Algorithm for \gdp Problem.}
For our DP algorithm, we start by defining the subproblems as follows.

\softpara{Subproblems $\mathcal{T}(S, k, \overrightarrow{EP})$.}
For each triplet
$\langle S,k,\overrightarrow{EP}\rangle$, we let
$\mathcal{T}(S,k,\overrightarrow{EP})$ denote a frontier of feasible annotated fusion trees. Each tree
$T\in\mathcal{T}(S,k,\overrightarrow{EP})$ generates a GHZ state over $S\cup\{v_1\}$ at a rate of at least $k$ while consuming an EP-rate vector bounded by $\overrightarrow{EP}$. The annotation of $T$ stores the GHZ-basis fidelity vector at every node; in particular, $\mathbf{P}_T$ denotes the vector stored at the root. We write
\[
    F^+(\mathbf{P}) = P[0,\mathbf{0}]
\]
for the ideal-component fidelity represented by $\mathbf{P}$. The optimal DP stores a frontier of annotated trees rather than a single tree $T(S,k,\overrightarrow{EP})$, because two trees with the same current value of $P[0,\mathbf{0}]$ can have different weights on the other GHZ-basis components and can therefore behave differently under later purification operations.

\softpara{Discrete Rate Values.} Above, the \emph{rate} parameter is drawn from a finite set of discrete values $\discreterate = \{K_1, K_2, \ldots, K_q\}$, and, as in the previous section, for an arbitrary rate value $r_i$, $\lfloor r_i \rfloor$ represents the largest $K_i \in \discreterate$ such that $K_i \leq r_i$, and $\lceil r_i \rceil$ denotes the smallest $K_i \in \discreterate$ satisfying $K_i \geq r_i$. 
In the recursive equations defined later, sometimes the rate parameter
may take a value larger than the highest discretized value $K_q$; to handle this situation, we define $\lceil r_i \rceil$ for $r_i > K_q$ 
to be a special value $\spvalue$, for which the corresponding annotated-tree frontier is empty.

\softpara{Base Case.}
The base case initializes the subproblems for sets $S$ with a single element using the source to generate 2-GHZs/EPs for an edge $(v_1,v_j)\in\stargraph$. In this case, the frontier
$\mathcal{T}(\{v_j\},k,\overrightarrow{EP})$ is initialized with the single-node tree corresponding to this 2-GHZ, whose root vector is the source vector $\mathbf{P}^{(2)}_j$, if $k$ and $\overrightarrow{EP}$ satisfy the
conditions given below.

\begin{eqnarray*}
\mathcal{T}(S, \spvalue, \overrightarrow{EP}) &=& \emptyset, \mbox{ where \spvalue is as defined above}. \\
\mathcal{T}(S, k, \overrightarrow{EP}) &=& \emptyset, \mbox{ if } \overrightarrow{EP}=\langle 0, \ldots, 0\rangle \\
\mathcal{T}(\{v_j\}, k, \overrightarrow{EP}) &=& \emptyset, \mbox{ if } k > \floor{r_j}\\
\mathcal{T}(\{v_j\}, k, \overrightarrow{EP}) &=& \{T_j\}, \mbox{ if } k \le \floor{r_j} \land \langle 0, \ldots, k, \ldots, 0\rangle\leq\overrightarrow{EP}
\end{eqnarray*} 
where the root vector of $T_j$ is $\mathbf{P}_{T_j}=\mathbf{P}^{(2)}_j$.

\softpara{Recursive Step.}
There are three cases to consider for the recursive step.
\begin{enumerate}
    \item The last operation in $T(S, k, \overrightarrow{EP})$ is a {\em fusion} operation, i.e., the GHZ state over $S\cup\{v_1\}$ is formed by fusing two smaller GHZ states over $S_1\cup\{v_1\}$ and $S_2\cup\{v_1\}$, where $S = S_1\cup S_2$.
    
    \item The last operation in $T(S,k,\overrightarrow{EP})$ is a {\em full bit-pattern purification} or {\em full phase purification} operation. In this case, the target and sacrificial inputs are both $r$-GHZ states over the same qubit set $S\cup\{v_1\}$. The two inputs may either be two copies generated by the same child tree, or copies generated by two independently generated child trees over the same qubit set.

    \item The last operation in $T(S,k,\overrightarrow{EP})$ is a {\em subset bit-pattern purification} operation. In this case, the target is an $r$-GHZ state over $S\cup\{v_1\}$, and the sacrifice is an $r'$-GHZ state over $S'\cup\{v_1\}$ for a strict subset $S'\subset S$.
\end{enumerate}
In all the above cases, the children of the root in $T(S, k, \overrightarrow{EP})$ are selected from the appropriate child frontiers. Thus, the \emph{optimal-substructure property} holds at the level of frontiers: a parent annotated tree is obtained by choosing feasible child annotated trees from the relevant subproblems and applying one root operation, whose output vector is determined by the child root vectors.
Incorporating the above three cases, we can write the following recurrence for the annotated-tree frontier $\mathcal{T}(S,k,\overrightarrow{EP})$ when $|S|>1$:
\begin{align}
    \mathcal{T}(S, k, \overrightarrow{EP})
    =
    \operatorname{Prune}
    \left(
    \mathcal{C}_f
    \cup
    \mathcal{C}_{p1}^{\mathsf{Bit}}
    \cup
    \mathcal{C}_{p1}^{\mathsf{Phase}}
    \cup
    \mathcal{C}_{p2}^{\mathsf{Bit}}
    \cup
    \mathcal{C}_{p2}^{\mathsf{Phase}}
    \right).
    \label{eqn:dp-GHZ}
\end{align}
\noindent
where $\mathcal{C}_f$, $\mathcal{C}_{p1}^{\mathsf{Bit}}$,
$\mathcal{C}_{p1}^{\mathsf{Phase}}$,
$\mathcal{C}_{p2}^{\mathsf{Bit}}$, and
$\mathcal{C}_{p2}^{\mathsf{Phase}}$ are defined as follows. The operator $\operatorname{Prune}$ removes repeated root vectors: if two annotated trees in the set have the same root vector, we keep one tree and discard the other. For $|S|=1$, the frontier includes feasible trees that use the source EP directly or purify copies of it. We use the notation $\overrightarrow{EP_1} \le \overrightarrow{EP}$ to indicate an element-wise comparison, i.e., where each element in $\overrightarrow{EP_1}$ is less than or equal to its counterpart in $\overrightarrow{EP}$.

\begin{footnotesize}
\begin{eqnarray*}
\mathcal{C}_f
&=&
\left\{\mathsf{root}_{\mathsf{Fuse}}(T_1,T_2)
\;\middle|\;
\begin{array}{l}
\emptyset\neq S_1\subset S,\ S_2=S\setminus S_1,\\ 
\overrightarrow{EP_1}+\overrightarrow{EP_2}\le\overrightarrow{EP},\\
T_1\in\mathcal{T}\!\left(S_1,\left\lceil \dfrac{3}{2p_f}k\right\rceil,\overrightarrow{EP_1}\right),\\
T_2\in\mathcal{T}\!\left(S_2,\left\lceil \dfrac{3}{2p_f}k\right\rceil,\overrightarrow{EP_2}\right)
\end{array}
\right\},
\\[1ex]
\mathcal{C}_{p1}^{\mathsf{op}}
&=&
\left\{\mathsf{root}_{\mathsf{op}}(T,T)
\;\middle|\;
\begin{array}{l}
\mathsf{op}\in\{\mathsf{Bit},\mathsf{Phase}\},\\
T\in\mathcal{T}\!\left(S,\left\lceil \dfrac{2}{p_p}k\right\rceil,\overrightarrow{EP}\right)
\end{array}
\right\},
\\[1ex]
\mathcal{C}_{p2}^{\mathsf{Bit}}
&=&
\left\{\mathsf{root}_{\mathsf{Bit}}(T_{\rm t},T_{\rm s})
\;\middle|\;
\begin{array}{l}
\emptyset\neq S'\subseteq S,\\
\overrightarrow{EP}_{\rm t}+\overrightarrow{EP}_{\rm s}\le\overrightarrow{EP},\\
T_{\rm t}\in\mathcal{T}\!\left(S,\left\lceil \dfrac{3}{2p_p}k\right\rceil,\overrightarrow{EP}_{\rm t}\right),\\
T_{\rm s}\in\mathcal{T}\!\left(S',\left\lceil \dfrac{3}{2p_p}k\right\rceil,\overrightarrow{EP}_{\rm s}\right)
\end{array}
\right\},
\\[1ex]
\mathcal{C}_{p2}^{\mathsf{Phase}}
&=&
\left\{\mathsf{root}_{\mathsf{Phase}}(T_{\rm t},T_{\rm s})
\;\middle|\;
\begin{array}{l}
\overrightarrow{EP}_{\rm t}+\overrightarrow{EP}_{\rm s}\le\overrightarrow{EP},\\
T_{\rm t}\in\mathcal{T}\!\left(S,\left\lceil \dfrac{3}{2p_p}k\right\rceil,\overrightarrow{EP}_{\rm t}\right),\\
T_{\rm s}\in\mathcal{T}\!\left(S,\left\lceil \dfrac{3}{2p_p}k\right\rceil,\overrightarrow{EP}_{\rm s}\right)
\end{array}
\right\}.
\end{eqnarray*}
\end{footnotesize}
Above, $\mathsf{root}_{\mathsf{op}}$ denotes the annotated tree obtained by attaching operation $\mathsf{op}$ to the child tree or trees. The root vector of $\mathsf{root}_{\mathsf{op}}$ is computed by the corresponding vector-update map in \S\ref{sec:model}; in bit-pattern purification, the first child is the target and the second child is the sacrifice, and the qubit set of the sacrificial child determines which bit positions are checked.
$p_f$ is the success probability of fusion, and $p_p$ is the success probability of the corresponding purification operation.\footnotemark[\getrefnumber{fn:pp}]
The terms $\mathcal{C}_{p1}^{\mathsf{op}}$ represent same-stream full purification: two copies are generated by the same child tree, so the EP-consumption vector is not duplicated and the rate factor is $p_p/2$. The terms
$\mathcal{C}_{p2}^{\mathsf{Bit}}$ and
$\mathcal{C}_{p2}^{\mathsf{Phase}}$ represent two-stream purification:
the input states are generated by separately allocated child trees, so their EP-consumption vectors add, and the synchronization factor is $2p_p/3$.

\softpara{DP Solution.} 
Given the fidelity threshold $\fth$, the maximum rate $k_m$ is the largest discretized rate for which at least one annotated tree in the final frontier has ideal-component fidelity above the threshold:

\begin{equation*}
k_m = \max \left\{
    k \in \discreterate
    \;\middle|\;
    \begin{aligned}
    & \exists \overrightarrow{EP} \leq
    \langle \floor{r_2}, \floor{r_3}, \ldots, \floor{r_m} \rangle,\\
    & \exists T\in
    \mathcal{T}(\{v_2, v_3, \ldots, v_m\}, k, \overrightarrow{EP}):\\
    &F^+(\mathbf{P}_T) \geq \fth
    \end{aligned}
\right\}.
\end{equation*}

\para{Performance Guarantee.}
We now show that the DP formulation is optimal for the \gdp problem under the adopted discretized-rate model.

\begin{thm-prf}
\label{thm:ghz-dp}
For any instance of the \gdp problem over the star graph $\stargraph$, the dynamic programming formulation based on the frontiers $\mathcal{T}(S,k,\overrightarrow{EP})$ and the recurrence in Eqn.~\ref{eqn:dp-GHZ} returns an optimal solution to the \gdp problem under the adopted discretized-rate model.
\end{thm-prf}

The proof is given in Appendix~\ref{app:ghz-dp-heuristics}.

\para{Time Complexity.}
To analyze the time complexity of the above algorithm, observe that the subproblem $\mathcal{T}(S,k,\overrightarrow{EP})$ is indexed by three quantities:
\begin{itemize}
\item the subset $S$, for which there are $O(2^{m-1})$ possibilities;
\item the discretized rate $k$, which has $O(|\discreterate|)$ possible values; and
\item the EP-consumption vector $\overrightarrow{EP}$, which has $O(|\discreterate|^{m-1})$ possible values.
\end{itemize}
Thus, the number of subproblems is $O(2^{m-1}\cdot |\discreterate|^m)$. Computing a single subproblem requires enumerating the possible subset decompositions and EP-consumption allocations that arise in the $\mathcal{C}_f$,
$\mathcal{C}_{p2}^{\mathsf{Bit}}$, and
$\mathcal{C}_{p2}^{\mathsf{Phase}}$ terms.

Moreover, each subproblem stores a frontier of root vectors. Therefore, the actual running time and memory also scale with the number of distinct vectors retained in these frontiers and with the cost of applying the GHZ-basis vector-update maps. Since an $r$-GHZ vector has $2^r$ entries, this vector-update cost is exponential in the GHZ size in the worst case.

This optimal DP has exponential dependence on the target GHZ size $m$ and is therefore intended for small-to-moderate GHZ instances. This motivates the computationally efficient DP formulations introduced next. These formulations trade optimality for substantially lower planning overhead while preserving the GHZ-basis vector update model, as confirmed in Section~\ref{sec:eval}.

\para{Computationally Efficient DP Formulations.}
The computational savings come from this reduced search state rather than from a reduced physical model.
In these rate-discretized formulations, each reduced DP cell stores one selected annotated tree instead of a full frontier; when multiple candidates map to the same cell, we retain the highest-ranked one according to $F^+(\mathbf{P})=P[0,\mathbf{0}]$.
These formulations still carry the full GHZ-basis vector $\mathbf{P}$ at every intermediate state and apply the vector update rules from \S\ref{sec:model}. 
In the equations below, $\mathsf{Fuse}$, $\mathsf{Bit}$, and $\mathsf{Phase}$ denote the vector-update maps for fusion retain, bit-pattern purification, and phase purification, respectively, as defined in \S\ref{sec:model}. For $\mathsf{Bit}(\mathbf{P}_{\rm t},\mathbf{P}_{\rm s})$, the first argument is the target and the second is the sacrifice; the sacrifice's qubit set determines the checked bit positions.

\begin{enumerate}
\item{}
\textbf{Bounded Purification (\DPOT).} 
In this scheme, we allow all purification operations but allow at most $C$ purification operations between two fusion operations.
The subproblem is to determine the selected tree $T(S,k,\overrightarrow{EP},c)$ and its root vector $\mathbf{P}(S,k,\overrightarrow{EP},c)$ that generate a GHZ state over $S\cup\{v_1\}$ at rate at least $k$ while consuming an EP-rate vector bounded by $\overrightarrow{EP}$. Here $c$ indicates the number of purification operations since the last fusion operation. We use the following recurrence; the base case is omitted for clarity of presentation.

\begin{align*}
    \mathbf{P}(S,k,\overrightarrow{EP},0)
    &=
    \mathbf{P}_f,\\
    \mathbf{P}(S,k,\overrightarrow{EP},c)
    &=
    \operatorname{Select}_{F^+}
    \left\{
    \begin{aligned}
    \mathbf{P}_{p1}^{\mathsf{Bit},c},
    \mathbf{P}_{p1}^{\mathsf{Phase},c},
    \\
    \mathbf{P}_{p2}^{\mathsf{Bit},c},
    \mathbf{P}_{p2}^{\mathsf{Phase},c}
    \end{aligned}
    \right\},
    1 \le c \le C .
\end{align*}
\vspace*{-0.2in}
\begin{footnotesize}
\begin{eqnarray*}
&& \hspace*{-0.4in} \mathbf{P}_f =
\underset{
\substack{
\emptyset\neq S_1\subset S\\
\overrightarrow{EP}_1+\overrightarrow{EP}_2\leq\overrightarrow{EP}\\
0\le c_1,c_2\le C
}}
{\operatorname{Select}_{F^+}}
\mathsf{Fuse}\!\left(
\begin{aligned}
&\mathbf{P}\!\left(
S_1,\left\lceil \tfrac{3}{2p_f}k \right\rceil,
\overrightarrow{EP}_1,c_1
\right),\\
&\mathbf{P}\!\left(
S\setminus S_1,\left\lceil \tfrac{3}{2p_f}k \right\rceil,
\overrightarrow{EP}_2,c_2
\right)
\end{aligned}
\right),
\\[1ex]
&& \hspace*{-0.4in} \mathbf{P}_{p1}^{\mathsf{Bit},c} =
\mathsf{Bit}\!\left(
\begin{aligned}
&\mathbf{P}\!\left(
S,\left\lceil \tfrac{2}{p_p}k \right\rceil,
\overrightarrow{EP},c-1
\right),\\
&\mathbf{P}\!\left(
S,\left\lceil \tfrac{2}{p_p}k \right\rceil,
\overrightarrow{EP},c-1
\right)
\end{aligned}
\right),
\\[1ex]
&& \hspace*{-0.4in} \mathbf{P}_{p1}^{\mathsf{Phase},c} =
\mathsf{Phase}\!\left(
\begin{aligned}
&\mathbf{P}\!\left(
S,\left\lceil \tfrac{2}{p_p}k \right\rceil,
\overrightarrow{EP},c-1
\right),\\
&\mathbf{P}\!\left(
S,\left\lceil \tfrac{2}{p_p}k \right\rceil,
\overrightarrow{EP},c-1
\right)
\end{aligned}
\right),
\\[1ex]
&& \hspace*{-0.4in} \mathbf{P}_{p2}^{\mathsf{Bit},c} =
\underset{
\substack{
\emptyset\neq S_2\subseteq S\\
\overrightarrow{EP}_0+\overrightarrow{EP}_2\leq\overrightarrow{EP}\\
0\le c_2\le C
}}
{\operatorname{Select}_{F^+}}
\mathsf{Bit}\!\left(
\begin{aligned}
&\mathbf{P}\!\left(
S,\left\lceil \tfrac{3}{2p_p}k \right\rceil,
\overrightarrow{EP}_0,c-1
\right),\\
&\mathbf{P}\!\left(
S_2,\left\lceil \tfrac{3}{2p_p}k \right\rceil,
\overrightarrow{EP}_2,c_2
\right)
\end{aligned}
\right),
\\[1ex]
&& \hspace*{-0.4in} \mathbf{P}_{p2}^{\mathsf{Phase},c} =
\underset{
\substack{
\overrightarrow{EP}_0+\overrightarrow{EP}_2\leq\overrightarrow{EP}\\
0\le c_2\le C
}}
{\operatorname{Select}_{F^+}}
\mathsf{Phase}\!\left(
\begin{aligned}
&\mathbf{P}\!\left(
S,\left\lceil \tfrac{3}{2p_p}k \right\rceil,
\overrightarrow{EP}_0,c-1
\right),\\
&\mathbf{P}\!\left(
S,\left\lceil \tfrac{3}{2p_p}k \right\rceil,
\overrightarrow{EP}_2,c_2
\right)
\end{aligned}
\right).
\end{eqnarray*}
\end{footnotesize}

\item{} \textbf{Self-Purification (\DPss).} 
In this scheme, we disallow purification of a GHZ state using a smaller GHZ state, and thus only allow purification of a GHZ state using another GHZ state over the same set of nodes.
It uses the same rate/resource-indexed cells as above, but without the purification counter, and only considers fusion, same-stream full bit-pattern purification, and same-stream full phase purification. The explicit recurrence is given in Appendix~\ref{app:ghz-dp-heuristics}.

\item{} \textbf{EP-based purification (\DPtwoG):}
In this scheme, in addition to fusion operations, we use the same $(S,k,\overrightarrow{EP})$ cells as \DPss{} but allow same-stream full phase purification and restrict bit-pattern purification to EP-sized checks. Thus, when the target has more than two qubits, the sacrificial input for bit-pattern purification must be a 2-GHZ state; for a 2-GHZ target, the same rule reduces to ordinary EP purification. The explicit recurrence is given in Appendix~\ref{app:ghz-dp-heuristics}.

\end{enumerate}

\para{Fidelity-Discretized DP Formulations.}
The three rate-discretized formulations \DPOT{}, \DPss{}, and \DPtwoG{} also have fidelity-discretized counterparts, denoted \GHZDPFBP{}, \GHZDPFSS{}, and \GHZDPFtwoG{}, respectively. These fidelity-discretized variants use the same physical operation sets
as their rate-discretized versions, and also use the full GHZ-basis vector from \S\ref{sec:model}.
However, instead of defining subproblems for each rate value $k$, they define subproblems for each discretized value $f$ of $F^+(\mathbf{P})=P[0,\mathbf{0}]$.
The bounded version \GHZDPFBP{} also keeps the purification counter $c$, and thus defines subproblems for each combination of $(S,f,c)$. Each subproblem's solution is the candidate fusion tree with the highest generation rate; we store its full GHZ-basis vector for subsequent vector updates and retain its EP-consumption vector for feasibility checks, rather than using a separate table index.

\subsection{\textbf{Linear Programming for Generalized GHZ Distribution}}
\label{sec:ghz-lp}

We now extend the GHZ formulation from a single purification-augmented fusion tree to a level-based structure that aggregates multiple concurrent fusion trees, and we formulate the corresponding generalized GHZ distribution problem.

\para{Level-Based Fusion Structure.}
As in the level-based swapping structure used to generate \epss{}, we can aggregate multiple concurrent purification-augmented fusion trees into a single structure. We refer to this aggregate as a \emph{level-based fusion structure}. The bottom level consists of \epss{} (i.e., 2-GHZ states) on the edges of the star graph \stargraph; the quantum network generates these \epss{} in the first stage of the GHZ generation process. Higher levels then generate larger GHZ states by applying fusion and purification operations to GHZ states produced at earlier levels.

\para{Generalized \gdp{} (\ggdp) Problem.}
Consider a star graph $\stargraph=(V,E)$ with
$V=\{v_1,v_2,\ldots,v_m\}$ and $E=\{(v_1,v_i)\mid 2\leq i\leq m\}$, where each edge $(v_1,v_i)$ is associated with an \eps{} generation rate $r_i$ and a 2-GHZ source vector $\mathbf{P}^{(2)}_i$, for $i=2,\ldots,m$. Given a fidelity threshold $\fth$, the \ggdp{} problem is to determine a level-based fusion structure that generates the $m$-GHZ state over $V(\stargraph)$ with ideal-component fidelity at least $\fth$ and maximum possible generation rate. \hfill $\Box$

\para{Optimal LP Formulation for \ggdp{} Problem (\GHZLPOpt).}
To solve the \ggdp{} problem, we use the directed-hypergraph flow viewpoint introduced for EP distribution, but with a GHZ-specific state space and operation set. In particular, a hypervertex represents an intermediate GHZ state over a terminal subset together with its GHZ-basis fidelity vector $\mathbf{P}$. The hyperarcs represent GHZ fusion, full bit-pattern purification, subset bit-pattern purification, and full phase purification; we compute their output vector and success probability using the vector-update rules from \S\ref{sec:model}. A feasible LP flow in this hypergraph corresponds to a level-based fusion structure, and maximizing the flow into the terminal vertex gives the maximum achievable generation rate of the target GHZ state subject to the fidelity threshold $P[0,\mathbf{0}]\geq\fth$. We defer the full hypergraph construction and LP constraints to Appendix~\ref{app:ghz-lp-details}.

\begin{theorem}
For any instance of the \ggdp{} problem over the star graph $\stargraph$, let $H_{\stargraph,\mathcal{D}}$ denote the associated hypergraph defined in Appendix~\ref{app:ghz-lp-details}. For the chosen collection $\mathcal{D}$ of GHZ-basis vector labels, the optimum LP objective value over $H_{\stargraph,\mathcal{D}}$ equals the maximum achievable generation rate for the \ggdp{} instance. Moreover, any optimal LP solution induces an optimal level-based fusion structure for that instance.
\label{thm:ghz-lp}
\end{theorem}
The proof appears in Appendix~\ref{app:ghz-lp-details}.

\begin{figure}[t]
\centering
\includegraphics[width=0.45\textwidth]{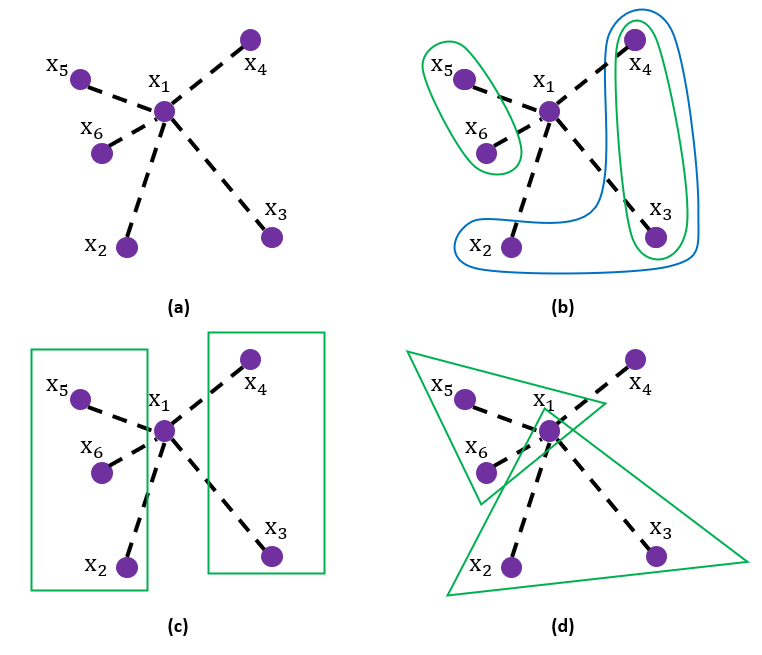}
\caption{Illustration of the computationally efficient \LP{} formulations. (a) Initial star graph for a 6-qubit GHZ state with nodes ${v_1,v_2,v_3,v_4,v_5,v_6}$, where $v_1$ is the central node. The node-labeling strategy determines the ordering of the remaining nodes. 
(b) \LPin{} constructs intermediate GHZ states only over contiguous intervals, such as $[v_3,v_4]$, $[v_5,v_6]$, and $[v_2,v_3,v_4]$. 
(c) \LPts{} partitions the non-central vertices into disjoint subsets, such as $[v_2,v_5,v_6]$ and $[v_3,v_4]$, constructs GHZ states within each subset, and then fuses them into the final GHZ state. (d) \LPms{} constructs GHZ states in multiple stages by progressively combining smaller GHZ states into larger ones; for example, it may first generate 3-GHZ states, such as $[v_1,v_2,v_3]$ and $[v_1,v_5,v_6]$, and then fuse them into a larger intermediate GHZ state, such as $[v_1,v_2,v_3,v_5,v_6]$.}
\vspace*{-0.1in}
\label{fig:lp-star}
\end{figure}

\para{Time Complexity.} For a target GHZ state of size $m$, the number of subset labels $S$ is exponential in $m$, and thus the number of hypervertices in the hypergraph is $O(2^m|\mathcal{D}|)$. The number of hyperarcs is $O(4^m|\mathcal{D}|^2)$, due to the possible combinations of operand subsets and vector labels in fusion and purification operations. Since the LP contains one variable per hyperarc and one flow-conservation equation per nontrivial hypervertex, it has $O(4^m|\mathcal{D}|^2)$ variables, $O(2^m|\mathcal{D}|)$ equality constraints, and $O(m)$ inequality constraints arising from the start-arc capacities.
Using a sparse interior-point method~\cite{lee2015efficient} with the above LP size bounds, the LP-solving time for fixed numerical encoding and precision is conservatively bounded by $\tilde{O}\!\left((4^m|\mathcal{D}|^2)^{3/2}\right)=\tilde{O}(8^m|\mathcal{D}|^3)$, where the tilde hides polylogarithmic factors. This bound excludes hypergraph construction. Thus, the LP formulation is exponential in the GHZ size $m$, motivating the computationally efficient LP formulations we introduce next.

\para{Computationally Efficient LP Formulations.}
To improve scalability, we introduce computationally efficient LP formulations that retain the same hypergraph-flow structure while limiting the family of intermediate GHZ states considered; see Fig.~\ref{fig:lp-star}. 
In all three formulations below, purification is restricted to same-stream full bit-pattern and full phase checks; subset bit-pattern purification and two-stream purification are omitted.
These restrictions reduce the LP search space. In particular, \LPin{} and \LPms{} use only polynomially many candidate intermediate states for a fixed $|\mathcal{D}|$, while \LPts{} provides a tunable trade-off whose complexity depends on the chosen number of partitions. These formulations trade full-model optimality for substantially lower planning overhead.

\begin{itemize}
    \item \textbf{Intermediate GHZ States for Intervals (\LPin).}
    Instead of allowing intermediate GHZ states for all possible subsets $S$, we restrict them to only {\em intervals} of the ordered list $\langle v_2,v_3,\ldots,v_m\rangle$, i.e., we only create intermediate GHZ states corresponding to the subsets of the form $[v_i,v_j]=\langle v_i,v_{i+1},\ldots,v_j\rangle$ for $2 \le i \le j \le m$. 
    (Note that $v_1$, being the center of the star graph, is always part of any GHZ state, as per our earlier convention.) This strategy reduces the number of candidate subsets from $O(2^m)$ to $O(m^2)$, and thus reduces the LP-solving time to $\tilde{O}(m^6 |\mathcal{D}|^3)$. The overall performance of the strategy depends on the choice of the central node and ordering of the remaining vertices; the node-labeling rule and possible alternatives are deferred to Appendix~\ref{app:ghz-lp-details}.
    \item \textbf{Partition-Based Two-Step Approach (\LPts).}
    Another viable strategy to reduce the number of intermediate states and thus the LP size---is to {\em partition} the $m-1$ edges of the star graph into $b$ (disjoint) subsets, each of size either $\lceil \frac{m-1}{b} \rceil$ or $\lfloor \frac{m-1}{b} \rfloor$. 
    We retain intermediate GHZ states whose non-center qubit sets are contained in one partition or are unions of whole partitions. A single LP jointly optimizes generation within the partitions and their subsequent combination into the target GHZ state.
    In essence, the above breaks the overall second stage for generating the target GHZ state from EPs over the star-graph edges into two sub-stages: first, creating GHZ states over partitioned subsets, and then fusing (and purifying) them. 
    With $b$ partitions, there are $O(b\,2^{m/b}+2^b)$ candidate subsets, giving a conservative LP-solving bound of $\tilde{O}\!\left((b\,2^{m/b}+2^b)^3|\mathcal{D}|^3\right)$.
    \item \textbf{Multi-Step Approach (\LPms).}
    The above two-step approach can be naturally extended to a multi-step approach in many ways. One systematic multi-step approach is as follows:
    Starting from singleton groups of non-center nodes, each step pairs consecutive groups in the fixed rate-based ordering. The last group at each step may contain fewer nodes than the others.
    The first step consists of generating intermediate GHZ states of size 3, and the second step creates states of size 5, the $t^{th}$ step creates states of size $2^t+1$, and so on, except for smaller groups arising from an unpaired remainder.
    Each higher step uses fusion operations to combine states from two consecutive groups at the previous step,
    and purification operations are restricted to GHZ states on the same qubit set. Thus, there will be $O(\log m)$ steps and the number of candidate subsets will be $O(m^2)$; the LP-solving time is conservatively bounded by $\tilde{O}(m^6(\log m) |\mathcal{D}|^3)$.
\end{itemize}
We note that each of the above formulations can be viewed as essentially considering an appropriate subgraph of the original hypergraph and developing an LP formulation based on the chosen subgraph.

\para{LP with Two-Scalar Fidelity Labels (\GHZLPTwoF{}).}
The restricted-state LP formulations above are the vector-labeled LP versions used in our GHZ evaluation. We also use the \GHZLPTwoF{} heuristic to test whether the LP planner can work with only the ideal component and its opposite-phase counterpart. This heuristic uses the same restrictions on allowed intermediate GHZ subsets as \LPin{}, \LPts{}, and \LPms{}, but projects the state label from the full vector $\mathbf{P}$ to two scalar components. For a GHZ-basis vector $\mathbf{P}$, define
\[
    F^+(\mathbf{P})=P[0,\mathbf{0}],
    \qquad
    F^-(\mathbf{P})=P[1,\mathbf{0}].
\]
A reduced LP hypervertex is then labeled by $(S,F^+,F^-)$ rather than by $(S,\mathbf{P})$. The terminal constraint is still imposed on $F^+$. Thus, this projection keeps the ideal component $P[0,\mathbf{0}]$ and its opposite-phase counterpart $P[1,\mathbf{0}]$; it discards how the remaining probability mass is distributed over the nonzero bit-error strings.

For fusion-retain, the two-component update gives the same $F^+$ and $F^-$ as the full-vector update. For purification, we retain the same operation restrictions as the vector-labeled heuristics and use approximate updates based on these two components. The update equations and success probabilities are given in Appendix~\ref{app:ghz-lp-details}.

\subsection{\textbf{Greedy Heuristic for GHZ Distribution}}
\label{sec:ghz-greedy}
In addition to the above DP- and LP-based schemes, we also evaluate a greedy heuristic for GHZ distribution. This scheme combines local construction rules with ASAP-style (``as soon as possible'') execution. It chooses the fusion order using estimated generation rates and applies a fixed purification rule, then executes the prescribed operations as their required inputs become available. Unlike the DP and LP schemes, it does not perform a global optimization. Related work has also shown that real-time ordering of entanglement-generation operations can significantly affect performance~\cite{sundaram2024optimized}.

\softpara{Greedy Rule for Fusion.}
In our setting, the greedy scheme starts from the source \epss{} on the star edges. To determine the fusion order, it maintains a list initially containing these two-qubit resources, each with its generation rate and GHZ-basis vector. At each step, it selects two entries with the highest rates and combines them by fusion retain. Both input entries are removed, and the resulting larger GHZ resource is inserted with its updated rate and vector. The next pair is selected from this updated list, so the fusion order depends on intermediate-resource rates as well as source rates. This continues until the list contains only the target GHZ state.

\softpara{Greedy Rule for Purification.}
The greedy scheme uses fixed purification rules: each source EP undergoes one full phase check before fusion, and the final GHZ state undergoes one full bit-pattern check.
\section{\bf General Graph State Distribution with Purification} 
\label{sec:gen-graphs}

We now extend the purification-aware framework from EPs and GHZ states to arbitrary graph states. The key additional challenge is that graph states lack the simple path or star structure, so we introduce graph-growth primitives and graph-basis vector fidelity tracking to support DP and LP formulations over a connected target subgraph.

\subsection{\textbf{Graph State Construction and Purification Model}}
To generalize our schemes for arbitrary graph states, we first develop and describe the following three key components.
\begin{enumerate}
    \item The graph-basis fidelity representation used to track noisy graph states.
    
    \item Schemes to generate arbitrary graph states from smaller graph states using local operations over target-edge EPs.\footnote{Since a Bell pair and the corresponding 2-qubit edge graph state differ only by local Clifford operations, we use the two terms interchangeably in this paper.}
    
    \item Purification protocols to improve the fidelity of noisy graph states.
\end{enumerate}
The above components form the basis of our efficient generation schemes for arbitrary graph states.

\para{1. Graph-Basis Fidelity Representation.}
For a graph state on graph $G=(V,E)$, we use the graph basis
\[
    \ket{\Psi_{G,\boldsymbol{\mu}}}
    =
    \prod_{v\in V} Z_v^{\mu_v}\ket{\Psi_G},
    \qquad
    \boldsymbol{\mu}\in\{0,1\}^{V}.
\]
Thus, the noisy state is represented by a probability vector $\mathbf{P}_G$, where $P_G[\boldsymbol{\mu}]$ is the probability of $\ket{\Psi_{G,\boldsymbol{\mu}}}$. The fidelity with the ideal graph state is the distinguished component $P_G[\mathbf{0}]$.

All graph-state growth and purification operations below are interpreted as transformations of this graph-basis vector. Operation-specific update rules are described with the corresponding primitives, with full details deferred to Appendix~\ref{app:graph-fidelity}.

\para{2. Graph-State Growth Operations.}
To the best of our knowledge, there is no known strategy using fusions to generate an {\em arbitrary} graph from the smallest graph states, i.e., EPs or edge graph states. Some of the closest construction-oriented approaches are:
\begin{itemize}
    \item In our recent work~\cite{fan2024optimized}, we propose efficient fusion-based methods to construct path and tree graph states using {\tt fusion-retain} and {\tt fusion-discard} operations. We extended the approach to other specialized graph states, such as grids and bipartite graphs. However, these methods do not apply to arbitrary graphs because the fusion operations may unintentionally delete edges among neighbors of the fusion qubits, thereby altering the intended structure of more complex graphs, such as cycles or dense subgraphs.
    \item Lee and Jeong~\cite{korean23} also propose a method for generating graph states using fusion operations from 3-qubit GHZ states. Their approach entails decomposing the target graph state into bipartite and clique subgraphs, and then further decomposing these subgraphs into a set of star graphs. Then, they used a fusion network to fuse the star subgraphs into the target graph state. Their method starts from 3-qubit GHZ resources, whereas our two-stage framework requires only link EPs as its initial entangled resources.
\end{itemize}
Depending on the structure of the desired graph state, we may use one of the above approaches when applicable, particularly for trees, paths, grids, or other graph states to which it applies. However, these approaches do not directly address graph-state generation with purification under a fidelity constraint.
Accordingly, for arbitrary graph states, we adopt a more explicit construction framework based on local operations over target-edge EPs. Starting from EPs corresponding to the target edges, the framework incrementally builds the desired graph state: at each step, it either merges two connected subgraphs, attaches a new vertex, or adds a missing internal target edge using local graph-state operations. These growth primitives are general enough to construct any connected target graph, while also providing the DP and LP formulations with an explicit state space over connected subgraphs.

\softpara{Two-Stage Framework.}
As in the GHZ case, we adopt a two-stage approach. In the first stage, for each target edge of the desired graph state, we generate an end-to-end EP resource between the corresponding network nodes using the EP-generation techniques developed earlier. 
Thus, for each target edge $e=(u,v)$ of the desired graph $G$, the first stage provides a generation rate $r_e$ and a two-qubit source vector $\mathbf{P}^{(2)}_e$. If the source is specified by a scalar Bell-pair fidelity $f_e$, we initialize $\mathbf{P}^{(2)}_e$ by the corresponding isotropic Bell/edge graph-basis vector.
In the second stage, we consume these target-edge resources to construct the desired graph state. This separation is natural: the first stage handles routing, swapping, and edge-level purification within the network, while the second stage focuses solely on combining the resulting edge resources into the desired target graph state.

\softpara{Intermediate States in the Second Stage.}
Let $G=(V(G),E(G))$ be the target graph state. Since the first stage generates an end-to-end EP resource for each target edge, the base intermediate states in the second stage are the 2-vertex edge graph states corresponding to the edges of $G$. More generally, an intermediate state is a connected target subgraph $X=(V(X),E(X))$ of $G$, together with its current graph-basis vector. Larger intermediate states are obtained by repeatedly applying graph state growth operations to smaller connected subgraphs.

\softpara{Graph State Growth Operations.}
Our construction uses only local operations, namely local CZ gates, single-qubit Pauli measurements, and local Clifford operations such as local complementation. As described in Appendix~\ref{app:graph-local-fusion} and illustrated there with a figure, these operations realize the underlying fusion step by incrementally merging, at each network node, locally held qubits into a designated {\em star center} that serves as the attachment point for multiple incident target edges. This local realization gives rise to the following three graph state growth primitives.

\begin{itemize}
    \item \emph{Fuse-Two-Components.}
    Let $X_1$ and $X_2$ be two disjoint connected target subgraphs, and let $e=(a,b)\in E(G)$ be a target edge such that $a\in V(X_1)$ and $b\in V(X_2)$. Using one copy of EP corresponding to $e$, we combine the graph states corresponding to $X_1$ and $X_2$ to obtain the graph state corresponding to the connected graph
    \[
    X=\bigl(V(X_1)\cup V(X_2),\; E(X_1)\cup E(X_2)\cup\{e\}\bigr).
    \]
    
    \item \emph{Extend-by-New-Vertex.}
    Let $X$ be a connected target subgraph, and let $e=(a,b)\in E(G)$ be a target edge such that $a\in V(X)$ and $b\notin V(X)$. Using one copy of EP corresponding to $e$, we extend $X$ by attaching the new vertex $b$, thereby obtaining the graph state corresponding to
    \[
    X'=\bigl(V(X)\cup\{b\},\; E(X)\cup\{e\}\bigr).
    \]
    
    \item \emph{Add-Internal-Edge.}
    Let $X$ be a connected target subgraph, and let $e=(a,b)\in E(G)$ be a target edge such that both $a,b\in V(X)$ and $e\notin E(X)$. Using one copy of EP corresponding to $e$, we add the missing internal edge to obtain the graph state corresponding to
    \[
    X'=\bigl(V(X),\; E(X)\cup\{e\}\bigr).
    \]
\end{itemize}

For each successful graph-growth primitive above, the measurement and feed-forward corrections induce a deterministic map from the input graph-basis labels to an output graph-basis label. The explicit vector updates for these growth primitives are given in Appendix~\ref{app:graph-fidelity}.

\para{3. Graph-State Purification.}
Prior work on the purification of graph states has primarily focused on two-colorable graph states~\cite{aschauer2005multiparticle}, in which recurrence-based protocols using CNOT operations have been shown to improve fidelity under local noise. These techniques, however, do not generalize to arbitrary graph states. To address this limitation,~\cite{kruszynska2006entanglement} develops a general protocol that applies to all $k$-colorable graph states by decomposing the purification process into $k$ sub-protocols. The method relies solely on local Clifford operations, multilateral CNOTs, and single-qubit measurements, making it implementable in a distributed setting. Therefore, we adopt the purification strategy from~\cite{kruszynska2006entanglement} as a general method for improving the fidelity of noisy graph states; we briefly describe it
below.

\softpara{Purifying a $k$-colorable Graph State.} 
To purify a $k$-colorable graph state on a graph\footnote{Here $G$ denotes the graph state currently being purified; in our second-stage model, it may be either the final target graph or an intermediate graph generated during the construction.} $G=(V,E)$, we fix a proper $k$-coloring
\[
    V=A_1\cup A_2\cup\cdots\cup A_k,
\]
where no edge has both endpoints in the same color class. The protocol in~\cite{kruszynska2006entanglement} introduces a set of $k$ sub-protocols $\{P_j\}$, where $P_j$ checks errors associated with the color class $A_j$.

For a selected color class $A_j$, let $\bar A_j=V\setminus A_j$. The auxiliary graph state used by $P_j$ is obtained from the edges of $G$ crossing the cut $A_j\,|\,\bar A_j$. More precisely, define
\[
    N_G(A_j)=\{u\in \bar A_j:\exists a\in A_j
    \text{ with }(a,u)\in E\},
\]
and let
\[
    W_j=A_j\cup N_G(A_j),
    \quad
    E_j=\{(a,u)\in E:a\in A_j,\ u\in \bar A_j\}.
\]
The sacrificial input is a noisy graph state on the two-colorable auxiliary graph
\[
    g_j=(W_j,E_j).
\]
Thus, the sacrifice keeps exactly the checked color class, its neighboring vertices, and the edges needed to test that color.

\softpara{One-Color Check Operation.}
The sub-protocol $P_j$ keeps the target copy on $G$ and consumes the auxiliary sacrifice state on $g_j$. For each vertex $u\in W_j$, the party holding $u$ applies a CNOT gate between the target qubit and the corresponding sacrificial qubit: for vertices in $A_j$, the CNOT is applied from the sacrifice to the target, and for vertices in $W_j\setminus A_j$, it is applied from the target to the sacrifice.
The sacrificial qubits are then measured locally: qubits in $A_j$ are measured in the $X$ basis and qubits in $W_j\setminus A_j$ are measured in the $Z$ basis. Let $\xi_a\in\{0,1\}$ denote the $X$-measurement outcome for $a\in A_j$, and let $\zeta_b\in\{0,1\}$ denote the $Z$-measurement outcome for $b\in W_j\setminus A_j$. For each qubit $a\in A_j$, the target stabilizer generator is
\[
    K_a
    =
    X_a
    \prod_{\substack{b\in \bar A_j\\ (a,b)\in E}}
    Z_b
\]
and the measured check syndrome is
\[
s_a
=
\xi_a
\bigoplus_{\substack{b\in \bar A_j\\ (a,b)\in E}}
\zeta_b .
\]
For graph-basis inputs, $s_a=\mu^{\rm t}_a\oplus\mu^{\rm s}_a$, where ${\rm t}$ and ${\rm s}$ denote the target and sacrificial labels, respectively. If all the computed values $s_a$ (for $a \in A_j$) are zero, the target copy is kept; otherwise, it is discarded.

\softpara{Purification Step and Fidelity Update.}
At the graph-basis label level, the one-color check accepts exactly when the target and sacrificial labels agree on the checked color class $A_j$. If the accepted target and sacrifice labels are written as $(\boldsymbol{x},\boldsymbol{y})$ and $(\boldsymbol{x},\boldsymbol{z})$, respectively, then the retained target label is $(\boldsymbol{x},\boldsymbol{y}\oplus\boldsymbol{z})$. The output vector is obtained by summing the product probabilities of all accepted input-label pairs that produce the same output label, and then normalizing by the total accepted probability. The explicit vector formula is given in Appendix~\ref{app:graph-fidelity}. Applying the sub-protocols $P_1,\ldots,P_k$ once checks all color classes. In practice, however, it may be advantageous to apply only a subset of the color checks to enhance purification efficiency.

\subsection{\textbf{Dynamic Programming for Graph State Distribution}}
\label{sec:graph-dp}
\para{Graph State Distribution with Purification (\graphdp) Problem.}
Consider a quantum network $Q=(V(Q),E(Q))$ and a target graph state $G=(V(G),E(G))$, whose vertices are mapped to network nodes via $\tau:V(G)\to V(Q)$. As in the above two-stage framework, we assume that the first stage has already generated, for each target edge $e\in E(G)$, an end-to-end EP resource with generation rate $r_e$ and source vector $\mathbf{P}^{(2)}_e$. Given a fidelity threshold $\fth$, the \graphdp problem is to determine a purification-augmented construction tree that generates the target graph state $G$ with the maximum generation rate and final ideal-component fidelity of at least $\fth$. \hfill $\Box$

\para{Optimal DP Formulation.}
We now describe an optimal DP formulation for the \graphdp problem under the graph state growth and purification model adopted in this paper. As in the GHZ case, it is natural to define subproblems by identifying the intermediate graph state to be generated, the target output rate, and the EP resources consumed over the target edges during construction. This yields a DP formulation that is useful for specifying the problem's recursive structure, but whose full state space is too large for practical use with general graph states. We therefore first describe this formulation and then derive computationally efficient DP formulations from it for evaluation.

\softpara{Subproblems $\mathcal{T}(X,k,\overrightarrow{EP})$.}
We define subproblems for each triplet $\langle X,k,\overrightarrow{EP}\rangle$, where $X$ is a connected target subgraph of $G$, $k$ is a discretized generation rate, and $\overrightarrow{EP}$ is a vector indexed by the target edges in $E(G)$, whose entry for an edge $e$ records an upper bound on the amount of EP rate consumed over $e$ to generate $X$. 
Let $\mathcal{T}(X,k,\overrightarrow{EP})$ denote a frontier of feasible annotated construction trees. Each tree $T\in\mathcal{T}(X,k,\overrightarrow{EP})$ generates the graph state corresponding to $X$ at a rate of at least $k$ while consuming an EP-rate vector bounded by $\overrightarrow{EP}$. Each annotated tree carries the graph-basis vectors of its intermediate states. For a tree \(T\) generating the graph state on \(X\), we write \(\mathbf{P}_T^{(X)}\) for the vector at its root. The frontier retains all distinct reachable root vectors, because two trees with the same ideal component $P_T^{(X)}[\mathbf{0}]$ may behave differently under later color-check purification.

\softpara{Base Case.}
The base case corresponds to $X$ being a 2-vertex edge graph state associated with a target edge $e\in E(G)$. Since the first stage already provides an end-to-end EP for $e$ with rate $r_e$ and source vector $\mathbf{P}^{(2)}_e$, the corresponding DP entry is initialized by that EP resource. Let $X_e$ denote this edge graph state and let $T_e$ be the single-node construction tree whose root vector is $\mathbf{P}^{(2)}_e$. Then
\[
\mathcal{T}(X_e,k,\overrightarrow{EP})
=
\begin{cases}
\{T_e\},&
k\le \floor{r_e}\ \text{and}\ 
\langle 0,\ldots,k,\ldots,0\rangle\leq\overrightarrow{EP},\\
\emptyset,&
\text{otherwise,}
\end{cases}
\]
where the nonzero entry in $\langle 0,\ldots,k,\ldots,0\rangle$
corresponds to the target edge $e$.

\softpara{Recursive Step.}
For any tree with more than one node represented in $\mathcal{T}(X,k,\overrightarrow{EP})$, the root operation must be one of the following four types:
\begin{enumerate}
    \item {\tt Fuse-Two-Components},
    \item {\tt Extend-by-New-Vertex},
    \item {\tt Add-Internal-Edge},
    \item One-color graph-state purification using an auxiliary sacrificial graph.
\end{enumerate}
Thus, the DP recursively enumerates all valid last-step decompositions of $X$, combines the corresponding predecessor subproblems, and computes the output graph-basis vector using the vector-update rules developed earlier. In particular, the first three cases correspond exactly to the three graph-state growth primitives introduced above, while the last case applies a one-color check to a graph state using the corresponding auxiliary sacrifice.

In all the above cases, the children of the root are selected from the appropriate child frontiers. Thus, the optimal-substructure property holds at the level of frontiers: a parent annotated construction tree is obtained by choosing feasible child annotated trees from predecessor subproblems and applying one root operation, whose output vector is determined by the child root vectors. For $X$ with more than one edge, incorporating the above four cases gives the recurrence
\[
\mathcal{T}(X,k,\overrightarrow{EP})
=
\operatorname{Prune}
\left(
\mathcal{C}_{\rm fuse}
\cup
\mathcal{C}_{\rm ext}
\cup
\mathcal{C}_{\rm int}
\cup
\mathcal{C}_{\rm pur}
\right),
\]
where $\operatorname{Prune}$ removes only duplicate root vectors, keeping one annotated tree for each distinct vector. For an edge graph $X_e$, the frontier includes feasible trees that use the source EP directly or purify copies of it. The sets $\mathcal{C}_{\rm fuse}$, $\mathcal{C}_{\rm ext}$, $\mathcal{C}_{\rm int}$, and $\mathcal{C}_{\rm pur}$ correspond to the four operation families above and are defined in Appendix~\ref{app:graph-dp-details}.

\softpara{DP Solution.}
Given the fidelity threshold $\fth$, the maximum rate $k^\star$ is the largest discretized rate for which at least one annotated tree in the final frontier has ideal-component fidelity above the threshold:
\begin{equation*}
k^\star
=
\max \left\{
    k\in\discreterate
    \;\middle|\;
    \begin{aligned}
    &\exists\,\overrightarrow{EP}
    \leq
    \langle \floor{r_e}\rangle_{e\in E(G)},\\
    &\exists\,T\in\mathcal{T}(G,k,\overrightarrow{EP}):\\
    &P_T^{(G)}[\mathbf{0}]\ge\fth
    \end{aligned}
\right\}.
\end{equation*}

\para{Computationally Efficient DP Formulations.}
Since the optimal DP explicitly tracks both the connected target subgraph and the full EP-consumption vector, its state space grows prohibitively large for general graph states.
We therefore use two computationally efficient DP formulations. As in the GHZ DP formulations, both still carry the full graph-basis vector $\mathbf{P}$ at every retained intermediate state and apply the vector-update rules from Appendix~\ref{app:graph-fidelity}. 
Each reduced DP cell keeps one selected annotated tree instead of a full frontier, and the EP-consumption vector is not enumerated as a separate table index.

\softpara{Rate-Discretized DP (\GraphDPV).}
\GraphDPV{} keeps the rate-indexed viewpoint of the optimal DP. For each connected target subgraph $X$ and discretized output-rate bucket $k$, the table stores at most one annotated construction tree $T(X,k)$ and its root vector $\mathbf{P}(X,k)$. Candidate trees are generated by the same operation families as in the optimal DP: the three graph-growth primitives and one-color graph-state purification from Appendix~\ref{app:graph-fidelity}. When several candidates fall in the same $(X,k)$ cell, the retained tree is the one with the largest ideal component $P[\mathbf{0}]$.

\softpara{Fidelity-Discretized DP (\GraphDPF).}
\GraphDPF{} starts from the same reduced vector-state representation, but changes the table index. Instead of indexing a cell by an output-rate bucket, it indexes the cell by a bucketed value $f$ of the ideal component $P[\mathbf{0}]$. Thus, for each connected target subgraph $X$ and fidelity bucket $f$, the table stores at most one annotated construction tree $T(X,f)$ and its root vector $\mathbf{P}(X,f)$. If several candidates map to the same $(X,f)$ cell, the retained tree is the one with the largest generation rate. The scalar fidelity bucket is only a search index; all operation outputs are still computed from and stored as full graph-basis vectors.

\subsection{\textbf{Linear Programming for Generalized Graph-State Distribution}}
\label{sec:graph-lp}
\para{Generalized Graph State Distribution with Purification (\graphlp) Problem.}
Consider a quantum network $Q=(V(Q),E(Q))$ and a target graph state $G=(V(G),E(G))$, whose vertices are mapped to network nodes via $\tau:V(G)\to V(Q)$. As in the above two-stage framework, we assume that the first stage has already generated, for each target edge $e\in E(G)$, an end-to-end EP resource with generation rate $r_e$ and source vector $\mathbf{P}^{(2)}_e$. Given a fidelity threshold $\fth$, the \graphlp problem is to determine a purification-augmented level-based construction structure that generates the target graph state $G$ with the maximum generation rate and final ideal-component fidelity of at least $\fth$. \hfill $\Box$

\para{Optimal LP Formulation (\GraphLPOpt).}
We now describe an LP formulation for the \graphlp problem under the graph state growth and purification model adopted in this paper. As in the GHZ case, the LP allocates the first-stage EP supplies among multiple concurrent constructions. It uses a directed hypergraph in which generation flows for the same available-state label can be combined from different operations and distributed among subsequent operations.

\softpara{Hypergraph-Based LP Solution.}
Let $\mathcal{X}$ denote the family of connected target subgraphs of $G$, including the 2-vertex edge graph states corresponding to the target edges. Let $\mathcal{D}$ denote the collection of graph-basis vectors used as labels in the LP. Whenever a vector label is paired with a connected subgraph $X$, it is understood to have the graph-basis dimension for $X$.
For each connected target subgraph $X\in\mathcal{X}$ and each compatible vector $\mathbf{P}\in\mathcal{D}$, we create an available-state vertex $\id{avail}(X,\mathbf{P})$ and operation vertices representing copies of $X$ produced by the various operations.
In particular, the hypergraph contains source arcs for the target edge EP resources produced by the first stage, hyperarcs for the three graph state growth primitives {\tt Fuse-Two-Components}, {\tt Extend-by-New-Vertex}, and {\tt Add-Internal-Edge}, as well as hyperarcs for color-check purification of graph states. A terminal hyperarc collects copies of the full target graph $G$ whose ideal component is at least $\fth$.

\softpara{LP Variables and Constraints.}
We associate a nonnegative flow variable with each hyperarc. Intuitively, the flow on a hyperarc represents the generation rate of the corresponding operation. The LP enforces:
\begin{enumerate}
\item  Flow conservation at the available-state vertices.
\item  Operation constraints that scale the incoming flow by the effective rate-scaling factor of the corresponding primitive.
\item  Capacity constraints in Start hyperarcs so that the consumed rate of each target edge EP does not exceed the first-stage rate $r_e$.
\item  Maximization of the total flow entering the terminal vertex. 
\end{enumerate}
Thus, the LP formulation maximizes the achievable generation rate of the target graph state $G$ subject to the fidelity threshold $\fth$ under the adopted graph-state construction and fidelity-tracking model. This formulation captures the corresponding level-based optimization problem, but its size is exponential in the number of target vertices. We defer the detailed hypergraph construction and LP constraints to Appendix~\ref{app:graph-lp-details}.

\para{Computationally Efficient LP Formulations.}
The hypergraph construction above can be instantiated with different state labels. To limit the number of labels, we use two computationally efficient graph LP formulations. Both retain the second-stage hypergraph-flow structure and graph-growth primitives, while simplifying the state labels as follows.
\begin{itemize}
    \item \textbf{Vector-Labeled Graph LP (\GraphLPV).}
    In this scheme, a hypervertex for an intermediate graph $X$ retains a graph-basis vector $\mathbf{P}$, as in the optimal LP formulation above. To limit the number of labels, candidate vectors are grouped by buckets of their ideal component $P[\mathbf{0}]$, and representative full vectors are retained for subsequent growth and color-check updates. Thus, each retained label still stores an error distribution, but the LP does not distinguish every reachable vector.

    \item \textbf{Fidelity-Labeled Graph LP (\GraphLPF).}
    In this scheme, a hypervertex for an intermediate graph $X$ is labeled only by a scalar fidelity bucket $f$ for the ideal component $P[\mathbf{0}]$, i.e., by $(X,f)$ instead of $(X,\mathbf{P})$. 
    Scalar growth and purification updates estimate the output fidelity from the input labels, and the estimate is rounded down to the fidelity grid.
    Thus, each state label keeps only the ideal-component bucket and no longer distinguishes the distribution over non-ideal graph-basis components. In the evaluated scalar variant, graph-state purification is restricted to a full-color check, implemented by applying the one-color checks over all color classes.
\end{itemize}

\subsection{\textbf{Greedy Heuristic for Graph State Distribution}}
The graph-state greedy heuristic follows the same local-construction and ASAP-style execution approach as the GHZ heuristic in \S\ref{sec:ghz-greedy}. Starting from the first-stage target-edge EP resources, it grows one connected subgraph, prioritizing vertex extension over adding missing internal edges. Purification follows fixed rules: each target-edge EP resource undergoes two successive purification rounds before graph growth, and the completed target graph undergoes one full-color check. No purification is performed between growth operations.

\section{\bf Incorporating Memory Decoherence Model}
\label{sec:decoherence}

The construction procedures above specify which quantum resources are combined and in which order. We now describe how memory decoherence caused by waiting is incorporated into the same model. A stored resource may be an EP, a GHZ state, or a general graph state. When the resource is represented by a basis vector, we model waiting as a stochastic map on that vector. If a resource $R$ with vector $\mathbf{P}_R$ waits for time $\tau$, we write the aged vector
as
\[
    \mathcal{D}^{(R)}_{\tau}(\mathbf{P}_R).
\]

\para{Waiting Channel.}
We model memory noise by applying an independent single-qubit channel to each physical qubit in the stored resource during its waiting time. Specifically, we use a single-qubit depolarizing channel with probability
\[
    \lambda(\tau)
    =
    1-\exp(-\tau/t_c),
\]
where $t_c$ is the memory coherence time. Thus $\mathcal{D}^{(R)}_{\tau}$ is obtained by applying this local channel to all qubits of the resource and rewriting the resulting state in the same basis: the GHZ basis for GHZ resources and the graph basis for graph-state resources. The aged vector is then used as the input to the corresponding fusion or purification update.

This notation is also compatible with scalar-fidelity formulas. When an isotropic Bell-pair state is represented only by its fidelity $F$, we use the scalar waiting model
~\cite{inesta2023optimal}

\[
F_\tau = \frac{1}{4} +
\left(F-\frac{1}{4}\right)e^{-\tau/t_c}.
\]
This updated scalar fidelity can then be substituted into the usual swapping or purification fidelity expressions.

\para{Waiting-Aware Operation Updates.}
For each operation, the waiting channel is applied to its inputs before the operation itself. For example, consider a two-input operation $\mathsf{Op}$ with input resources $R_1$ and $R_2$, vectors $\mathbf{P}_1$ and $\mathbf{P}_2$, and storage times $\tau_1$ and $\tau_2$. The retained output $R_{\rm out}$ remains in memory while the required measurement outcomes are communicated classically to the relevant nodes. Let $\tau_{\rm cc}$ denote this additional storage time. In the planner, its vector after communication is modeled as
\[
    \mathbf{P}_{\rm out}
    =
    \mathcal{D}^{(R_{\rm out})}_{\tau_{\rm cc}}
    \left(
    \mathsf{Op}
    \left(
        \mathcal{D}^{(R_1)}_{\tau_1}(\mathbf{P}_1),
        \mathcal{D}^{(R_2)}_{\tau_2}(\mathbf{P}_2)
    \right)
    \right).
\]
For GHZ and graph states, the DP and greedy rate estimates also include communication waiting by adding it to the expected duration of each operation.
The remaining task is to estimate the storage times $\tau_1$ and $\tau_2$. In the DP planner, if two resources are generated by independent Poisson processes with rates $r_1$ and $r_2$, the storage time of a resource is the time it waits after being generated until the other required resource is also available. Let $T_i\sim{\rm Exp}(r_i)$ be the generation time of resource $i$.  Then
\[
    \Pr[T_1<T_2]
    =
    \int_0^\infty
    \Pr[T_2>t]\,r_1e^{-r_1t}\,dt
    =
    \frac{r_1}{r_1+r_2},
\]
and similarly $\Pr[T_2<T_1]=r_2/(r_1+r_2)$.  Conditioned on $T_1<T_2$, resource 1 is ready and waits in memory while resource 2 has not yet arrived. By the memoryless property of the exponential distribution,
\[
    \Pr[T_2>T_1+s\mid T_2>T_1]
    =
    \Pr[T_2>s],
\]
so the remaining time until resource 2 arrives is distributed as a fresh ${\rm Exp}(r_2)$ random variable and has mean $1/r_2$. Thus, including the local operation time $t_{\rm op}$, the expected storage times are
\[
    \tau_1
    =
    \frac{r_1}{r_2(r_1+r_2)}
    +
    t_{\rm op},
    \qquad
    \tau_2
    =
    \frac{r_2}{r_1(r_1+r_2)}
    +
    t_{\rm op}.
\]
When $r_1=r_2=r$, both inputs wait $1/(2r)+t_{\rm op}$ on average. For same-stream purification, the newly available copy waits only for the local operation time, while the stored copy has one additional inter-arrival waiting time, so its storage time is modeled as $1/r+t_{\rm op}$.

The LP uses the same decoherence map, but a flow solution does not by itself specify a single synchronization history. For both GHZ and graph states, we use a two-pass approximation: the positive operation flows from the first LP pass determine representative waiting times for the stored resources, and the second pass rebuilds and solves the LP using these waiting times. Within each LP solve, the rate coefficients remain fixed; communication-induced decoherence is incorporated through the state labels and the associated purification success probabilities.
The greedy scheme also uses estimated waiting times when evaluating its local construction and purification choices. During physical execution, all schemes apply the waiting channel using actual resource storage times, including the time spent waiting for classical messages.
 
\section{\bf Evaluation}
\label{sec:eval}

\begin{figure*}[t]
\vspace*{-0.1in}
    \centering
    \includegraphics[width=\textwidth]{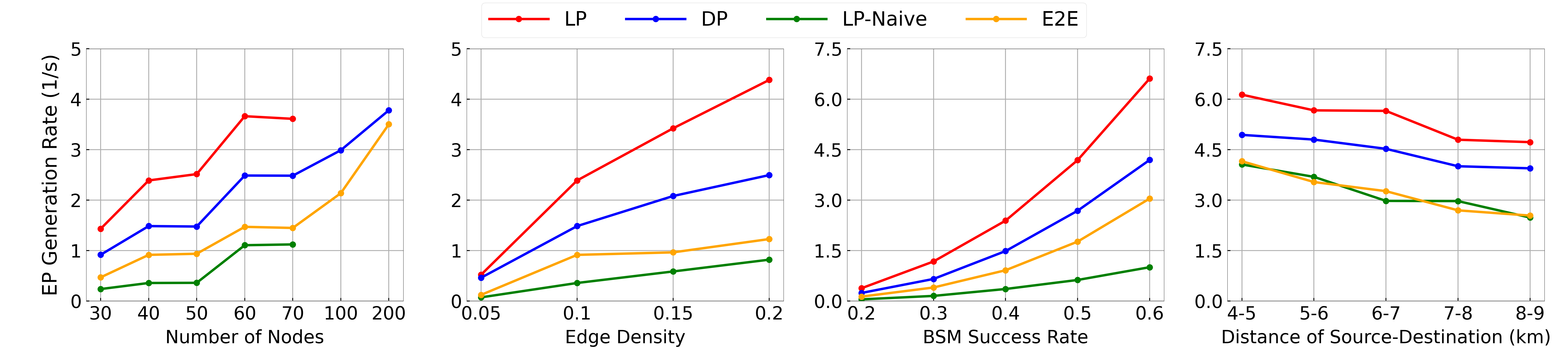}
    \caption{Generation Rates for EPs for various schemes, for varying parameter values, from NetSquid simulations.}
    \label{fig:ep-4-parameters}
\end{figure*}

We evaluate the solutions generated by our algorithms and the protocols from existing works using the quantum network simulator NetSquid~\cite{netsquid2020}.
We report generation rate and algorithmic computation overhead for EP, GHZ, and general graph-state generation. For GHZ and general graph states, we also report the average output fidelity.

\subsection{\bf Experiment Setup and Compared Methods}

\para{NetSquid Protocol.}
For the EP experiments, we build our protocols on top of the link-layer protocol of~\cite{sigcomm19}, delegated to continuously generate \epss on a link at a desired rate. The key aspect of our protocols is that a swapping or a fusion operation or a graph state growth operation is done only when {\it all} required input states have been generated, and a purification operation is done only when {\it both} appropriate target and sacrificial states are generated. In all experiments, swapping/fusion and purification operations were implemented to transmit classical information to all relevant nodes upon success.
On failure, the qubits of the entanglement state are discarded, allowing the protocols to generate new link EPs from scratch. During the purification process, the sacrificial state is consumed through measurements. 

\para{Simulation Setting.}
For the EP experiments, we generate random quantum networks in a similar way to recent works~\cite{swapping-tqe-22, sigcomm20, fan2024optimized}. By default, we use a network spread over an area of $100 km \times 100 km$. We use the Waxman model~\cite{waxman}, used to create Internet topologies, distribute the nodes, and create links. We vary the number of nodes from 30 to 70 (with a default of 40; additional scalability experiments extend the network size up to 200 nodes) and the edge density from 0.05 to 0.2 with a default value of 0.1.
For EP generation, the fidelity of each link EP is a random number between 0.7 and 0.95, and the fidelity threshold for the goal EP state is 0.8, the same as the most similar work~\cite{zhao2022e2e}. 
For the GHZ and general graph-state experiments, the first stage is represented by remote EP sources on the star edges or target graph edges, and the second-stage algorithms consume these sources to construct the target state. The target graphs in the graph-state experiments are randomly generated spanning trees. The terminal-count sweep uses target sizes from 3 to 9 vertices, with default target size 6 for the remaining parameter sweeps. The final target-state fidelity threshold is 0.8. Unless it is the swept parameter, the fusion success probability is 0.9, the average remote-EP generation rate is 70/s, the coherence time is 10 seconds, the local operation time is $10^{-5}$ seconds, the per-node quantum memory capacity is 100 qubits.
Each EP simulation runs for 100 seconds, while each GHZ and general graph-state simulation runs for 300 seconds.

\softpara{Parameter Values.}
For the EP experiments, we use parameter values similar to the ones used in~\cite{caleffi, swapping-tqe-22}. In particular, we use swapping/fusion probability of success (\fp) to be 0.4 and latency (\ft) to be 10 $\mu$ secs; in some plots, we vary $\fp$ from 0.2 to 0.6.
The atomic-BSM probability of success (\bp) and latency (\bt) always equal their fusion counterparts \fp and \ft. 
The optical-BSM probability of success (\php) is half of \bp. 
For generating link-level \epss, we use atom-photon generation times (\gt) and probability of success (\gp) as 50 $\mu$sec and 0.33, respectively. 
Finally, we use the photon transmission success probability 
as $e^{-d/(2L)}$~\cite{caleffi} where $L$ is the channel attenuation length
(chosen as 20 $km$ for optical fiber) and $d$ is the distance between the nodes.
In the EP NetSquid experiments, we pick the depolarizing channel as our noise model and set the depolarization rate of 0.01. For GHZ and general graph-state experiments, the remote-EP fidelities are sampled uniformly from $[0.8,1.0]$. Memory decoherence during storage and local operations follows \S\ref{sec:decoherence}, with no additional local gate or measurement noise.
We also compare GHZ and graph-state generation with and without classical-communication delays under otherwise matched settings. In this test, we include only points with generation rates of at least 0.05 states/s in both settings. Across the four parameter sweeps, the median absolute relative change in generation rate is $1.51\%$ for GHZ states and $3.22\%$ for graph states, based on 327 and 96 matched points, respectively.
\begin{table*}[t]
\centering
\scriptsize
\setlength{\tabcolsep}{5pt}
\begin{tabular}{|l|l|l|l|}
\hline
\textbf{Target state} & \textbf{Type} & \textbf{Methods} & \textbf{Main distinction} \\
\hline
\multirow{6}{*}{GHZ states}
& \multirow{2}{*}{DP} & \DPss{}, \DPOT{}, \DPtwoG{} & Rate-discretized vector DP heuristics \\
\cline{3-4}
& & \GHZDPFSS{}, \GHZDPFBP{}, \GHZDPFtwoG{} & Fidelity-discretized vector DP heuristics \\
\cline{2-4}
& \multirow{3}{*}{LP} & \GHZLPVIn{}, \GHZLPVTs{}, \GHZLPVMs{} & Vector-labeled LP heuristics \\
\cline{3-4}
& & \GHZLPTwoFIn{}, \GHZLPTwoFTs{}, \GHZLPTwoFMs{} & Scalar-fidelity-labeled LP heuristics \\
\cline{3-4}
& & \GHZLPOpt{} & Optimal vector-labeled LP \\
\cline{2-4}
& Greedy & \Greedy{} & Local construction and purification rules \\
\hline
\multirow{6}{*}{\makecell[cl]{General\\graph states}}
& \multirow{2}{*}{DP} & \GraphDPV{} & Rate-discretized vector DP heuristic \\
\cline{3-4}
& & \GraphDPF{} & Fidelity-discretized vector DP heuristic \\
\cline{2-4}
& \multirow{3}{*}{LP} & \GraphLPV{} & Vector-labeled LP heuristic \\
\cline{3-4}
& & \GraphLPF{} & Scalar-fidelity-labeled LP heuristic \\
\cline{3-4}
& & \GraphLPOpt{} & Optimal vector-labeled LP \\
\cline{2-4}
& Greedy & \Greedy{} & Local construction and purification rules \\
\hline
\end{tabular}
\caption{GHZ and graph-state algorithms compared in the evaluation.}
\label{tab:evaluation-algorithms}
\end{table*}

\para{Prior Algorithms Compared.}
For comparison with prior work, we implement two schemes for EP: a recent end-to-end-fidelity-based scheme based on~\cite{zhao2022e2e} (called \ee) and a naive approach (called \LPNaive). 
\ee has three steps: 
First, find the top ten shortest (fidelity-based weights) paths for each pair of source and destination nodes; then, for each path, purify all the links above a link fidelity threshold to guarantee that the end-to-end is above the fidelity threshold, using the right-deep swapping trees; finally, using linear programming to allocate limited resources among different paths, maximizing the generation rate.
The \LPNaive scheme uses LP-based methods to generate EPs, restricting purification to the top level only. In \LPNaive, only the desired EPs can serve as sacrificial states.

\para{Our Algorithms.}
For the generation of EPs, we implement the \LP and \DP schemes. 
To compare the results for multiple source-destination pairs, we use a modified \DP scheme (called \DPIterative), which first employs the original \DP method to identify a purification-augmented swapping tree. It then removes the resources utilized by this tree and iteratively applies the \DP scheme to identify additional purification-augmented swapping trees.

Table~\ref{tab:evaluation-algorithms} summarizes the algorithms evaluated for GHZ and general graph-state generation.

\subsection{\bf Evaluation Results for EPs Generation}

Fig.~\ref{fig:ep-4-parameters} shows the generation rates of various schemes for EPs as determined by the NetSquid simulations of at least 100-second duration. In particular, we vary one parameter at a time while keeping the other parameters constant (to their default values). We observe that our \LP scheme outperforms the \LPNaive and \ee schemes nearly by a factor of five and 2.5, respectively, on average. And, our \DP scheme consistently outperforms the \LPNaive and \ee schemes, even in cases where our \DP scheme only finds one purification-augmented swapping tree while the \LPNaive and \ee schemes find multiple trees.

\begin{figure}[h]
    \centering
    \includegraphics[width=0.3\textwidth]{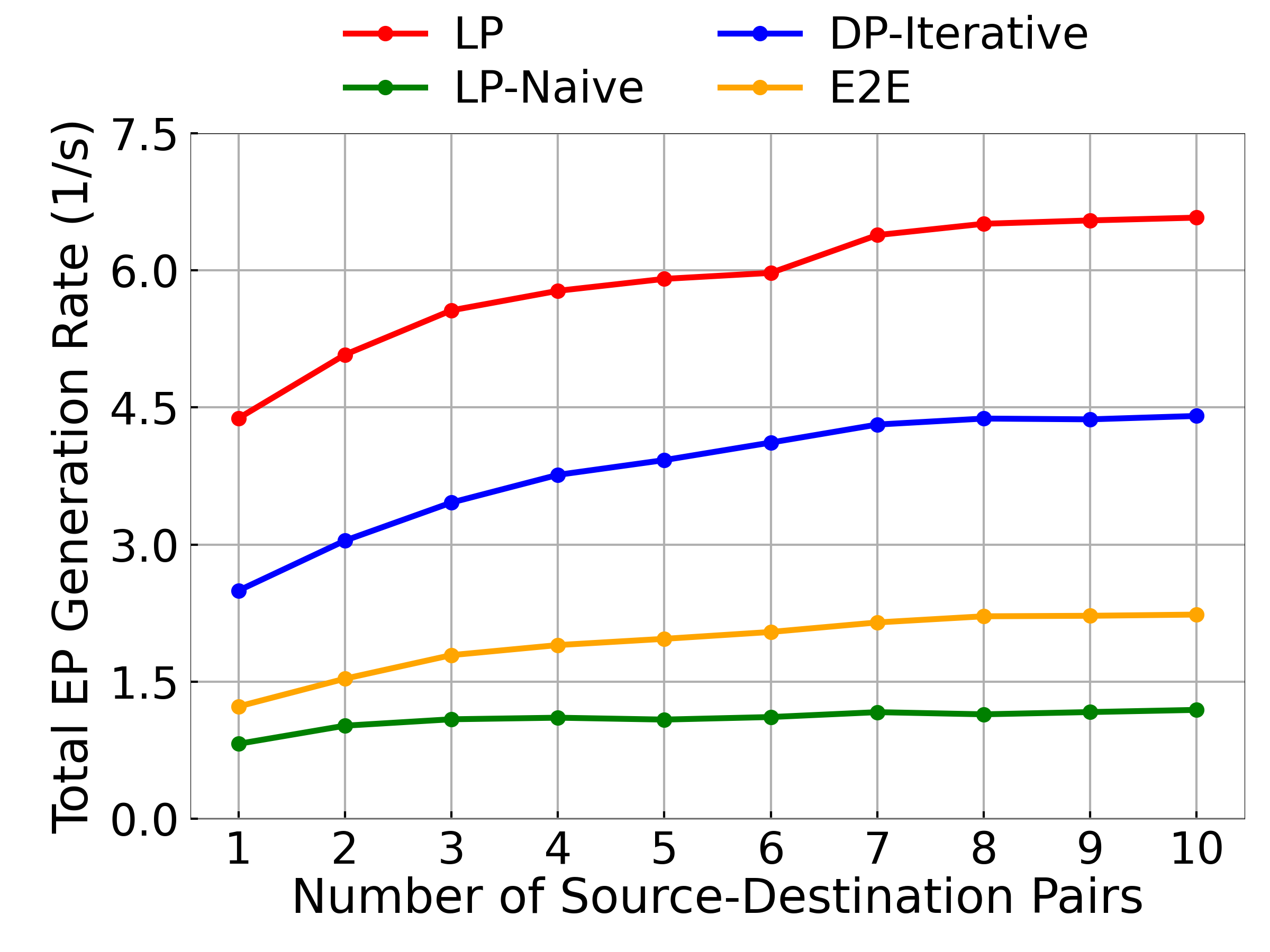}
    \caption{EP generation rates for a varying number of source-destination pairs (the edge density is 0.2, larger than the default, ensuring we can find purification-augmented swapping trees between all sources and destinations).}
    \vspace{-0.1in}
    \label{fig:ep-pairs}
\end{figure}

Fig.~\ref{fig:ep-pairs} shows the results for multiple source-destination pairs; the \ee scheme consistently outperforms \LPNaive, and our \DPIterative scheme outperforms \ee by approximately 100\% on average. Our \LP scheme outperforms \ee by approximately 200\% on average.

\begin{figure}[h]
    \centering
    \includegraphics[width=0.3\textwidth]{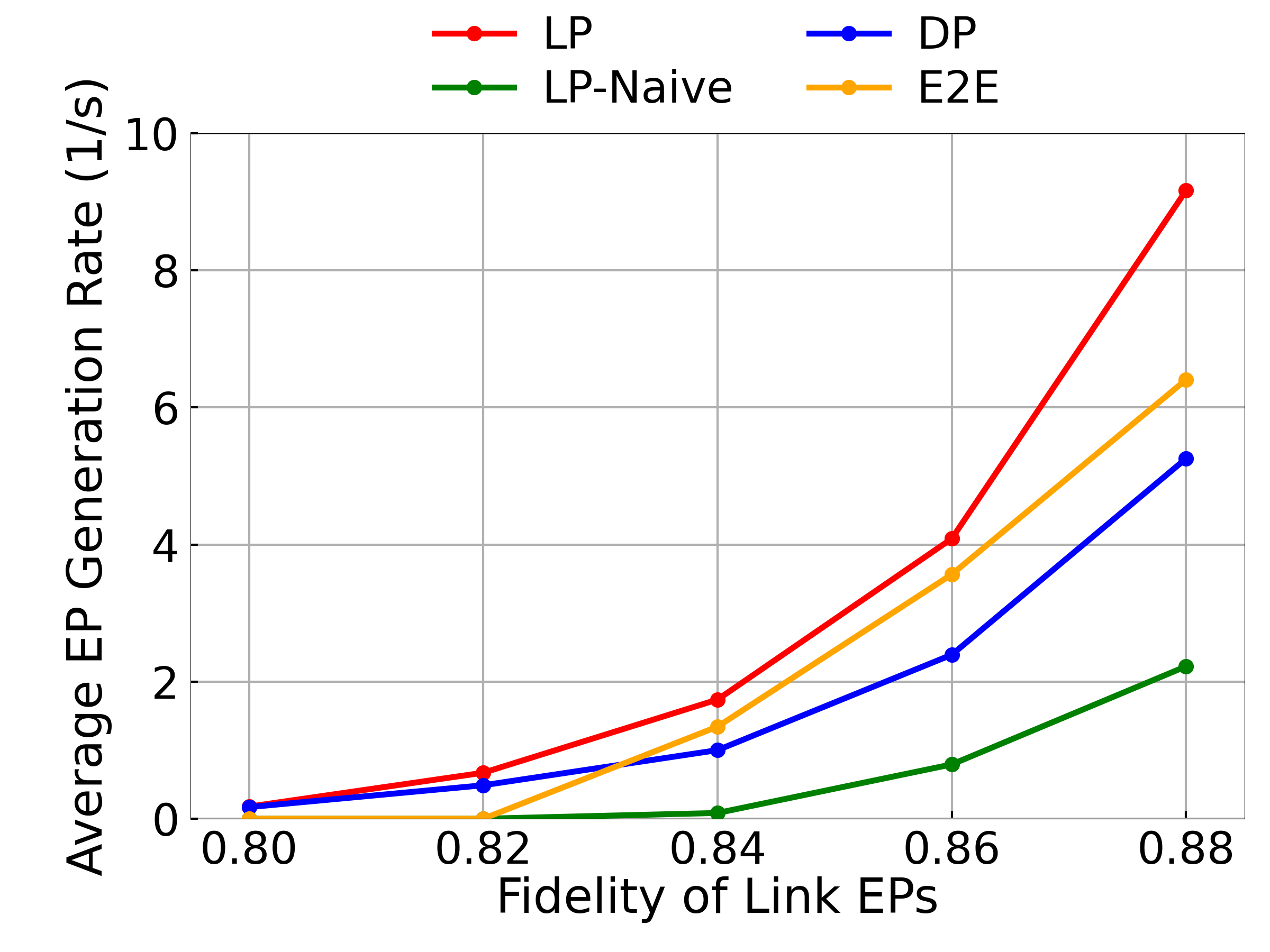}
    \vspace{0.05in}
    \caption{EP generation rates for varying the fidelity of link EPs. }
    \vspace{-0.1in}
    \label{fig:ep-link-fidelity}
\end{figure}

Fig.~\ref{fig:ep-link-fidelity} shows the results varying the fidelity for link EPs. Our \LP scheme still outperforms other schemes, and the performance of \ee and \LPNaive is highly influenced by link fidelity, and both perform poorly when link fidelity is low.

\begin{figure}[h]
    \centering
    \includegraphics[width=0.48\textwidth]{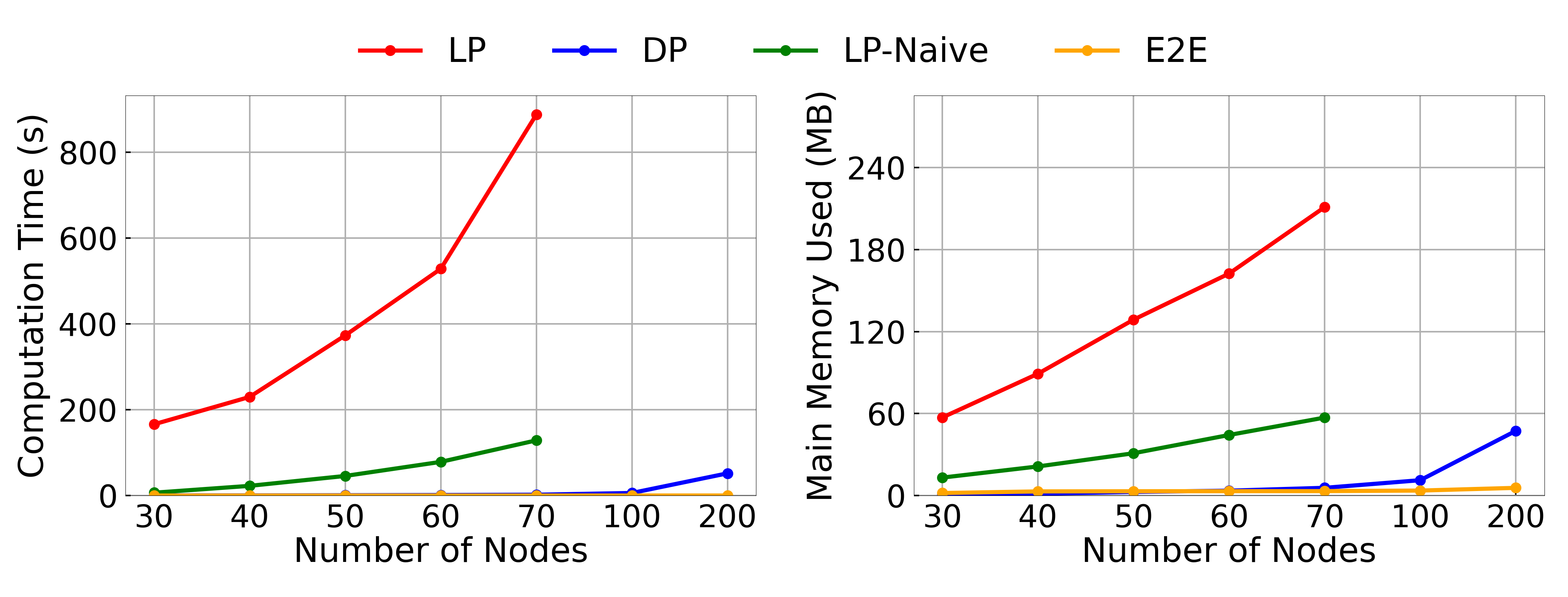}
    \caption{Computation time and main memory usage incurred in computing the solutions for EP generation schemes. (a) Time incurred, (b) Memory usage.}
    \vspace*{-0.1in}
    \label{fig:time_memory1}
\end{figure}

\begin{figure}[h]
    \centering
    \includegraphics[width=0.35\textwidth]{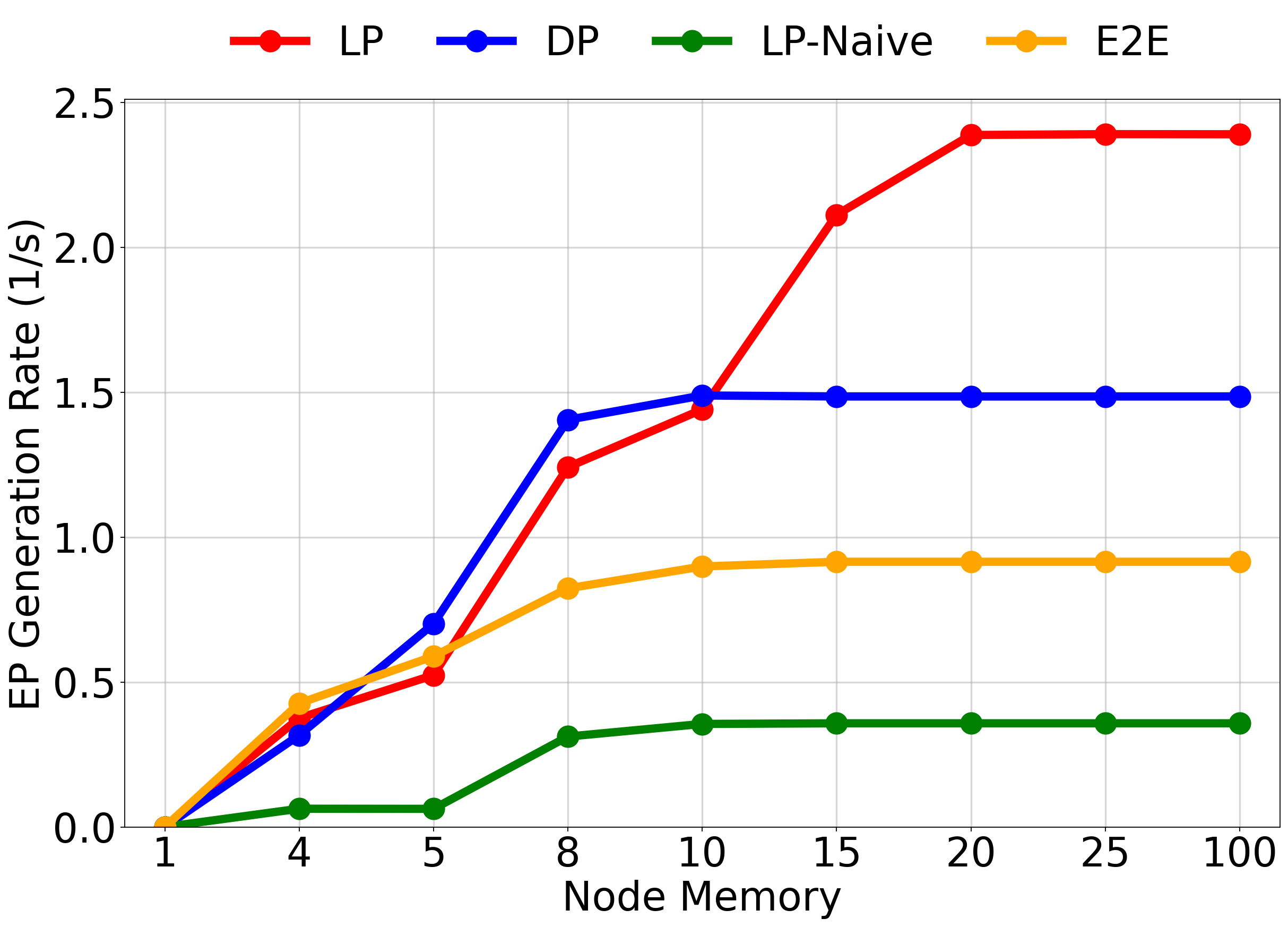}
    \vspace{0.05in}
    \caption{EP generation rates for varying the memory available at each node. }
    \vspace{-0.1in}
    \label{fig:ep-memory}
\end{figure}

Fig.~\ref{fig:ep-memory} shows the EP generation rates as we vary the per-node memory budget for the same source-destination pair. Our \LP scheme achieves the highest saturated rate, whereas \DP saturates earlier because it uses a single path, while \LP spreads across multiple paths that contend for scarce memory.

\begin{figure}[h]
    \centering
    \includegraphics[width=0.45\textwidth]{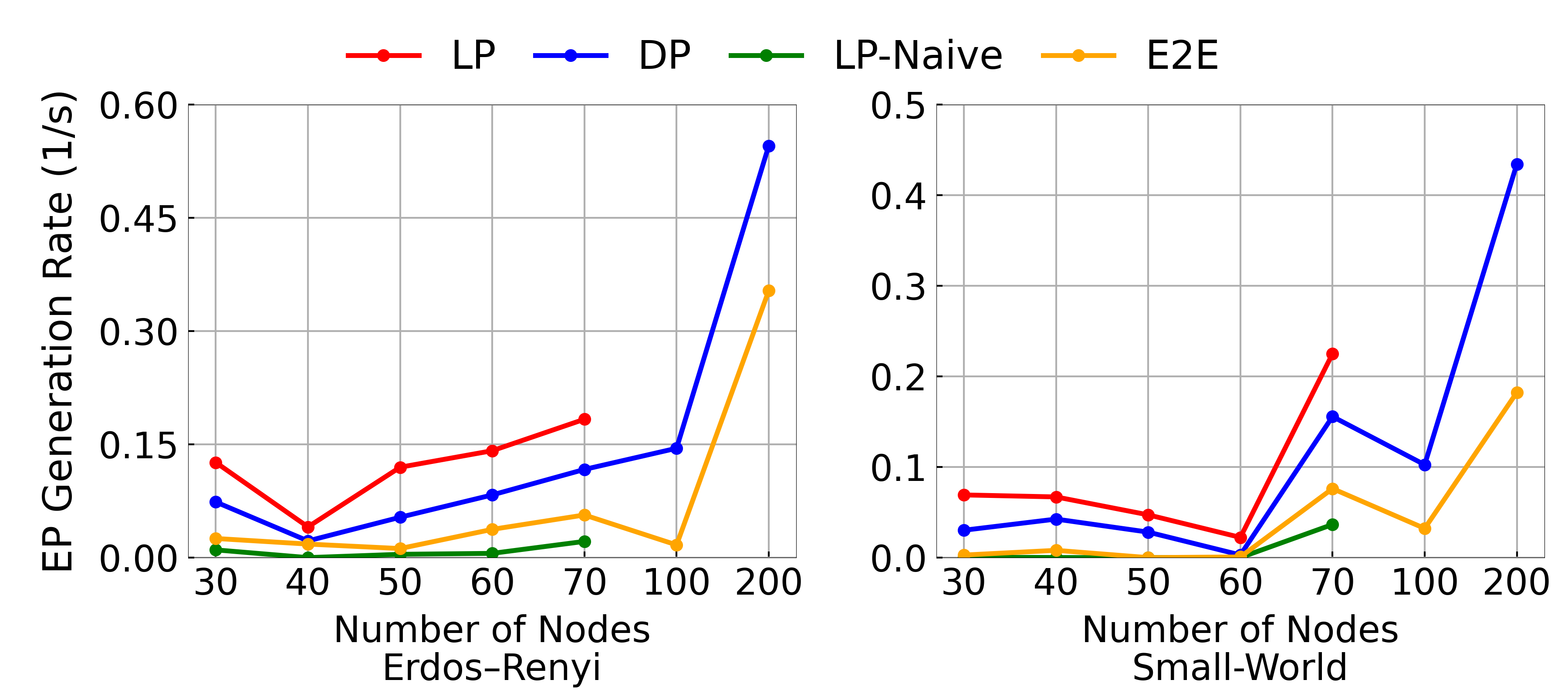}
    \vspace{0.05in}
    \caption{EP generation rates for varying the number of nodes under different topologies. }
    \vspace{-0.1in}
    \label{fig:ep-topo}
\end{figure}

Fig.~\ref{fig:ep-topo} shows the EP generation rates as we vary the number of nodes under the Erdős--Rényi and Small-World topologies. Our \LP scheme still achieves the highest rate across both topologies.

Overall, we observe that the LP scheme achieves the highest entanglement generation rate in most evaluated settings, while DP can outperform LP under tight per-node memory constraints.

\softpara{Scalability of the Schemes.} 
We evaluated the time and memory performance of our schemes for EP generation on an AMD Ryzen 9 7950X3D CPU machine. See Fig.~\ref{fig:time_memory1}. We observe that our \DP algorithm took about 52 seconds for a 200-node network with low memory consumption. In contrast, the \LP and \LPNaive schemes exhibit steep increases in both computation time and memory usage with an increase in the network size. In particular,
the LP scheme for EP generation took approximately 2.8 to
14.8 minutes for networks with 30 to 70 nodes and required substantial memory usage.

\softpara{Summary.}
Thus, in conjunction with the overall performance of the EP generation schemes shown in Fig.~\ref{fig:ep-4-parameters}, while \LP achieves the highest entanglement generation rates, \DP offers better scalability and is more practical for larger networks, making it the most suitable choice in most realistic deployment scenarios.

We note that the LP-based approach remains viable in settings where computational and memory resources are not limiting factors. The underlying optimization is highly parallelizable and well-suited to high-performance computing. LP solvers (e.g., Gurobi, used in our experiments) support multi-threaded solving, and our LP models exhibit structure and sparsity that make them amenable to parallel acceleration. On more powerful centralized systems or supercomputers, we expect the same optimization to complete in seconds to minutes, making this approach practical for offline scheduling or periodic re-optimization in large-scale networks—particularly in applications where maximizing entanglement throughput is the primary objective.

\begin{figure*}[t]
    \centering
    \includegraphics[width=0.95\textwidth]{figures_gg/evaluation/ghz_dp_greedy_rate_fidelity_2x4.pdf}
    \caption{GHZ generation rates (a)--(d) and average output fidelities (e)--(h) for the six DP heuristics and \Greedy{}. Columns vary terminal count, fusion success probability, average remote-EP generation rate, and per-node quantum memory capacity. Points average five NetSquid runs. Dotted horizontal lines mark the fidelity threshold of 0.8; measured fidelities below it are retained.}
    \vspace*{-0.1in}
    \label{fig:ghz-generation}
\end{figure*}

\begin{figure*}[t]
    \centering
    \includegraphics[width=0.95\textwidth]{figures_gg/evaluation/ghz_lp_rate_fidelity_2x4.pdf}
    \caption{GHZ generation rates (a)--(d) and average output fidelities (e)--(h) for the vector-labeled LP heuristics, the scalar-fidelity-labeled LP heuristics, and \GHZLPOpt{}. Columns show the same four sweeps as Fig.~\ref{fig:ghz-generation}. Points average five NetSquid runs. Dotted horizontal lines mark the fidelity threshold of 0.8. \GHZLPOpt{} is shown only at three terminals; larger cases are omitted due to the memory limit. Very low fidelities fall below the plotted range; fidelity estimates are omitted when too few outputs are available.}
    \vspace*{-0.1in}
    \label{fig:ghz-lp-generation}
\end{figure*}

\subsection{\bf Evaluation Results for GHZ Generation}

Figs.~\ref{fig:ghz-generation} and~\ref{fig:ghz-lp-generation} compare the DP heuristics and \Greedy{}, and the LP methods, respectively. Panels (a)--(d) show generation rates and panels (e)--(h) show average output fidelities.
The columns vary terminal count, fusion success probability, average\footnote{In particular, we assign heterogeneous EP generation rates to the star graph edges: edge rates increase in steps of 5 (e.g., 60, 65, 70, 75, 80), with an average of 70; this asymmetry more clearly exposes efficiency differences among schemes. 
We note that, in our preliminary version~\cite{purification-qcnc}, we instead used a uniform rate of 50 on all edges; in that setting of low and uniform edge rates, the \DP schemes performed slightly worse than \LPNaive. In the present GHZ comparison, \Greedy{} serves a similar baseline role to the conference-version \LPNaive{}: both use predetermined purification stages: \LPNaive{} purifies only the final state, whereas \Greedy{} purifies the source EPs and the final state.} EP generation rate on the star graph edges, and per-node quantum memory capacity.

\softpara{Generation Rate and Fidelity.}
\begin{itemize}
\item The rate- and fidelity-discretized DP counterparts produce similar generation rates and output fidelities across most tested settings. Differences are more pronounced for larger targets, where rate discretization meets the fidelity threshold more consistently.
\item Compared with these methods, \Greedy{} generally achieves lower generation rates but higher output fidelity.
\item Among the LP methods, the vector-labeled heuristics generally achieve higher generation rates than their two-scalar counterparts, whose rates decline sharply for larger targets. However, the vector-labeled methods also exhibit fidelity shortfalls as the target size increases.
\item The memory sweep shows that the DP and vector-labeled LP methods saturate once sufficient memory is available; further memory increases neither improve their rates nor eliminate the remaining fidelity shortfalls.
\end{itemize}

\softpara{Representation and Discretization.}
\begin{itemize}
\item Both DP heuristics retain full GHZ-basis vectors to track the error distribution through fusion and purification, and generally achieve mean output fidelities above the threshold. Although the DP can discretize either rate or fidelity, the results favor rate discretization.
\item The vector-labeled methods achieve higher rates than their two-scalar counterparts at almost every tested point. This supports retaining the error distribution needed to evaluate subsequent purification, rather than approximating its effect from two components.
\item The DP estimates waiting times directly from the input rates of each candidate construction. In the LP, however, rates are optimization variables, so the waiting times are not known when constructing the fidelity labels. The two-pass approximation therefore estimates waiting times from the first-pass flows and uses them to construct the labels for the second solve. Since the second solve can change the flow allocation, these waiting-time estimates may not match the final solution, which can contribute to the fidelity shortfalls observed for the LP methods.
\item The high fidelity and low rate of \Greedy{} are consistent with its fixed purification rules. \Greedy{} purifies every source EP and the final GHZ state, whereas DP and LP select purification steps to maximize the generation rate subject to the fidelity threshold. Its fixed policy does not omit checks when fewer would suffice, potentially spending more resources on fidelity improvements beyond the threshold.
\end{itemize}

\begin{figure}[t]
    \centering
    \includegraphics[width=0.45\textwidth]{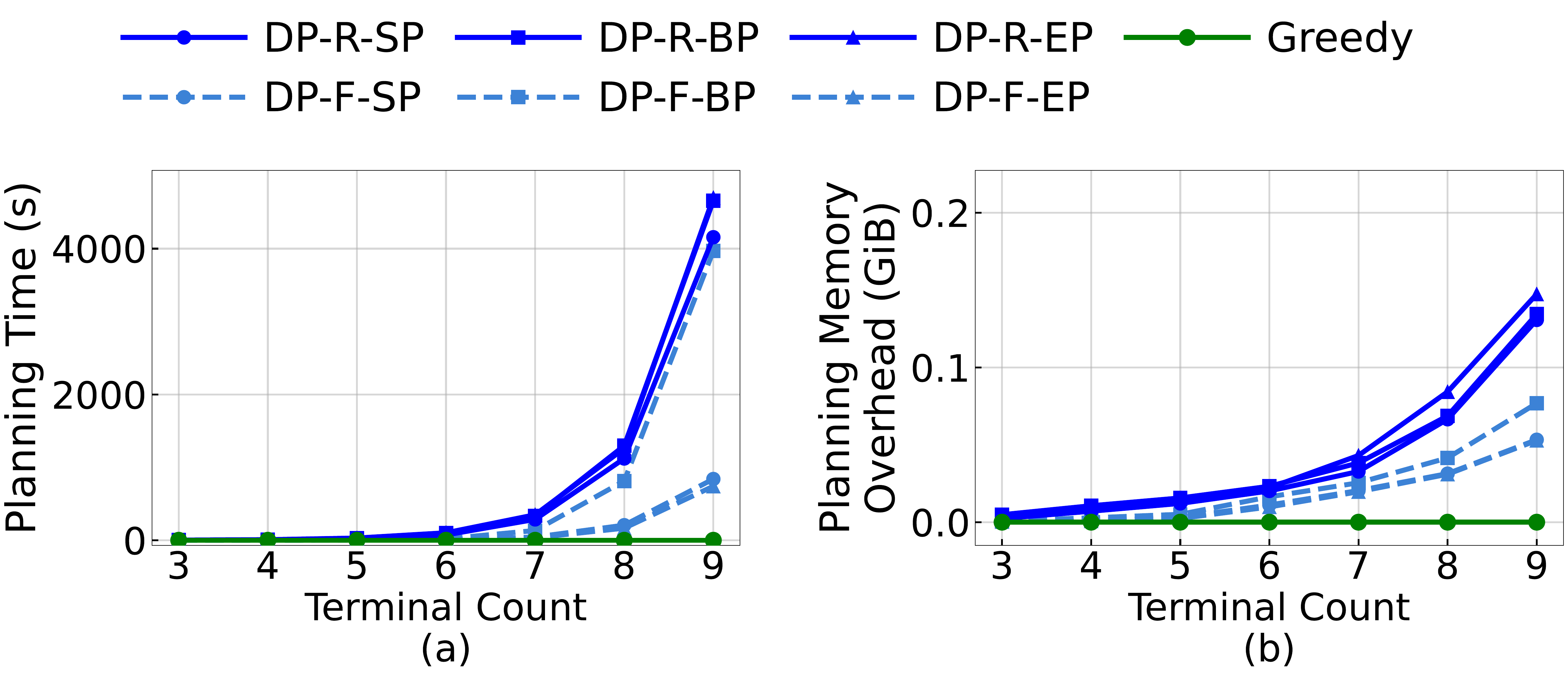}
    \caption{Planning time (a) and planning-memory overhead (b) for the six GHZ DP heuristics and \Greedy{}, as terminal count varies. Overhead includes planning and solution extraction but excludes NetSquid simulation; memory overhead is the increase in resident memory over the planning baseline. Both overheads are zero for \Greedy{}, since it has no offline planning stage.}
    \vspace*{-0.1in}
    \label{fig:ghz-time-memory}
\end{figure}

\begin{figure}[t]
    \centering
    \includegraphics[width=0.45\textwidth]{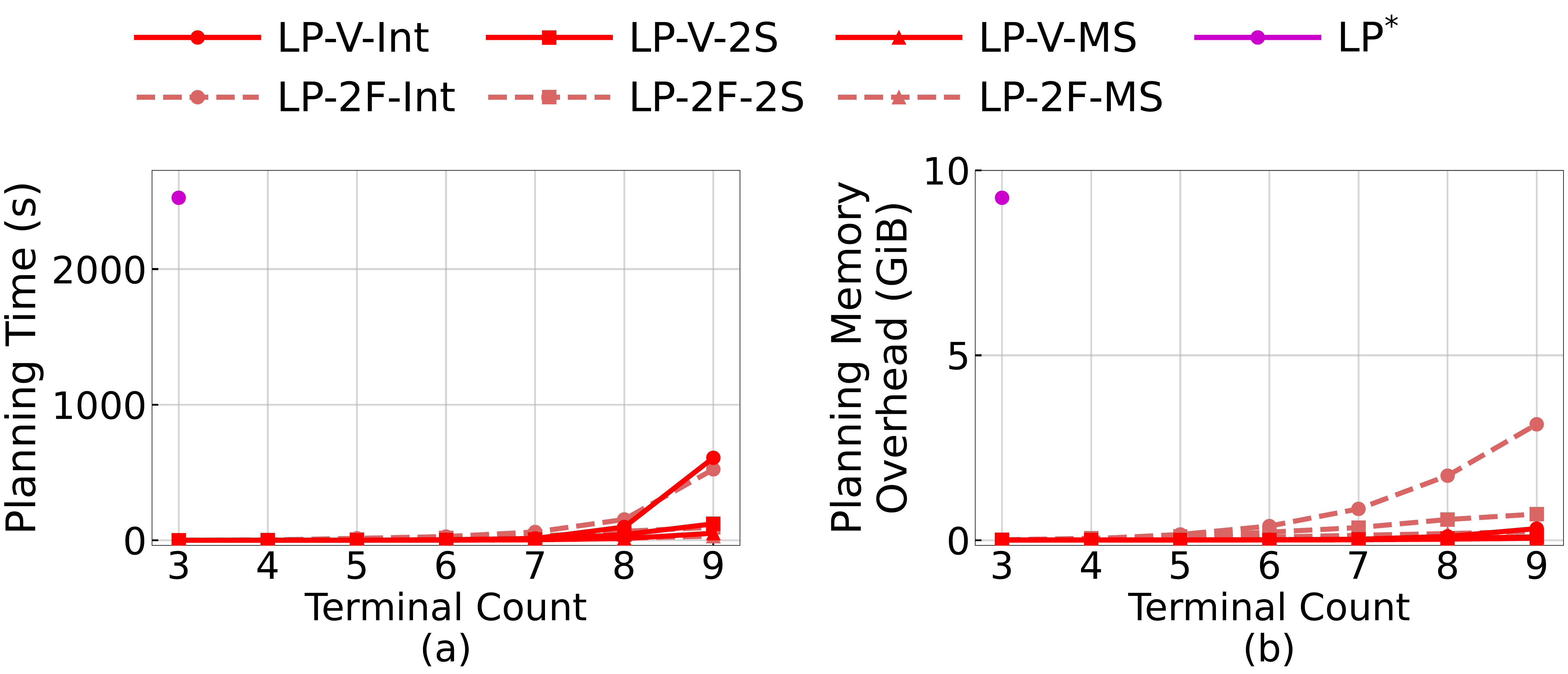}
    \caption{Planning time (a) and planning-memory overhead (b) for the GHZ LP methods, as terminal count varies. Overhead is measured as in Fig.~\ref{fig:ghz-time-memory}. \GHZLPOpt{} is shown only at three terminals; larger cases are omitted due to the memory limit.}
    \vspace*{-0.1in}
    \label{fig:ghz-lp-time-memory}
\end{figure}

\softpara{Scalability of GHZ Schemes.}
Figs.~\ref{fig:ghz-time-memory} and~\ref{fig:ghz-lp-time-memory} compare planning time and memory overhead as the terminal count increases.
\begin{itemize}
\item The fidelity-discretized DP heuristics require less planning time and memory than their rate-discretized counterparts throughout this sweep, but these savings come with the fidelity trade-offs discussed above.
\item The two-scalar LP heuristics use more planning memory and do not consistently reduce planning time, yet generally achieve lower generation rates and fidelities than their vector-labeled counterparts. Taken together, the rate and overhead results further support the use of full-vector fidelity labels.
\item \GHZLPOpt{} completes at three terminals but reaches the memory limit at four. At three terminals, all three vector-labeled LP heuristics match its measured rate and fidelity with substantially lower planning overhead, demonstrating their effectiveness.
\end{itemize}

\softpara{Summary.} 
Overall, the GHZ results favor rate discretization for DP and full-vector fidelity labels for LP. The rate-discretized DP heuristics offer the most favorable rate--fidelity trade-off in our experiments, but incur high planning time for larger targets. The vector-labeled LP heuristics plan faster, although they meet the fidelity threshold less consistently. The greedy baseline provides high fidelity with no offline planning overhead, but generally lower generation rates.

\begin{figure*}[t]
    \centering
    \includegraphics[width=0.95\textwidth]{figures_gg/evaluation/graph_rate_fidelity_2x4.pdf}
    \caption{Graph-state generation rates (a)--(d) and average output fidelities (e)--(h) for \GraphDPV{}, \GraphDPF{}, \GraphLPV{}, \GraphLPF{}, \GraphLPOpt{}, and \Greedy{}. Columns vary terminal count, fusion success probability, average target-edge EP generation rate, and per-node quantum memory capacity. Points average five NetSquid runs. Dotted horizontal lines mark the fidelity threshold of 0.8; measured fidelities below it are retained. \GraphLPOpt{} is shown only at three terminals; larger cases are omitted due to the planning-time limit. Fidelity estimates are omitted when too few outputs are available.}
    \vspace*{-0.1in}
    \label{fig:graph-generation}
\end{figure*}

\subsection{\bf Evaluation Results for Graph States}
Fig.~\ref{fig:graph-generation} shows graph-state generation rates and average output fidelities as we vary terminal count, fusion success probability, average target-edge EP generation rate, and per-node quantum memory capacity.

\softpara{Generation Rate and Fidelity.}
\begin{itemize}
\item Both DP heuristics achieve mean output fidelities above the threshold. The fidelity-discretized DP generally achieves higher rates, whereas the rate-discretized DP generally provides higher fidelity.
\item The vector-labeled LP generally achieves generation rates comparable to or higher than the DP heuristics, but its output fidelity is unstable and often falls below the threshold. The scalar LP offers a poor rate--fidelity trade-off: even for small targets, its high output fidelity is accompanied by low generation rates.
\item Unlike in the GHZ experiments, \Greedy{} has much lower throughput.
\item The memory sweep shows that the DP and vector-labeled LP methods saturate once sufficient memory is available; further memory increases neither improve their rates nor eliminate the remaining fidelity shortfalls.
\end{itemize}

\softpara{Representation and Discretization.}
\begin{itemize}
\item Unlike in the GHZ experiments, the graph-state results favor fidelity-discretized DP: both DP variants achieve average output fidelities above the threshold, but the fidelity-discretized variant generally achieves higher generation rates.
\item For the LP methods, the vector-labeled heuristic offers a better rate--fidelity trade-off than its scalar counterpart. This supports the importance of retaining the full fidelity vector.
\item The waiting-time limitation discussed for the GHZ LP also applies here: the second solve can change the first-pass flows used to estimate storage times, contributing to fidelity shortfalls.
\item The fixed purification strategy of \Greedy{} works well as a high-fidelity baseline for GHZ generation, but does not generalize well to graph-state generation, possibly because construction and purification choices depend more strongly on the target topology.
\end{itemize}

\begin{figure}[t]
    \centering
    \includegraphics[width=0.45\textwidth]{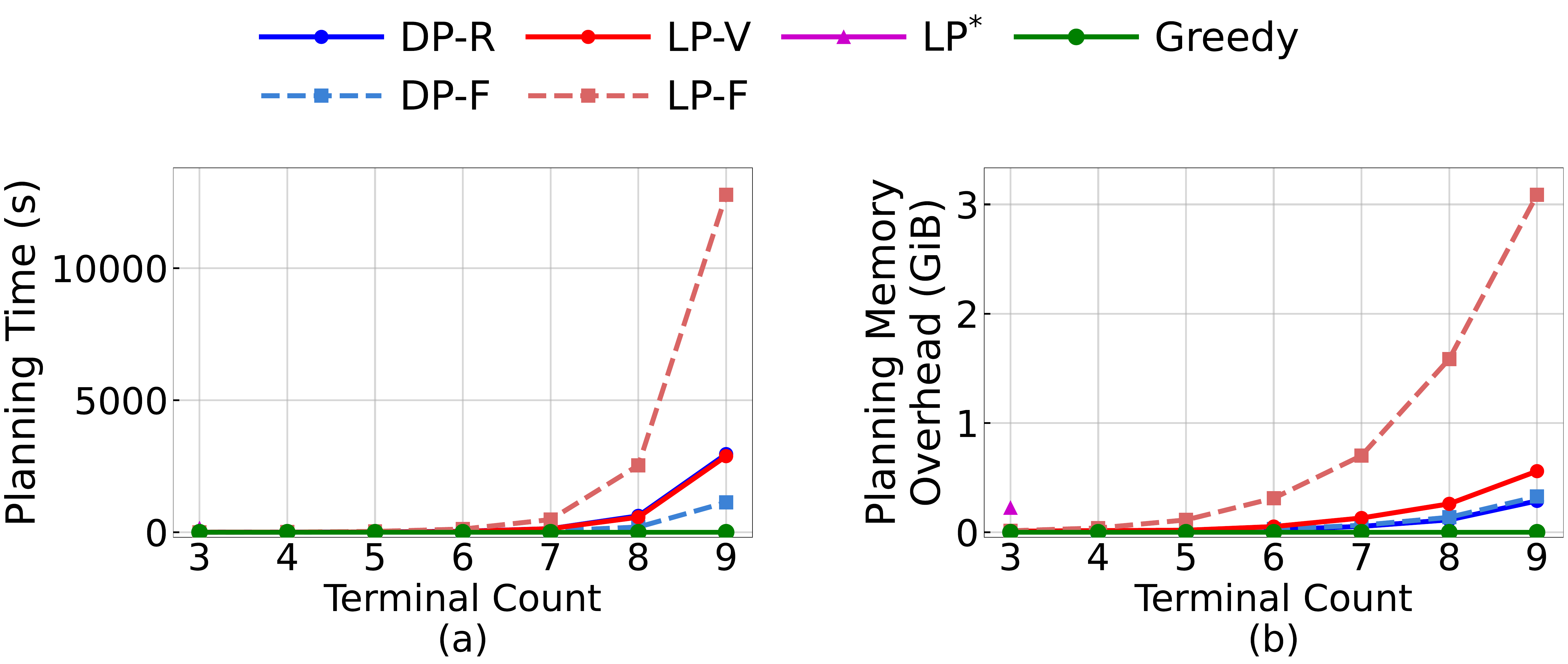}
    \caption{Planning time (a) and planning-memory overhead (b) for the graph-state methods, as terminal count varies. Overhead includes planning and solution extraction but excludes NetSquid simulation; memory overhead is the increase in resident memory over the planning baseline. Both overheads are zero for \Greedy{}, since it has no offline planning stage. \GraphLPOpt{} is shown only at three terminals; larger cases are omitted due to the planning-time limit.}
    \vspace*{-0.1in}
    \label{fig:graph-time-memory}
\end{figure}

\softpara{Scalability of the Graph State Schemes.} 
Fig.~\ref{fig:graph-time-memory} compares planning time and memory overhead as the terminal count increases.
\begin{itemize}
\item The fidelity-discretized DP requires less planning time than the rate-discretized DP throughout this sweep. Unlike in the GHZ experiments, its time savings are not accompanied by memory savings.
\item The vector-labeled LP requires less planning time and memory than the scalar LP. Thus, the scalar LP does not provide an overhead advantage to offset its poorer rate--fidelity trade-off.
\item \GraphLPOpt{} completes at three terminals but reaches the planning-time limit at four. At three terminals, \GraphLPV{} achieves a comparable rate and fidelity with substantially lower planning overhead.
\end{itemize}

\softpara{Summary.} 
Overall, fidelity-discretized DP offers a favorable rate--fidelity trade-off with faster planning, while rate-discretized DP generally provides higher fidelity. Vector-labeled LP performs better than scalar LP with lower overhead, but exhibits fidelity shortfalls. The greedy heuristic is uncompetitive.

The DP trade-offs in the GHZ and graph-state experiments are consistent with their candidate-selection rules: rate-discretized DP retains the highest-fidelity candidate in each rate cell, whereas fidelity-discretized DP retains the highest-rate candidate in each fidelity cell. They favor fidelity or generation rate, respectively. This helps explain the different performance trends observed for GHZ and graph states.

\section{\bf Deployment Considerations and Other Generalizations}
\label{sec:gen}

We now discuss deployment considerations and two extensions of our framework: maximizing fidelity subject to a generation-rate threshold, and maximizing the expected number of outputs meeting a fidelity threshold from a finite supply of EPs.

\para{Implementation and Deployment Considerations.} 
Although our proposed algorithms are evaluated in simulation, we have designed them with near-term deployment feasibility in mind. The purification-augmented swapping/fusion trees and level-based structures are inherently compatible with repeater-based quantum network architectures, which currently represent the most promising approach for wide-area quantum communication~\cite{briegel1998quantum,van2008system}. Our framework does not rely on hardware-specific assumptions and can incorporate fidelity models calibrated from physical devices. In particular, our fidelity abstractions can be readily adapted to leading hardware platforms, including nitrogen-vacancy centers, trapped ions, and photonic systems employing fusion-based gates. While NetSquid provides high-level abstractions of quantum noise and operations, our modeling is intentionally modular and can be mapped to fidelity profiles measured from physical devices.

Regarding computational cost, the entanglement generation plans produced by our algorithms depend only on the network topology and a set of input parameters (e.g., source-destination pairs, fidelity thresholds), both of which are relatively stable in many deployment scenarios. As a result, generation plans can be pre-computed offline for a wide range of expected inputs—for example, EP generation between a small number of source-destination pairs, GHZ states over a few terminal nodes, or frequently requested small graph states.

For cases where pre-computation is not feasible, our optimal schemes can still be deployed effectively in real networks, particularly when integrated into a control plane that periodically re-optimizes entanglement paths. As discussed earlier, our LP-based approach for EP generation completed in under an hour on a standard workstation for a 70-node network, and we expect it to run in seconds to minutes on high-performance computing platforms. If sub-optimal generation rates are acceptable, our efficient heuristics and approximate schemes are well-suited for use in real-time settings with moderate computational resources.

Our algorithms ultimately produce execution plans and are thus best integrated into path/route selection layers of quantum network stacks, analogous to the network layer in the classical OSI model. We view this work as contributing foundational techniques for fidelity-aware entanglement orchestration, capable of informing the design and deployment of control protocols in future quantum repeater networks and metropolitan-scale testbeds.

\para{Maximizing Fidelity Under Rate Constraint.} 
We note that our dynamic programming (DP) schemes easily generalize to problems where the optimization objective is to maximize fidelity under a minimum generation rate constraint. 
For GHZ and graph states (\S\ref{sec:ghz-dp} and \S\ref{sec:graph-dp}), the optimal DP stores frontiers of annotated trees under rate and EP-resource constraints. We select the tree with the largest ideal-component fidelity among the final frontiers satisfying the required rate and source-capacity constraints. Only the final selection rule changes; the DP recurrences remain unchanged.
Similarly, the DP formulation for the generation of EPs, i.e., for the \fed problem in Eqns.~\ref{eqn:dp-recursion-link-latency}-\ref{eqn:dp_purification_latency},
can be converted to maximization of fidelity for a given latency (and thus rate) constraint.

\para{Quantity Model.}
The authors of~\cite{santos2023shortest} consider a quantity model in which the available resources are specified as discrete counts of link-level EPs, rather than as generation rates. Under this model, we aim to maximize the expected number of end-to-end EPs generated with fidelity above a threshold. Our DP-based schemes can be adapted naturally to this setting. We describe the adaptation for EP generation below; the same idea extends to GHZ and general graph-state generation. This model is also closely related to resource-efficiency metrics such as elementary-pair consumption and output yield.

\softpara{Quantity-Model Problem for EPs.}
Consider a quantum network in which each link $(v_i,v_j)$ is associated with a finite number $n_{ij}$ of available EPs, each with link fidelity $f_{ij}$. Given a source-destination pair $(s,d)$ and a fidelity threshold $\fth$, the quantity-model problem is to construct a purification-augmented swapping tree that maximizes the expected number of EPs generated over $(s,d)$ with fidelity at least $\fth$, using only the available link-level EPs.

\softpara{DP-based Solution.}
Our DP formulation for EP generation in \S\ref{sec:ep} computes optimal generation latency through the recurrences in Eqs.~\ref{eqn:dp-recursion-link-latency}--\ref{eqn:dp_purification_latency}. For the quantity model, the DP would instead track integer resource counts and evaluate expected output counts by accounting for operation success and failure. The waiting-based synchronization factors used for generation rates do not scale counts of already-available EPs. Thus, the DP can still combine smaller construction trees through swapping or purification. Instead of computing generation latency, its recurrences must account for how many EPs each operation consumes and how many it produces upon success or failure, weighted by the probabilities of these outcomes.

The same approach extends to the multipartite settings: the GHZ and graph-state DPs can retain their qubit-set or connected-subgraph decompositions while tracking available resource counts and using count-based transitions to evaluate the expected output quantity.

\softpara{LP-based Extension.}
A quantity-model extension of the LP-based schemes is also possible, but it requires a different optimization formulation. The current LPs use continuous flow variables to represent generation rates. In the quantity model, an integer or mixed-integer formulation could represent operation attempts and available resource counts; the formulation must also account for operation success and failure. We leave the development of such an integer or mixed-integer LP formulation to future work.

\section{\bf Conclusion}
\label{sec:conc}

The efficient generation of high-fidelity entangled states is a fundamental requirement for realizing the full potential of quantum networks. In this work, we developed a comprehensive framework for fidelity-aware entanglement distribution using purification-enhanced swapping, fusion, and graph-state growth operations. Our DP and LP formulations maximize generation rate under a fidelity threshold for EP distribution and second-stage GHZ and general graph-state generation. We provide optimal formulations under the stated models, discretization, and imposed purification bounds, together with computationally efficient heuristics. GHZ-basis and graph-basis fidelity vectors track the error distributions needed to evaluate construction and purification. The framework remains adaptable to different operation and fidelity models.

Through analytical modeling and extensive simulations using the NetSquid quantum network simulator, we demonstrated that our LP approach improves EP generation rates over the compared prior and naive schemes, with an average increase of 160\% and gains of up to 330\% in some scenarios.
For GHZ and graph states, we are not aware of directly comparable prior algorithms beyond our preliminary work~\cite{purification-qcnc}, so our comparisons use greedy baselines and reduced scalar-label LP variants. The DP heuristics generally achieve better performance than the greedy baselines, while vector-labeled LP heuristics outperform their scalar-label counterparts. These results underscore the potential of our techniques to support efficient entanglement services in future quantum networks.

As future work, we aim to extend our framework to support real-time, adaptive, and fault-tolerant generation strategies, and to enable integration with quantum network protocol stacks--paving the way for deployment in emerging quantum network infrastructures.

\section*{\bf Acknowledgment}
This work was supported by NSF awards NQVL-2410725, FET-2106447 and CNS-2128187.

\section*{\bf Code and Data Availability}
The custom algorithm, simulation, and analysis code is available at \url{https://github.com/yukun-yang/Entanglement_generation_purification}. 
(NetSquid and Gurobi must be obtained separately under their respective licensing terms.)
The numerical data underlying the EP, GHZ, and graph-state evaluation figures are also available at the same repository. 

\section*{\bf Author Contributions}
Y.Y.\ led the algorithm design, theoretical analysis, and manuscript writing for the GHZ and graph-state components, implemented the methods, and performed the simulations and analysis. X.F.\ implemented the EP methods, performed the corresponding simulations, and analyzed the results. C.R.\ contributed to the GHZ methods in the preliminary conference version of the work. H.G.\ supervised the overall research, developed or guided the algorithm design and theoretical analysis, and reviewed the proofs and manuscript.

\section*{\bf AI Assistance}
The authors used ChatGPT as an editorial aid for manuscript preparation, and OpenAI's Codex to assist with code development. The authors remain responsible for the scientific content, code, and reported results.

\section*{\bf Competing Interests}
The authors declare no competing interests.

\bibliographystyle{IEEEtran}
\bibliography{all_shorten}

\appendices
\section{\bf Steady-State Waiting Factors in the Analytical Model}
\label{app:waiting-model}

Our rate analysis uses the steady-state waiting model adopted in~\cite{swapping-tqe-22}. The basic idea is that when an operation requires multiple independently generated operand states, it can begin only after the last required operand arrives. Thus, the relevant waiting time is the maximum of the operand-generation times.

In our setting, link-EP generation is modeled as a Poisson process; hence, the inter-arrival times are exponential. Accordingly, if $X_1,\ldots,X_k$ denote the generation times of $k$ independently generated operand states, and these random variables are independent and identically distributed exponential random variables with a common mean $\mu$, then the expected waiting time is
\[
\mathbb{E}[\max(X_1,\ldots,X_k)] = H_k \mu,
\]
where $H_k=\sum_{i=1}^{k}\frac{1}{i}$ is the $k$-th harmonic number.

For the case of two operands, which is the case used in the swapping-latency model in Eqn.~\eqref{eqn:EP_swaplatency}, we have
\[
\mathbb{E}[\max(X_1,X_2)] = \Bigl(1+\frac12\Bigr)\mu = \frac32 \mu.
\]
This is the source of the coefficient $\frac32$ in Eqn.~\eqref{eqn:EP_swaplatency}. Equivalently, when converting waiting time into an effective rate-scaling factor, the corresponding factor is the reciprocal, namely $\frac{2}{3}$.

Similarly, for three operands, we have
\[
\mathbb{E}[\max(X_1,X_2,X_3)] = \Bigl(1+\frac12+\frac13\Bigr)\mu = \frac{11}{6}\mu.
\]
Thus, when synchronizing three equal-rate inputs under the same steady-state waiting model, the corresponding effective rate-scaling factor is $\frac{6}{11}$, which is used for synchronizing the two component graph states and the connecting EP in {\tt Fuse-Two-Components}; see Appendix~\ref{app:graph-dp-details}.

The waiting factors used in our rate analysis are therefore exact in the idealized i.i.d.\ exponential case, and serve as a standard approximation when operand-generation processes are approximately Poisson/exponential with comparable means. More detailed hardware- or protocol-specific waiting models can be incorporated by replacing the corresponding latency and rate-scaling expressions, without changing the overall DP/LP structure.

\section{\bf Optimality Proofs for EP Distribution}
\label{app:ep-optimality-proofs}

\para{Proof of Theorem~\ref{thm:ep-dp}.}
\begin{prf}
We prove by induction on $h$ that, for every pair of network nodes $(x,y)$ and every discretized fidelity value $f$, the value $L(x,y,f,h)$ equals the minimum expected generation latency among all purification-augmented swapping trees of height at most $h$ that generate an \eps over $(x,y)$ with fidelity at least $f$.

\softpara{Base Case.}
For $h=0$, the only feasible tree is a single-node tree corresponding to a direct link-\eps. Thus, Eqn.~\ref{eqn:dp-recursion-link-latency} yields the minimum achievable latency.

\softpara{Inductive Step.}
Assume that the claim holds for height at most $h-1$. Consider any purification-augmented swapping tree of height at most $h$ that generates an \eps over $(x,y)$ with fidelity at least $f$. Its root must be of one of the following three types: (i) the tree already has height at most $h-1$, (ii) the root is a purification node whose child has height at most $h-1$, or (iii) the root is a swapping node whose two children each have height at most $h-1$. These are exactly the three cases captured by Eqns.~\ref{eqn:dp-recursion-latency}--\ref{eqn:dp_purification_latency}. By the induction hypothesis, the optimal child subtree or subtrees in each case are given by the corresponding $L(\cdot,\cdot,\cdot,h-1)$ values. Hence $L(x,y,f,h)$ is the minimum achievable latency among all purification-augmented swapping trees of height at most $h$ that generate an \eps over $(x,y)$ with fidelity at least $f$.

By Lemma~\ref{lemma:nocycle}, there exists an optimal purification-augmented swapping tree of height at most $2n$. Therefore, $L(s,d,\lceil f^\tau\rceil,2n)$ is the optimal latency value for the given \fed instance, and the corresponding tree $T(s,d,\lceil f^\tau\rceil,2n)$ is an optimal purification-augmented swapping tree. Under the adopted no-pipelining model, maximizing generation rate is equivalent to minimizing generation latency; therefore, $T(s,d,\lceil f^\tau\rceil,2n)$ is an optimal solution to the \fed problem.\footnote{The generation rate is the inverse of the generation latency when \eps are generated one at a time, i.e., without pipelining, as in~\cite{swapping-tqe-22}. Pipelining requires additional memory units at the nodes along the network path.}
\end{prf}

\para{Proof of Theorem~\ref{thm:ep-lp}.}
\begin{prf}
We show that feasible level-based swapping structures for the \gfed problem and feasible solutions of the LP correspond to each other without changing the achieved target generation rate.

First, consider any feasible level-based swapping structure for the given instance. Each generated \eps type in that structure is identified by its endpoint pair and discretized fidelity value, and thus corresponds to a hypervertex in $H_{Q,\Delta}$. Likewise, each swapping or purification operation used in the structure corresponds to one of the hyperarcs in $H_{Q,\Delta}$, and each direct generation of a link-\eps corresponds to a Start hyperarc. Assign to each hyperarc the rate at which the corresponding operation is used in the level-based swapping structure. Then the Start-arc capacity constraints are satisfied because no link-\eps resource is used above its available generation rate; the flow-conservation constraints at the $\texttt{avail}(\cdot)$ hypervertices are satisfied because the total rate at which a given \eps type is produced equals the total rate at which it is either consumed by later operations or delivered to the terminal demand; and the operation constraints at the $\texttt{swap}(\cdot)$, $\texttt{purify1}(\cdot)$, and $\texttt{purify2}(\cdot)$ hypervertices are satisfied because they are exactly the adopted rate-update rules for swapping and purification, including the corresponding success-probability and waiting factors. Thus, every feasible level-based swapping structure induces a feasible LP solution with the same total target generation rate.

Conversely, for any feasible LP solution, we first eliminate positive-flow cycles among the available-state vertices. When identifying cycles, we follow each operation from its available input states to its available output state. On a simple directed cycle, reduce every operation's input flow by the same amount until one becomes zero. Since each operation produces no more flow than it consumes from each input, this reduction cannot create a shortage of any EP type. Repeating this procedure removes all cycles. Any surplus is then removed by reducing source and producer flows, working backward from outputs to sources while keeping terminal flows fixed. The resulting flow satisfies the original constraints, preserves every terminal rate, and uses no greater source capacities. We use $z$ to denote this acyclic flow. We construct a level-based swapping structure by unfolding $z$ backward from the $\texttt{term}$ hypervertex.

For each hyperarc entering $\texttt{term}$, assign it the same rate as in $z$. We then recursively process the required rates. Whenever a required rate reaches an $\texttt{avail}(u,v,f)$ hypervertex, we use the LP flow-conservation constraint at that hypervertex to split the required rate among its incoming hyperarcs, without exceeding their LP values. If such an incoming hyperarc is a Start hyperarc, this corresponds directly to generating link-\epss at that rate. If such an incoming hyperarc is a consolidation hyperarc from a $\texttt{swap}(\cdot)$, $\texttt{purify1}(\cdot)$, or $\texttt{purify2}(\cdot)$ hypervertex, then we instantiate the corresponding swapping or purification operation at that same rate, and recursively request from the corresponding operand $\texttt{avail}(\cdot)$ hypervertices the input rates required by the associated LP constraint.

By construction, every instantiated operation obeys exactly the adopted rate-update rule, and no Start hyperarc is used above its LP value. Moreover, each swapping or purification step incurs a multiplicative factor strictly smaller than $1$ in the forward direction. As shown above, this property allows cycles to be removed without reducing the terminal generation rates or increasing source usage. In the resulting acyclic structure, following operations backward cannot revisit the same available-state vertex along a single branch. Since the hypergraph contains only finitely many vertices, each branch contains only finitely many operation steps. By flow conservation, each branch ends at a Start hyperarc that generates a link EP. Therefore, the recursive construction terminates at the link-EP sources, yielding a finite level-based swapping structure.

Finally, LP flow-conservation constraints ensure that rates are consistent across all intermediate \eps types, and the terminal generation rate produced by the resulting level-based swapping structure equals the LP objective value. Thus, every feasible LP solution induces a feasible level-based swapping structure with the same achieved generation rate.

Therefore, feasible LP solutions and feasible level-based swapping structures realize exactly the same achievable target generation rates. Hence, the optimum LP value equals the optimum \gfed value, and any optimal LP solution induces an optimal level-based swapping structure.
\end{prf}

\section{\bf Fidelity Updates for GHZ Operations}
\label{app:purification-detail}

\para{GHZ-Diagonal Representation.}
For an $N$-qubit GHZ state, we use the basis
\[
\ket{\psi^{(N)}_{p,\mathbf{b}}}
=
\frac{
    \ket{0,\mathbf{b}}
    +
    (-1)^p\ket{1,\overline{\mathbf{b}}}
}{\sqrt{2}},
\quad
p\in\{0,1\},
\mathbf{b}\in\{0,1\}^{N-1}.
\]
Equivalently,
\begin{equation}
    \ket{\psi^{(N)}_{p,\mathbf{b}}}
    =
    \frac{1}{\sqrt{2}}
    \sum_{x\in\{0,1\}}
        (-1)^{px}
        \ket{x,\mathbf{b}\oplus x\mathbf{1}},
    \label{eqn:app-ghz-binary-form}
\end{equation}
where $\mathbf{1}$ is the all-one string. A GHZ-diagonal state is
\[
    \rho
    =
    \sum_{p,\mathbf{b}}
        P[p,\mathbf{b}]
        \ket{\psi^{(N)}_{p,\mathbf{b}}}
        \!\bra{\psi^{(N)}_{p,\mathbf{b}}},
    \qquad
    \sum_{p,\mathbf{b}}P[p,\mathbf{b}]=1.
\]
The fidelity with the ideal GHZ state is the distinguished component
$F^+=P[0,\mathbf{0}]$. No symmetry among the remaining vector entries is assumed in the derivations below.

\para{Fusion Retain.}
Consider the basis states
$\ket{\psi^{(n_1)}_{p_1,\mathbf{b}_1}}$ and
$\ket{\psi^{(n_2)}_{p_2,\mathbf{b}_2}}$. Let $q_1$ and $q_2$ be the selected qubits of the first and second states, respectively. Without loss of generality, take $q_1$ and $q_2$ to be the first qubits of their respective GHZ states. Qubit $q_1$ controls a CNOT whose target is $q_2$. Let $m\in\{0,1\}$ be the subsequent $Z$-basis measurement outcome on $q_2$. Using
Eqn.~\ref{eqn:app-ghz-binary-form}, let $x,y\in\{0,1\}$ denote the computational-basis values of $q_1$ and $q_2$, respectively, before the
CNOT. The joint input state is
\[
\frac{1}{2}
\sum_{x,y\in\{0,1\}}
(-1)^{p_1x+p_2y}
\ket{x,\mathbf{b}_1\oplus x\mathbf{1}}
\ket{y,\mathbf{b}_2\oplus y\mathbf{1}}.
\]
The CNOT transforms the value of $q_2$ as $y\mapsto x\oplus y$. Conditioning on measurement outcome $m$ imposes $x\oplus y=m$, or equivalently $y=x\oplus m$. After removing the measured qubit $q_2$, the normalized conditional state before feed-forward is
\[
\ket{\phi_m}
=
\frac{1}{\sqrt{2}}
\sum_{x\in\{0,1\}}
(-1)^{p_1x+p_2(x\oplus m)}
\ket{x,\mathbf{b}_1\oplus x\mathbf{1}}
\ket{\mathbf{b}_2\oplus(x\oplus m)\mathbf{1}}.
\]
When $m=1$, applying Pauli-$X$ to every retained qubit from the second state removes the $m\mathbf{1}$ term. The remaining factor $(-1)^{p_2m}$ is a global phase. Thus, both measurement branches produce
\[
    p_{\rm out}=p_1\oplus p_2,
    \qquad
    \mathbf{b}_{\rm out}
    =
    \Pi_{\mathcal{L}_1,\mathcal{L}_2
    \rightarrow\mathcal{L}_{\rm out}}
    (\mathbf{b}_1,\mathbf{b}_2).
\]
Here, $\mathcal{L}_i$ is the ordered list of physical qubits of input $i$ other than $q_i$, $\mathcal{L}_{\rm out}$ is their canonical output order, and $\Pi$ places each syndrome bit at the position of its corresponding physical leaf. This permutation is necessary because the two input leaf lists need not already form the canonical output order.

For independent input vectors $\mathbf{P}_1$ and $\mathbf{P}_2$, both measurement outcomes are accepted and mapped, by feed-forward correction, to the same output GHZ-basis label. Since their probabilities sum to one, the resulting output vector is already normalized. The exact output vector
is
\begin{align*}
    P_{\rm FR}[p,\mathbf{c}]
    &=
    \sum_{\substack{
        p_1,p_2,\mathbf{b}_1,\mathbf{b}_2:\\
        p_1\oplus p_2=p,\\
        \Pi_{\mathcal{L}_1,\mathcal{L}_2
        \rightarrow\mathcal{L}_{\rm out}}
        (\mathbf{b}_1,\mathbf{b}_2)=\mathbf{c}
    }}
        P_1[p_1,\mathbf{b}_1]
        P_2[p_2,\mathbf{b}_2].
\end{align*}

\para{Fusion Discard.}
Fusion discard first performs fusion retain and then measures and discards the retained qubit $q_1$ in the $X$ basis. The resulting state is an $(n_1+n_2-2)$-GHZ state. Let
\[
    \mathcal{R}
    =
    (u_0,u_1,\ldots,u_{n_1+n_2-3})
\]
be the canonical ordering of the remaining physical qubits. For fixed input labels $\mathbf{b}_1$ and $\mathbf{b}_2$, let $s(u)$ denote the syndrome bit assigned to qubit $u$ after combining the two input qubit lists, and define
\[
    \mathbf{s}
    =
    \bigl(s(u_0),s(u_1),\ldots,s(u_{n_1+n_2-3})\bigr).
\]
Before measuring $q_1$, the fusion-retain state has the form
\[
    \frac{1}{\sqrt{2}}
    \left(
        \ket{0}_{q_1}\ket{\mathbf{s}}
        +
        (-1)^{p_1\oplus p_2}
        \ket{1}_{q_1}\ket{\overline{\mathbf{s}}}
    \right).
\]
Let $r\in\{0,1\}$ be the $X$-basis measurement outcome, with $r=0$ for $\ket{+}$ and $r=1$ for $\ket{-}$. Projecting $q_1$ onto this outcome gives the normalized state
\[
    \frac{1}{\sqrt{2}}
    \left(
        \ket{\mathbf{s}}
        +
        (-1)^{p_1\oplus p_2\oplus r}
        \ket{\overline{\mathbf{s}}}
    \right).
\]
Thus, the outcome $r=1$ introduces an additional relative minus sign. A conditional Pauli-$Z$ operation on any one of the remaining qubits removes this sign. After correction, both outcomes produce
\[
    \frac{1}{\sqrt{2}}
    \left(
        \ket{\mathbf{s}}
        +
        (-1)^{p_1\oplus p_2}
        \ket{\overline{\mathbf{s}}}
    \right).
\]
We choose $u_0$ as the first qubit in the output GHZ-basis ordering. This is only a labeling convention and gives $u_0$ no special physical role. Relative to $u_0$, define the output bit-error string by
\[
    \Delta_{\mathcal{R}}(\mathbf{b}_1,\mathbf{b}_2)
    =
    \bigl(
        s(u_1)\oplus s(u_0),
        \ldots,
        s(u_{n_1+n_2-3})\oplus s(u_0)
    \bigr).
\]
Rewriting the corrected state in this GHZ-basis ordering gives, up to an irrelevant global phase,
\[
    \frac{1}{\sqrt{2}}
    \left(
        \ket{0,\Delta_{\mathcal R}(\mathbf b_1,\mathbf b_2)}
        +
        (-1)^{p_1\oplus p_2}
        \ket{1,\overline{\Delta_{\mathcal R}
        (\mathbf b_1,\mathbf b_2)}}
    \right).
\]
Therefore, the output label is
\[
    p_{\rm out}=p_1\oplus p_2,
    \qquad
    \mathbf{b}_{\rm out}
    =
    \Delta_{\mathcal R}(\mathbf b_1,\mathbf b_2).
\]
For independent input vectors $\mathbf{P}_1$ and $\mathbf{P}_2$, the exact fusion-discard output vector is consequently
\[
    P_{\rm FD}[p,\mathbf{c}]
     =
    \sum_{\substack{
        p_1,p_2,\mathbf{b}_1,\mathbf{b}_2:\\
        p_1\oplus p_2=p,\\
        \Delta_{\mathcal R}(\mathbf{b}_1,\mathbf{b}_2)=\mathbf{c}
    }}
        P_1[p_1,\mathbf{b}_1]
        P_2[p_2,\mathbf{b}_2].
\]

For the special case $n_1=n_2=2$, each input is a Bell-diagonal state described by the four labels $(p,b)\in\{0,1\}^2$. The fusion-discard label
map becomes
\[
    p_{\rm out}=p_1\oplus p_2,
    \qquad
    b_{\rm out}=b_1\oplus b_2,
\]
which is the usual Bell-state update for entanglement swapping after the measurement-dependent Pauli correction. The output is the ideal Bell state, with $(p_{\rm out},b_{\rm out})=(0,0)$, exactly when the two input labels are identical. Hence, the output fidelity is
\[
    F_{\rm FD}
    =
    P_{\rm FD}[0,0]
    =
    \sum_{p,b}P_1[p,b]P_2[p,b].
\]

For isotropic Bell-state inputs with fidelities $F_1$ and $F_2$, the
probability vectors satisfy
\[
    P_i[0,0]=F_i,
    \quad
    P_i[p,b]=\frac{1-F_i}{3}
    \quad\text{for }(p,b)\neq(0,0).
\]
Substituting these probabilities and simplifying, we obtain
\begin{align*}
    F_{\rm FD}
    &=
    F_1F_2
    +
    3
    \left(\frac{1-F_1}{3}\right)
    \left(\frac{1-F_2}{3}\right)
    \notag\\
    &=
    F_1F_2+\frac{(1-F_1)(1-F_2)}{3}
    \notag\\
    &=
    \frac{1}{4}
    \left(
        1+
        3\frac{4F_1-1}{3}\frac{4F_2-1}{3}
    \right).
\end{align*}
This agrees with the entanglement-swapping result in Ref.~\cite{ghaderibaneh2023generation}. For general non-isotropic inputs, however, the output fidelity is
\[
    F_{\rm FD}
    =
    \sum_{p,b}P_1[p,b]P_2[p,b],
\]
and therefore depends on all four components of each Bell-state probability vector; it cannot, in general, be determined from $F_1$ and $F_2$ alone.

\para{Full Bit-Pattern Purification.}
Let the target and sacrifice be equal-size $N$-qubit GHZ-basis states with labels 
$(p_{\rm t},\mathbf{b}_{\rm t})$ and
$(p_{\rm s},\mathbf{b}_{\rm s})$, respectively. Using
Eqn.~\ref{eqn:app-ghz-binary-form}, their joint state is
\begin{align*}
&\ket{\psi^{(N)}_{p_{\rm t},\mathbf{b}_{\rm t}}}
 \ket{\psi^{(N)}_{p_{\rm s},\mathbf{b}_{\rm s}}}
\notag\\
&\quad=
\frac{1}{2}
\sum_{x,y\in\{0,1\}}
(-1)^{p_{\rm t}x+p_{\rm s}y}
\ket{x,\mathbf{b}_{\rm t}\oplus x\mathbf{1}}_{\rm t}
\ket{y,\mathbf{b}_{\rm s}\oplus y\mathbf{1}}_{\rm s},
\end{align*}
where $x$ and $y$ are the binary branch variables of the target and sacrifice states. Apply a transversal CNOT from every target qubit to the corresponding sacrifice qubit. The joint state after the CNOTs is 
\begin{align*}
\frac{1}{2}
\sum_{x,y\in\{0,1\}}
&(-1)^{p_{\rm t}x+p_{\rm s}y}
\ket{x,\mathbf{b}_{\rm t}\oplus x\mathbf{1}}_{\rm t}
\notag\\
&\otimes
\ket{
x\oplus y,\,
\mathbf{b}_{\rm t}\oplus\mathbf{b}_{\rm s}
\oplus(x\oplus y)\mathbf{1}
}_{\rm s}.
\end{align*}

All sacrifice qubits are then measured in the $Z$ basis. Let $d=x\oplus y$ denote the measurement outcome on the first sacrifice qubit. The outcomes on the remaining sacrifice qubits form the string
\[
\mathbf{b}_{\rm t}\oplus\mathbf{b}_{\rm s}
\oplus d\mathbf{1}.
\]
The purification succeeds when all sacrifice measurement outcomes are equal. Since the first outcome is $d$, every entry of the remaining string must also equal $d$. Thus, the acceptance condition is
\begin{align*}
\mathbf{b}_{\rm t}\oplus\mathbf{b}_{\rm s}
\oplus d\mathbf{1}
&=d\mathbf{1},
\notag\\
\mathbf{b}_{\rm t}\oplus\mathbf{b}_{\rm s}
&=\mathbf{0},
\notag\\
\mathbf{b}_{\rm t}
&=\mathbf{b}_{\rm s}.
\end{align*}
Hence, the bit-pattern check accepts exactly when the target and sacrifice have the same bit-error string. The value of $d$ does not affect this acceptance condition.

For an accepted measurement branch, $y=x\oplus d$. Substituting this relation into the phase factor gives
\[
(-1)^{p_{\rm t}x+p_{\rm s}y}
=
(-1)^{(p_{\rm t}\oplus p_{\rm s})x}
(-1)^{p_{\rm s}d}.
\]
The factor $(-1)^{p_{\rm s}d}$ is independent of $x$ and is therefore an irrelevant global phase. After discarding the measured sacrifice state, the normalized target state is
\[
\frac{1}{\sqrt{2}}
\sum_{x\in\{0,1\}}
(-1)^{(p_{\rm t}\oplus p_{\rm s})x}
\ket{x,\mathbf{b}_{\rm t}\oplus x\mathbf{1}}_{\rm t}
=
\ket{\psi^{(N)}_{p_{\rm t}\oplus p_{\rm s},
\mathbf{b}_{\rm t}}}.
\]
Consequently, the basis-label update is
\[
p_{\rm out}=p_{\rm t}\oplus p_{\rm s},
\qquad
\mathbf{b}_{\rm out}=\mathbf{b}_{\rm t},
\qquad
\text{provided that }
\mathbf{b}_{\rm t}=\mathbf{b}_{\rm s}.
\]

Now let $\mathbf{P}_{\rm t}$ and $\mathbf{P}_{\rm s}$ be independent GHZ-basis probability vectors. An accepted output with label $(p,\mathbf{b})$ can arise from any pair of phase labels satisfying $p_{\rm t}\oplus p_{\rm s}=p$, provided that both inputs have the same bit-error string $\mathbf{b}$. The corresponding unnormalized output probability is
\[
\widetilde{P}_{\rm B}[p,\mathbf{b}]
=
\sum_{\substack{
p_{\rm t},p_{\rm s}\in\{0,1\}:\\
p_{\rm t}\oplus p_{\rm s}=p
}}
P_{\rm t}[p_{\rm t},\mathbf{b}]
P_{\rm s}[p_{\rm s},\mathbf{b}].
\]
Equivalently, its two phase components are
\begin{align*}
\widetilde{P}_{\rm B}[0,\mathbf{b}]
&=
P_{\rm t}[0,\mathbf{b}]P_{\rm s}[0,\mathbf{b}]
+
P_{\rm t}[1,\mathbf{b}]P_{\rm s}[1,\mathbf{b}],
\notag\\
\widetilde{P}_{\rm B}[1,\mathbf{b}]
&=
P_{\rm t}[0,\mathbf{b}]P_{\rm s}[1,\mathbf{b}]
+
P_{\rm t}[1,\mathbf{b}]P_{\rm s}[0,\mathbf{b}].
\end{align*}

Summing over all accepted output labels gives the success probability
\[
q_{\rm B}
=
\sum_{p,\mathbf{b}}
\widetilde{P}_{\rm B}[p,\mathbf{b}]
=
\sum_{\mathbf{b}}
\left(
\sum_{p_{\rm t}}P_{\rm t}[p_{\rm t},\mathbf{b}]
\right)
\left(
\sum_{p_{\rm s}}P_{\rm s}[p_{\rm s},\mathbf{b}]
\right).
\]
The normalized conditional output vector is then
\[
P_{\rm B}[p,\mathbf{b}]
=
\frac{\widetilde{P}_{\rm B}[p,\mathbf{b}]}{q_{\rm B}}.
\]
In particular, the fidelity of the accepted state is
\[
F_{\rm B}
=
P_{\rm B}[0,\mathbf{0}]
=
\frac{
P_{\rm t}[0,\mathbf{0}]P_{\rm s}[0,\mathbf{0}]
+
P_{\rm t}[1,\mathbf{0}]P_{\rm s}[1,\mathbf{0}]
}{
q_{\rm B}
}.
\]

As a consistency check, consider the special case in which both the target and sacrifice states are Bell pairs with identical isotropic fidelity $F_{\rm t}=F_{\rm s}=F$. Their Bell-basis probability vectors are
\begin{align*}
P_{\rm t}[0,0]&=P_{\rm s}[0,0]=F,
\\
P_{\rm t}[p,b]&=P_{\rm s}[p,b]=\frac{1-F}{3}
\quad\text{for }(p,b)\neq(0,0).
\end{align*}
Substituting these probabilities into the full bit-pattern purification vector update and normalizing by its success probability gives
\[
F'
=
P_{\rm B}[0,0]
=
\frac{
F^2+\frac{1}{9}(1-F)^2
}{
F^2+\frac{2}{3}F(1-F)+\frac{5}{9}(1-F)^2
}.
\]
This is precisely the fidelity recurrence of the original Bennett entanglement-purification protocol~\cite{bennett-95}.

\para{Subset Bit-Pattern Purification.}
Let the $N$ target qubits be written as
\[
    (q_0,q_1,\ldots,q_k,q_{k+1},\ldots,q_{N-1}),
\]
where $q_0$ is the first qubit in the GHZ-basis ordering. For clarity of notation, we derive the subset check for the selected target qubits $q_0,q_1,\ldots,q_k$, i.e., the first qubit together with the next $k$ qubits. This choice is only for notational clarity; any other choice of $k$ checked target qubits together with $q_0$ can be treated by writing those checked qubits as $q_1,\ldots,q_k$ in the derivation.

Let the $(k+1)$-qubit sacrifice state be ordered as $(s_0,s_1,\ldots,s_k)$, with label 
$(p_{\rm s},\mathbf{b}_{\rm s})$ and 
$\mathbf{b}_{\rm s}=(b_{{\rm s},1},\ldots,b_{{\rm s},k})$.
The subset bit-pattern check applies a CNOT from $q_j$ to $s_j$ for every $j\in\{0,\ldots,k\}$. Let $x$ and $y$ be the binary branch variables of the target and sacrifice states, respectively. Before the CNOTs, the selected target qubits in one branch have computational values
\[
    q_0=x,\qquad
    q_j=b_{{\rm t},j}\oplus x
    \quad (1\leq j\leq k),
\]
while the sacrifice qubits have computational values
\[
    s_0=y,\qquad
    s_j=b_{{\rm s},j}\oplus y
    \quad (1\leq j\leq k).
\]
After the CNOTs, the target qubits are unchanged and each sacrifice qubit stores its old value plus the corresponding target value. Define $d=x\oplus y$. The $Z$-basis outcome on $s_0$ is therefore 
\[
y\oplus x=d.
\]
For $1\leq j\leq k$, the outcome on $s_j$ is
\[
    (b_{{\rm s},j}\oplus y)
    \oplus(b_{{\rm t},j}\oplus x)
    =
    b_{{\rm t},j}\oplus b_{{\rm s},j}\oplus d.
\]
Thus the outcomes on $(s_1,\ldots,s_k)$ form the string
\[
    (b_{{\rm t},1},\ldots,b_{{\rm t},k})
    \oplus\mathbf{b}_{\rm s}
    \oplus d\mathbf{1}.
\]
The protocol accepts when every measured sacrifice qubit has the same outcome. Since the outcome on $s_0$ is $d$, every outcome on $s_1,\ldots,s_k$ must also equal $d$. Equivalently, the above equation must be $d\mathbf{1}$. This cancels the common branch-dependent value $d$ and leaves exactly the
componentwise condition
\[
    (b_{{\rm t},1},\ldots,b_{{\rm t},k})
    =
    \mathbf{b}_{\rm s}.
\]
Since $y=x\oplus d$, the phase factor of the two input branches is therefore
\[
    (-1)^{p_{\rm t}x+p_{\rm s}y}
    =
    (-1)^{(p_{\rm t}\oplus p_{\rm s})x}
    (-1)^{p_{\rm s}d}.
\]
The factor $(-1)^{p_{\rm s}d}$ is independent of $x$ and is therefore a global phase for this accepted branch. The remaining $x$-dependent phase is exactly the GHZ phase label $p_{\rm t}\oplus p_{\rm s}$. The target qubits are not measured and are not changed by the CNOTs, so the target computational string remains
$\ket{x,\mathbf{b}_{\rm t}\oplus x\mathbf{1}}$. Hence, every accepted event has the output label
\[
    p_{\rm out}=p_{\rm t}\oplus p_{\rm s},
    \qquad
    \mathbf{b}_{\rm out}=\mathbf{b}_{\rm t}.
\]
Now write the output bit-error string as $(b_1,\ldots,b_{N-1})$. By the above equation, this is the same string as the target input bit-error string. The sacrifice is accepted only when its bit-error string equals the checked target entries $(b_1,\ldots,b_k)$. Therefore, the unnormalized accepted vector is
\begin{align*}
   &\widetilde{P}_{{\rm B},k}
    [p,(b_1,\ldots,b_{N-1})]
    \\
    &=
    \sum_{p_{\rm t}\oplus p_{\rm s}=p}
        P_{\rm t}[p_{\rm t},(b_1,\ldots,b_{N-1})]
        P_{\rm s}
        [p_{\rm s},(b_1,\ldots,b_k)],
\end{align*}
The success probability is obtained by summing over all accepted output labels,
\[
    q_{{\rm B},k}
    =
    \sum_{p,\mathbf{b}}
        \widetilde{P}_{{\rm B},k}[p,\mathbf{b}],
\]
and the conditional output vector is the normalized accepted vector,
\[
    P_{{\rm B},k}[p,\mathbf{b}]
    =
    \frac{\widetilde{P}_{{\rm B},k}[p,\mathbf{b}]}
    {q_{{\rm B},k}}.
\]

\para{Full Phase Purification.}
Let the target and sacrifice be equal-size $N$-qubit basis states with labels 
$(p_{\rm t},\mathbf{b}_{\rm t})$ and
$(p_{\rm s},\mathbf{b}_{\rm s})$. Reverse the transversal-CNOT direction used for full bit-pattern purification, so every sacrifice qubit controls the corresponding target qubit. Before the CNOT, let $x$ and $y$ denote the target and sacrifice branch variables. After the CNOT, the target reference bit is $x\oplus y$, and its remaining bits form
\[
    \mathbf{b}_{\rm t}\oplus\mathbf{b}_{\rm s}
    \oplus(x\oplus y)\mathbf{1}.
\]
The sacrifice computational string is unchanged by the CNOT. Setting $z=x\oplus y$, so that $x=z\oplus y$, the joint state after the CNOT is, up to normalization,
\[
    \sum_{z,y\in\{0,1\}}
        (-1)^{p_{\rm t}z}
        (-1)^{(p_{\rm t}\oplus p_{\rm s})y}
        \ket{z,
        \mathbf{b}_{\rm t}\oplus\mathbf{b}_{\rm s}
        \oplus z\mathbf{1}}_{\rm t}
        \ket{y,\mathbf{b}_{\rm s}\oplus y\mathbf{1}}_{\rm s}.
\]
Next, measure all sacrifice qubits in the $X$ basis. Encode a 
$\ket{+}$ outcome by $r_j=0$ and a 
$\ket{-}$ outcome by $r_j=1$, and let
$\mathbf{r}=(r_0,\ldots,r_{N-1})$ denote the complete measurement record. The measured $X$-basis vector is
\[
    \ket{\mathbf{r}}_{\scriptscriptstyle X}
    =
    \bigotimes_{j=0}^{N-1}
    \frac{\ket{0}+(-1)^{r_j}\ket{1}}{\sqrt{2}},
\]
and define
\[
    \operatorname{par}(\mathbf{r})
    =
    r_0\oplus r_1\oplus\cdots\oplus r_{N-1}.
\]
The sacrifice computational string is not changed by the transversal CNOT. Therefore, for a fixed sacrifice branch $y$, its computational string is
\[
    \ket{y,\mathbf{b}_{\rm s}\oplus y\mathbf{1}}_{\rm s}
    =
    \ket{
        y,\,
        b_{{\rm s},1}\oplus y,\,
        \ldots,\,
        b_{{\rm s},N-1}\oplus y
    }_{\rm s}.
\]
We use the one-qubit identity
\[
    {}_{\scriptscriptstyle X}\!\left\langle r \middle| a \right\rangle
    =
    2^{-1/2}(-1)^{ra},
    \qquad
    r,a\in\{0,1\},
\]
which follows directly from
$\ket{r}_{\scriptscriptstyle X}=(\ket{0}+(-1)^r\ket{1})/\sqrt{2}$. For a fixed sacrifice branch $y$, the overlap with the measured $X$-basis state is
\begin{align*}
    {}_{\scriptscriptstyle X}\!\left\langle \mathbf r \middle| y,\mathbf b_{\rm s}\oplus y\mathbf 1 \right\rangle_{\rm s}
    &=
    2^{-N/2}
    (-1)^{
        r_0y+
        \sum_{j=1}^{N-1}r_j(b_{{\rm s},j}\oplus y)
    }
    \notag\\
    &=
    2^{-N/2}
    (-1)^{\sum_{j=1}^{N-1}r_jb_{{\rm s},j}}
    (-1)^{y\operatorname{par}(\mathbf{r})}.
\end{align*}
The factor
$(-1)^{\sum_{j=1}^{N-1}r_jb_{{\rm s},j}}$ depends on the observed measurement record and on the sacrifice bit-error string, but it is independent of the retained target branch $z$. It is therefore only an overall conditional phase. The branch-dependent measurement phase is
$(-1)^{y\operatorname{par}(\mathbf{r})}$. To make the projection explicit, write the post-CNOT state as
\[
    \ket{\Omega}
    \propto
    \sum_{z,y\in\{0,1\}}
        (-1)^{p_{\rm t}z}
        (-1)^{(p_{\rm t}\oplus p_{\rm s})y}
        \ket{z,
        \mathbf{b}_{\rm t}\oplus\mathbf{b}_{\rm s}
        \oplus z\mathbf{1}}_{\rm t}
        \ket{y,\mathbf{b}_{\rm s}\oplus y\mathbf{1}}_{\rm s}.
\]
Projecting the sacrifice qubits onto the observed $X$-basis record $\mathbf{r}$ gives
\begin{align*}
    \hspace*{-0.4in} &{}_{\mathrm{s},{\scriptscriptstyle X}}\!
    \left\langle \mathbf{r} \middle| \Omega \right\rangle
    \\
    \hspace*{-0.4in}\propto&
    \sum_{z,y\in\{0,1\}}
        (-1)^{p_{\rm t}z}
        (-1)^{(p_{\rm t}\oplus p_{\rm s})y}
        \left(
            {}_{\scriptscriptstyle X}\!
            \left\langle \mathbf{r}
            \middle|
            y,\mathbf{b}_{\rm s}\oplus y\mathbf{1}
            \right\rangle_{\rm s}
        \right)
        \ket{z,
        \mathbf{b}_{\rm t}\oplus\mathbf{b}_{\rm s}
        \oplus z\mathbf{1}}_{\rm t}
    \\
    \hspace*{-0.4in}\propto&
    \sum_{z,y\in\{0,1\}}
        (-1)^{p_{\rm t}z}
        (-1)^{
            y\left(
                p_{\rm t}\oplus p_{\rm s}
                \oplus\operatorname{par}(\mathbf{r})
            \right)}
        \ket{z,
        \mathbf{b}_{\rm t}\oplus\mathbf{b}_{\rm s}
        \oplus z\mathbf{1}}_{\rm t}
    \\
    \hspace*{-0.4in}=&
    \sum_{z\in\{0,1\}}
        (-1)^{p_{\rm t}z}
        \left[
            \sum_{y\in\{0,1\}}
            (-1)^{
                y\left(
                    p_{\rm t}\oplus p_{\rm s}
                    \oplus\operatorname{par}(\mathbf{r})
                \right)}
        \right]
        \ket{z,
        \mathbf{b}_{\rm t}\oplus\mathbf{b}_{\rm s}
        \oplus z\mathbf{1}}_{\rm t}.
\end{align*}
The inner sum is
\[
    \sum_{y\in\{0,1\}}
        (-1)^{
        y\left(
            p_{\rm t}\oplus p_{\rm s}
            \oplus\operatorname{par}(\mathbf{r})
        \right)}.
\]
It is equal to $2$ when
$\operatorname{par}(\mathbf{r})=p_{\rm t}\oplus p_{\rm s}$ and equal to $0$ otherwise. Hence, for fixed pure GHZ-basis input labels, only measurement outcomes with parity
$\operatorname{par}(\mathbf{r})=p_{\rm t}\oplus p_{\rm s}$ occur. The protocol accepts even measurement parity, so the acceptance condition is
equivalent to
\[
    p_{\rm t}=p_{\rm s}.
\]
Under this condition, the retained target state is
\[
    \frac{1}{\sqrt{2}}
    \sum_{z\in\{0,1\}}
        (-1)^{p_{\rm t}z}
        \ket{z,
        (\mathbf{b}_{\rm t}\oplus\mathbf{b}_{\rm s})
        \oplus z\mathbf{1}}_{\rm t}.
\]
It therefore has the GHZ-basis label
\[
    p_{\rm out}=p_{\rm t},
    \qquad
    \mathbf{b}_{\rm out}
    =
    \mathbf{b}_{\rm t}\oplus\mathbf{b}_{\rm s}.
\]
The unnormalized accepted vector, success probability, and conditional output are thus
\begin{align*}
    \widetilde{P}_{\rm H}[p,\mathbf{b}]
    &=
    \sum_{\mathbf{b}_{\rm t}\oplus\mathbf{b}_{\rm s}=\mathbf{b}}
        P_{\rm t}[p,\mathbf{b}_{\rm t}]
        P_{\rm s}[p,\mathbf{b}_{\rm s}],
    \\
    q_{\rm H}
    &=
    \sum_{p,\mathbf{b}}
        \widetilde{P}_{\rm H}[p,\mathbf{b}],
    \\
    P_{\rm H}[p,\mathbf{b}]
    &=
    \frac{\widetilde{P}_{\rm H}[p,\mathbf{b}]}{q_{\rm H}}.
\end{align*}

\para{Output Fidelity and Iterated Purification.}
For any accepted operation, the output fidelity is the ideal component of the normalized vector. In particular,
\[
    F^+_{\rm out}
    =
    P_{\rm out}[0,\mathbf{0}]
    =
    \frac{\widetilde{P}_{\rm out}[0,\mathbf{0}]}
    {\sum_{p,\mathbf{b}}\widetilde{P}_{\rm out}[p,\mathbf{b}]}.
\]
For iterated purification, let
$\mathbf{P}_{\rm t}^{(0)}$ be the initial target vector and let
$\mathbf{P}_{\rm s}^{(i)}$ be the sacrifice vector used in round $i$.
The operation in round $i$ is
$\mathsf{Op}_i\in
\{\mathsf{Bit},\mathsf{Phase}\}$, 
corresponding to a bit-pattern check or a full phase check above. Each operation first forms an unnormalized accepted vector
$\widetilde{\mathbf{P}}_{\mathsf{Op}_i}$ by keeping only the measurement outcomes accepted by that check. The scalar
\[
    q_i
    =
    \sum_{p,\mathbf{b}}
        \widetilde{P}_{\mathsf{Op}_i}[p,\mathbf{b}]
\]
is therefore the success probability of round $i$. Equivalently, using the shorthand $q_{\mathsf{Op}_i}$ for the appropriate success-probability formula, we have
\begin{align*}
    q_i
    &=
    q_{\mathsf{Op}_i}
    \left(
        \mathbf{P}_{\rm t}^{(i-1)},
        \mathbf{P}_{\rm s}^{(i)}
    \right),\\
    \mathbf{P}_{\rm t}^{(i)}
    &=
    \mathsf{Op}_i
    \left(
        \mathbf{P}_{\rm t}^{(i-1)},
        \mathbf{P}_{\rm s}^{(i)}
    \right).
\end{align*}
Here $\mathsf{Op}_i(\cdot,\cdot)$ denotes the normalized conditional output vector,
$\mathbf{P}_{\rm t}^{(i)}
=\widetilde{\mathbf{P}}_{\mathsf{Op}_i}/q_i$, after the round is accepted. Thus the target vector produced by one accepted round becomes the target input to the next round. The probability that all $a$ rounds succeed is
\[
    q_{\rm tot}
    =
    \prod_{i=1}^{a}q_i,
\]
and the final fidelity after the accepted sequence is
$\mathbf{P}_{\rm t}^{(a)}[0,\mathbf{0}]$.

\section{\bf Additional DP Details for GHZ States}
\label{app:ghz-dp-heuristics}

This appendix provides additional details for the GHZ DP formulations in \S\ref{sec:ghz-dp}. We first prove Theorem~\ref{thm:ghz-dp}, then present the recurrences for \DPss{} and \DPtwoG{}, whose equations are omitted from the main text.

\para{Proof of Theorem~\ref{thm:ghz-dp}.}
\begin{prf}
We prove the following invariant: for every triplet
$\langle S,k,\overrightarrow{EP}\rangle$,
$\mathcal{T}(S,k,\overrightarrow{EP})$ contains one annotated tree for every distinct root vector that is achievable by a purification-augmented fusion tree generating a GHZ state over $S\cup\{v_1\}$ at a rate of at least $k$ while using an EP-rate vector bounded by $\overrightarrow{EP}$.

We use induction in the following partial order on subproblems: first by increasing $|S|$, and, for a fixed $S$, by decreasing the discretized-rate value $k$.

\softpara{Base Case.}
We start the induction with $S=\{v_j\}$ and $k=K_q$. No purification tree can achieve this rate within the grid, since its root operation would require a child rate above $K_q$. The frontier therefore contains only the source tree, provided the source can supply rate $K_q$ within the EP budget, and is empty otherwise. Its root vector is $\mathbf{P}^{(2)}_j$. Purification trees at lower rates are covered by the inductive step below.

\softpara{Inductive Step.}
Consider a subproblem $\langle S,k,\overrightarrow{EP}\rangle$ with $|S|\geq1$, assuming that all preceding frontiers already contain a feasible tree for every distinct achievable root vector. Any feasible tree is either a source tree already included by the initialization, or ends in one of the root operations listed in the recursive step:
(i) a fusion operation, (ii) a same-stream full purification operation, (iii) a two-stream full bit-pattern purification or full phase purification operation, or (iv) a strict subset bit-pattern purification operation. These possibilities are exactly the candidate families
$\mathcal{C}_f$,
$\mathcal{C}_{p1}^{\mathsf{Bit}}$,
$\mathcal{C}_{p1}^{\mathsf{Phase}}$,
$\mathcal{C}_{p2}^{\mathsf{Bit}}$,
and $\mathcal{C}_{p2}^{\mathsf{Phase}}$ in Eqn.~\ref{eqn:dp-GHZ}.

In the fusion case, the child subproblems involve strict subsets of $S$, and therefore precede the current subproblem in the induction order. In the purification cases, each same-subset child must be generated at a strictly larger input rate, namely
$\left\lceil \frac{2}{p_p}k \right\rceil$ for same-stream purification or
$\left\lceil \frac{3}{2p_p}k \right\rceil$ for two-stream purification. In subset bit-pattern purification, the sacrificial child is over a strict subset $S'\subset S$. Hence these purification child subproblems precede the current subproblem in the induction order, because the DP is filled in decreasing order of the discretized-rate parameter for a fixed $S$. Therefore, all child frontiers required by Eqn.~\ref{eqn:dp-GHZ} are already available.

By the induction hypothesis, the child frontiers contain annotated trees for all achievable child root vectors. Applying the appropriate vector update from \S\ref{sec:model} to every feasible combination of child trees, together with the source initialization, therefore produces every achievable parent root vector. Conversely, every operation candidate is obtained by combining feasible child trees with a physically allowed root operation, and is therefore itself feasible. The pruning operator removes only repeated root vectors while retaining one annotated tree for each remaining vector. Hence the invariant holds for $\mathcal{T}(S,k,\overrightarrow{EP})$.

Finally, the \gdp solution chooses the maximum discretized rate $k_m$ for which there exists some EP-consumption vector $\overrightarrow{EP}$ and some final annotated tree
\[
T\in\mathcal{T}(\{v_2,v_3,\ldots,v_m\},k_m,\overrightarrow{EP})
\]
such that
\[
F^+(\mathbf{P}_T)\ge \fth.
\]
By the frontier invariant proved above, this is precisely the largest achievable discretized generation rate subject to the fidelity threshold. The corresponding tree is therefore an optimal solution to the \gdp problem under the adopted discretized-rate model.
\end{prf}

\para{Self-Purification (\DPss).}
In this formulation, purification is restricted to same-stream full bit-pattern and full phase purification. The recurrence is
\begin{align*}
\mathbf{P}(S,k,\overrightarrow{EP})
=
\operatorname{Select}_{F^+}
\left\{
\mathbf{P}_f,\,
\mathbf{P}_{p1}^{\mathsf{Bit}},\,
\mathbf{P}_{p1}^{\mathsf{Phase}}
\right\}.
\end{align*}
\vspace*{-0.2in}
\begin{footnotesize}
\begin{eqnarray*}
&& \hspace*{-0.4in} \mathbf{P}_f =
\underset{
\substack{
\emptyset\neq S_1\subset S\\
\overrightarrow{EP}_1+\overrightarrow{EP}_2\leq\overrightarrow{EP}
}}
{\operatorname{Select}_{F^+}}
\mathsf{Fuse}\!\left(
\begin{aligned}
&\mathbf{P}\!\left(
S_1,\left\lceil \tfrac{3}{2p_f}k \right\rceil,
\overrightarrow{EP}_1
\right),\\
&\mathbf{P}\!\left(
S\setminus S_1,\left\lceil \tfrac{3}{2p_f}k \right\rceil,
\overrightarrow{EP}_2
\right)
\end{aligned}
\right),
\\[1ex]
&& \hspace*{-0.4in} \mathbf{P}_{p1}^{\mathsf{Bit}} =
\mathsf{Bit}\!\left(
\begin{aligned}
&\mathbf{P}\!\left(
S,\left\lceil \tfrac{2}{p_p}k \right\rceil,
\overrightarrow{EP}
\right),\\
&\mathbf{P}\!\left(
S,\left\lceil \tfrac{2}{p_p}k \right\rceil,
\overrightarrow{EP}
\right)
\end{aligned}
\right),
\\[1ex]
&& \hspace*{-0.4in} \mathbf{P}_{p1}^{\mathsf{Phase}} =
\mathsf{Phase}\!\left(
\begin{aligned}
&\mathbf{P}\!\left(
S,\left\lceil \tfrac{2}{p_p}k \right\rceil,
\overrightarrow{EP}
\right),\\
&\mathbf{P}\!\left(
S,\left\lceil \tfrac{2}{p_p}k \right\rceil,
\overrightarrow{EP}
\right)
\end{aligned}
\right).
\end{eqnarray*}
\end{footnotesize}

\para{EP-based Purification (\DPtwoG).}
In this formulation, bit-pattern purification is restricted to EP-sized checks, while same-stream full phase purification is still allowed. The recurrence is
\begin{align*}
    \mathbf{P}(S,k,\overrightarrow{EP})
    =
    \operatorname{Select}_{F^+}
    \left\{
    \begin{aligned}
    \mathbf{P}_f,\,
    \mathbf{P}_{p1}^{\mathsf{Bit}},
    \\
    \mathbf{P}_{p1}^{\mathsf{Phase}},
    \mathbf{P}_{p2}^{\mathsf{Bit}}
    \end{aligned}
    \right\}.
\end{align*}
\vspace*{-0.2in}
\begin{footnotesize}
\begin{eqnarray*}
&& \hspace*{-0.4in} \mathbf{P}_f =
\underset{
\substack{
\emptyset\neq S_1\subset S\\
\overrightarrow{EP}_1+\overrightarrow{EP}_2\leq\overrightarrow{EP}
}}
{\operatorname{Select}_{F^+}}
\mathsf{Fuse}\!\left(
\begin{aligned}
&\mathbf{P}\!\left(
S_1,\left\lceil \tfrac{3}{2p_f}k \right\rceil,
\overrightarrow{EP}_1
\right),\\
&\mathbf{P}\!\left(
S\setminus S_1,\left\lceil \tfrac{3}{2p_f}k \right\rceil,
\overrightarrow{EP}_2
\right)
\end{aligned}
\right),
\\[1ex]
&& \hspace*{-0.4in} \mathbf{P}_{p1}^{\mathsf{Bit}} =
\mathsf{Bit}\!\left(
\begin{aligned}
&\mathbf{P}\!\left(
S,\left\lceil \tfrac{2}{p_p}k \right\rceil,
\overrightarrow{EP}
\right),\\
&\mathbf{P}\!\left(
S,\left\lceil \tfrac{2}{p_p}k \right\rceil,
\overrightarrow{EP}
\right)
\end{aligned}
\right),\quad \text{where } |S|=1,
\\[1ex]
&& \hspace*{-0.4in} \mathbf{P}_{p1}^{\mathsf{Phase}} =
\mathsf{Phase}\!\left(
\begin{aligned}
&\mathbf{P}\!\left(
S,\left\lceil \tfrac{2}{p_p}k \right\rceil,
\overrightarrow{EP}
\right),\\
&\mathbf{P}\!\left(
S,\left\lceil \tfrac{2}{p_p}k \right\rceil,
\overrightarrow{EP}
\right)
\end{aligned}
\right),
\\[1ex]
&& \hspace*{-0.4in} \mathbf{P}_{p2}^{\mathsf{Bit}} =
\underset{
\substack{
j\in S\\
\overrightarrow{EP}_0+\overrightarrow{EP}_j\leq\overrightarrow{EP}
}}
{\operatorname{Select}_{F^+}}
\mathsf{Bit}\!\left(
\begin{aligned}
&\mathbf{P}\!\left(
S,\left\lceil \tfrac{3}{2p_p}k \right\rceil,
\overrightarrow{EP}_0
\right),\\
&\mathbf{P}\!\left(
\{j\},\left\lceil \tfrac{3}{2p_p}k \right\rceil,
\overrightarrow{EP}_j
\right)
\end{aligned}
\right).
\end{eqnarray*}
\end{footnotesize}
 
\color{black}
\section{\bf LP Formulation for Generalized GHZ Distribution}
\label{app:ghz-lp-details}
In this appendix, we present the full LP formulation for the \ggdp problem.

\para{Hypergraph-Based LP Solution.}
As in the \eps case in \S\ref{sec:ep-lp}, we construct a hypergraph so that a feasible hyperflow specifies how intermediate GHZ states are generated and consumed across the levels of a feasible level-based fusion structure. Here, the LP operates only on the second stage of the overall GHZ-generation pipeline: for each star edge $(v_1,v_j)$, the first stage has already produced a remote \eps resource with generation rate $r_j$ and 2-GHZ source vector $\mathbf{P}^{(2)}_j$. The LP then determines how to combine these stage-1 resources via fusion and purification operations to maximize the generation rate of the target GHZ state while maintaining fidelity above the threshold.

Let $\mathcal{D}$ denote the collection of GHZ-basis vectors used as labels in the hypergraph. Whenever a vector label is paired with a subset $S$, it is understood to have the GHZ-basis dimension for a state over $S\cup\{v_1\}$, i.e., it is indexed by labels $(p,\mathbf{b})$ with $p\in\{0,1\}$ and $\mathbf{b}\in\{0,1\}^{|S|}$. The ideal component is 
$F^+(\mathbf{P})=P[0,\mathbf{0}]$.
For each subset $S$ and each compatible vector label $\mathbf{P}\in\mathcal{D}$, the hypergraph contains a vertex
\[
    \id{avail}(S,\mathbf{P}),
\]
representing an available GHZ state over $S\cup\{v_1\}$ with vector label $\mathbf{P}$, produced by the different allowed operations.
Hyperarcs represent feasible fusion and purification operations between such states. Thus, a feasible hyperflow in the hypergraph corresponds to a feasible level-based fusion structure, and the LP objective maximizes the total flow that reaches the final target GHZ state.
Let $H_{\stargraph,\mathcal{D}}$ denote the resulting hypergraph.

\softpara{$V(H_{\stargraph,\mathcal{D}})$ consists of:}
\begin{enumerate}
    \item Two distinguished vertices: \emph{start} and \emph{term}.
    \item For every subset $S\subseteq\{v_2,\ldots,v_m\}$ and every compatible vector $\mathbf{P}\in\mathcal{D}$, the following hypervertices:
    
    \begin{enumerate}
        \item $\id{avail}(S,\mathbf{P})$, representing a GHZ state over $S \cup \{v_1\}$ with vector $\mathbf{P}$ that is available for further use.
        \item $\id{fuse}(S,\mathbf{P})$, representing such a GHZ state produced by a fusion operation.
        \item $\id{purify}_1(S,\mathbf{P})$, representing such a GHZ state produced by same-stream purification.
        \item $\id{purify}_2(S,\mathbf{P})$, representing such a GHZ state produced by purification using two independently generated inputs.
    \end{enumerate}
\end{enumerate}

\softpara{$E(H_{\stargraph,\mathcal{D}})$ consists of:}
\begin{enumerate}
    \item{} [Start]: Hyperarcs that represent the generation of the remote \epss $\{v_1,v_j\}$.
    \begin{small}
    \begin{itemize}
        \item $\bigl(\{\id{start}\},
        \{\id{avail}(\{v_j\},
        \mathbf{P}^{(2)}_j)\}\bigr)$, for each $j \in \{2,\ldots,m\}$.
    \end{itemize}
    \end{small}

    \item{}[Fusion]: Hyperarcs that represent the fusion-retain operation.
    \begin{small}
    \begin{itemize}
        \item 
        $\bigl(\{\id{avail}(S_1,\mathbf{P}_1),
        \id{avail}(S_2,\mathbf{P}_2)\},
        \{\id{fuse}(S_1 \cup S_2,\mathbf{P}_{\rm out})\}\bigr)$,\\
        where $S_1 \cap S_2 = \emptyset$,
        $\mathbf{P}_1,\mathbf{P}_2\in\mathcal{D}$, and
        $\mathbf{P}_{\rm out}
        =
        \mathsf{Fuse}(\mathbf{P}_1,\mathbf{P}_2)
        \in\mathcal{D}$.
        \item 
        $\bigl(\{\id{fuse}(S,\mathbf{P})\},
        \{\id{avail}(S,\mathbf{P})\}\bigr)$, for each $S$ and
        $\mathbf{P}\in\mathcal{D}$.
    \end{itemize}
    \end{small}

    \item{}[SelfPurify]: Hyperarcs that represent same-stream purification:    
    \begin{small}
    \begin{itemize}
        \item 
        $\bigl(\{\id{avail}(S,\mathbf{P})\},
        \{\id{purify}_1(S,\mathbf{P}_{\rm out})\}\bigr)$, where
        $\mathbf{P}\in\mathcal{D}$,
        $\mathsf{Op}\in\{\mathsf{Bit},\mathsf{Phase}\}$, and
        $\mathbf{P}_{\rm out}
        =
        \mathsf{Op}(\mathbf{P},\mathbf{P})
        \in\mathcal{D}$.
        \item $\bigl(\{\id{purify}_1(S,\mathbf{P})\},
        \{\id{avail}(S,\mathbf{P})\}\bigr)$, for each $S$ and
        $\mathbf{P}\in\mathcal{D}$.
    \end{itemize}
    \end{small}

    \item{}[PurifyDistinct]: Hyperarcs that represent purification using two independently generated inputs:
    \begin{small}
    \begin{itemize}
        \item 
        $\bigl(\{\id{avail}(S,\mathbf{P}_{\rm t}),
        \id{avail}(S',\mathbf{P}_{\rm s})\},
        \{\id{purify}_2(S,\mathbf{P}_{\rm out})\}\bigr)$,\\
        where $S'\subseteq S$,
        $\mathbf{P}_{\rm t},\mathbf{P}_{\rm s}\in\mathcal{D}$,
        $(S,\mathbf{P}_{\rm t})\neq(S',\mathbf{P}_{\rm s})$,
        $\mathsf{Op}$ is an allowed purification map for this target--sacrifice pair,
        and
        $\mathbf{P}_{\rm out}
        =
        \mathsf{Op}(\mathbf{P}_{\rm t},\mathbf{P}_{\rm s})
        \in\mathcal{D}$.
        \item $\bigl(\{\id{purify}_2(S,\mathbf{P})\},
        \{\id{avail}(S,\mathbf{P})\}\bigr)$, for each $S$ and
        $\mathbf{P}\in\mathcal{D}$.
    \end{itemize}
    \end{small}

    \item{}[Term]: Hyperarcs that collect target GHZ states satisfying the fidelity threshold:
    \begin{small}
    \begin{itemize}
        \item $\bigl(\{\id{avail}(\{v_2,\ldots,v_m\},\mathbf{P})\},
        \{\id{term}\}\bigr)$, for each $\mathbf{P}\in\mathcal{D}$
        such that $P[0,\mathbf{0}]\ge \fth$.
    \end{itemize}
    \end{small}
\end{enumerate}

\para{Optimal LP Formulation (\GHZLPOpt).}
We introduce one nonnegative variable $z_e$ for each hyperarc $e$ in $H_{\stargraph,\mathcal{D}}$; intuitively, $z_e$ denotes the GHZ-state generation rate at the tail of $e$. Let $in(v)$ and $out(v)$ denote the sets of incoming and outgoing hyperarcs of a hypervertex $v$, respectively. The LP formulation consists of the following components:
\begin{itemize}
    \item \emph{Flow Constraints.} These constraints depend on the type of hypervertex.
    \begin{enumerate}
        \item For each $v$ such that $v=\id{avail}(\cdot)$:
        \[
        \sum_{e \in in(v)} z_e = \sum_{e' \in out(v)} z_{e'}.
        \]
        This expresses that a GHZ state made available at $v$ can be aggregated from different incoming operations and then split among different outgoing operations.
        \item For each $v$ such that $v=\id{fuse}(\cdot)$:
        \[
        \sum_{e \in in(v)} \tfrac{2}{3} z_e \fp = \sum_{e' \in out(v)} z_{e'}.
        \]
        Here, $\fp$ denotes the success probability of fusion, and the factor $\frac{2}{3}$ captures the effective rate loss due to synchronizing two independent input streams.
        \item For each $v$ such that $v=\id{purify}_1(\cdot)$:
        
        \[
        \sum_{e \in in(v)} \tfrac{1}{2} z_e p_p(e) = \sum_{e' \in out(v)} z_{e'}.
        \]
        Here, $p_p(e)$ denotes the success probability of the purification operation represented by hyperarc $e$, and the factor $\frac{1}{2}$ reflects that same-stream purification consumes two copies from the same input stream to produce one retained copy.
        \item For each $v$ such that $v=\id{purify}_2(\cdot)$:
        \[
        \sum_{e \in in(v)} \tfrac{2}{3} z_e p_p(e) = \sum_{e' \in out(v)} z_{e'}.
        \]
        As above, $p_p(e)$ captures the purification success, and the factor $\frac{2}{3}$ accounts for the synchronization loss when two independently generated inputs are required.
    \end{enumerate}

    \item \emph{Capacity Constraints.} For each start hyperarc
    $e=\bigl(\{\id{start}\},
    \{\id{avail}(\{v_j\}, \mathbf{P}^{(2)}_j)\}\bigr)$, we impose:   
    \[
    z_e \le r_j.
    \]
    This ensures that the usage of each stage-1 \eps resource does not exceed its available generation rate.

    \item \emph{Objective.} We maximize:
    \[
    \sum_{e \in in(\textit{term})} z_e,
    \]
    which is exactly the total generation rate of the target $m$-GHZ state with fidelity at least $\fth$.
\end{itemize}

\para{Proof of Theorem~\ref{thm:ghz-lp}.}
We show that the feasible level-based fusion structures for the \ggdp problem and the feasible solutions of the LP over $H_{\stargraph,\mathcal{D}}$ correspond to each other without changing the achieved generation rate.

First, consider any feasible level-based fusion structure for the given \ggdp instance. Each type of GHZ-state generated in that structure is identified by its subset label $S$ and vector label $\mathbf{P}$, and hence corresponds to a hypervertex in $H_{\stargraph,\mathcal{D}}$. Similarly, each valid second-stage fusion or purification operation used in the structure corresponds to a hyperarc in $H_{\stargraph,\mathcal{D}}$, and each stage-1 \eps resource corresponds to a Start hyperarc.
Assign to each hyperarc the rate at which the corresponding operation is used in the level-based fusion structure. Then the Start-arc capacity constraints are satisfied because no stage-1 \eps resource is used above its available generation rate; the flow-conservation equations at the $\id{avail}(\cdot)$ hypervertices are satisfied because the total rate at which a GHZ-state type is produced equals the total rate at which it is either consumed by later operations or delivered to the terminal demand; and the operation constraints at the $\id{fuse}(\cdot)$, $\id{purify}_1(\cdot)$, and $\id{purify}_2(\cdot)$ hypervertices are satisfied because they are exactly the adopted vector-update and rate-update rules for fusion and purification, including the corresponding success-probability and synchronization factors. Thus, every feasible level-based fusion structure induces a feasible LP solution with the same achieved generation rate.

Conversely, let $z$ be any feasible LP solution. Before unfolding the LP solution, we remove positive-flow cycles. Here, a cycle is a sequence of operations with positive flows that connects an available-state vertex back to itself. For each operation, the output generation rate is no greater than the rate consumed from each input stream. On a simple directed cycle, reduce every operation's input flow by the same amount until one becomes zero, reducing its output by the corresponding rate coefficient. At each cycle vertex, this removes at least as much consumption as production; consumption at other input vertices can only decrease. Thus, no state type has a shortage. Repeating removes all cycles, since each step eliminates at least one positive operation flow. We then reduce surplus production, working backward from outputs to sources while keeping terminal flows fixed. This restores the flow equalities without increasing source usage. We continue to use $z$ for the resulting acyclic flow.

We construct a level-based fusion structure by unfolding $z$ backward from the $\id{term}$ hypervertex. For each hyperarc entering $\id{term}$, assign it the same rate as in $z$, and then recursively process the required rates. Whenever a required rate reaches an $\id{avail}(\cdot)$ hypervertex, we use the corresponding flow-conservation equation to split that required rate among its incoming hyperarcs, without exceeding their LP values. If such an incoming hyperarc is a Start hyperarc, this corresponds directly to using the associated stage-1 \eps resource at that rate. 
If such an incoming hyperarc is a consolidation hyperarc, we trace it back to the corresponding operation vertex. Using that vertex's LP rate-balance equation, we split the required output rate among its incoming operation hyperarcs. For each such hyperarc, we instantiate the corresponding fusion or purification operation and recursively request from the operand $\id{avail}(\cdot)$ hypervertices the input rates required by the associated LP constraint.

By construction, every instantiated operation obeys exactly the adopted vector-update and rate-update rule, and no Start hyperarc is used above its LP value. Moreover, each fusion or purification step multiplies the forward rate by a factor strictly smaller than $1$, that is, $\frac{2}{3}\fp$, $\frac{1}{2}p_p(e)$, or $\frac{2}{3}p_p(e)$.
These factors justify the cycle removal above. In the resulting acyclic flow, following operations backward cannot revisit an available-state vertex along a branch. Since the hypergraph is finite, every branch contains only finitely many operation steps and ends at a Start hyperarc. Therefore, the above recursive construction yields a finite level-based fusion structure.

Finally, flow-conservation equations ensure consistency of the rates for all intermediate GHZ-state types, and the total rate delivered to $\id{term}$ is exactly the objective value of LP. Thus, every feasible LP solution induces a feasible level-based fusion structure with the same achieved generation rate.

Therefore, feasible LP solutions and feasible level-based fusion structures realize exactly the same achievable target generation rates. Hence, maximizing total flow into $\id{term}$ is exactly equivalent to maximizing the achievable generation rate of the target GHZ state above the fidelity threshold $\fth$, and the LP solution is an optimal level-based fusion structure for the \ggdp problem. \hfill \blackbox

\para{Node Labeling Strategy for Restricted LPs.}
The interval, partition-based, and multi-step LP formulations depend on how the star graph is labeled. We choose the labeling to improve the performance of these restricted LP formulations. In particular, we first select the central node $v_1$, and then order the remaining non-center nodes $\langle v_2,v_3,\ldots,v_m\rangle$.
\begin{itemize}
    \item {\em Generation-Latency Based Median to Select $v_1$.} One simple approach is to choose $v_1$ as the node that minimizes the total ``distance'' to all other nodes $\langle v_2, v_3, \ldots, v_m \rangle$, i.e., the node with the smallest $\sum_{v \neq v_1} d(v_1, v)$, where $d(\cdot, \cdot)$ denotes the shortest path distance in the quantum network graph. However, in our context, the latency to generate a remote EP over a pair of nodes is more appropriate than the shortest-path distance, since the central node determines the set of remote EPs to be generated in the first stage of our GHZ state generation scheme. Thus, we select the central node $v_1$ as the node that minimizes the aggregate EP generation latency, i.e., the terminal node $u$ with the smallest $\sum_{v} \text{Latency}(u, v)$ where $v$ is over all the terminal nodes. To estimate EP generation latencies, we must account for resource contention among different node pairs; we do so using the LP-based scheme~\cite{swapping-tqe-22} to generate EPs with the maximum aggregate rate.
    
    \item {\em Rate-Based Labeling of Remaining Vertices.} 
    In the interval-based LP approach, we label the remaining nodes in the increasing order of their \eps generation rates $r_j$; e.g., the label $v_2$ is assigned to the node with the lowest \eps generation rate from the first stage. This labeling scheme places nodes with similar EP generation rates next to each other, aiming to reduce ``waiting delays'' caused by skewed generation rates.
    In the partition-based two-step approach, we order and group nodes into disjoint subsets so that nodes with similar rates fall within the same partition; this ensures that the GHZ generation process within each partition proceeds roughly uniformly. The multi-step approach uses the same increasing-rate ordering to fix the consecutive groups merged at each step, with ties broken by the original node index.  
\end{itemize}
Labeling the remaining nodes by source-vector quality, such as $F^+(\mathbf{P}^{(2)}_j)$ for the corresponding edge to the central node $v_1$, could also be a viable approach. For example, if nodes with similar source qualities are grouped together, subsequent fusion and purification steps could be simplified, and the overall quality of the GHZ state could be improved. We defer the design and evaluation of such schemes to future work.

\para{Update Rules for \GHZLPTwoF{}.}
We now give the update rules for \GHZLPTwoF{} introduced in \S\ref{sec:ghz-lp}. Recall that its label $(F^+,F^-)$ consists of the ideal component $F^+=P[0,\mathbf{0}]$ and its opposite-phase counterpart $F^-=P[1,\mathbf{0}]$. Under fusion retain, these two output components depend only on their values in the two inputs and can therefore be computed exactly:
\begin{align*}
    F^+_{\rm out}
    &=
    F^+_1F^+_2+F^-_1F^-_2,
    \\
    F^-_{\rm out}
    &=
    F^+_1F^-_2+F^-_1F^+_2.
\end{align*}

For the same-stream purification updates below, the subscripts ${\rm t}$ and ${\rm s}$ denote the target and sacrificial inputs, respectively. For full bit-pattern purification, define
\[
    B=F^++F^-,
    \qquad
    R=1-F^+-F^- .
\]
Here $B$ is the total probability of the zero bit-error string, while $R$ is the total probability of all nonzero bit-error strings. The reduced estimate of the success probability of the full bit-pattern check is
\[
    q_{\rm B}
    =
    B_{\rm t}B_{\rm s}
    +
    R_{\rm t}R_{\rm s}.
\]
The conditional reduced update after acceptance is
\[
    F^+_{\rm out}
    =
    \frac{
        F^+_{\rm t}F^+_{\rm s}
        +
        F^-_{\rm t}F^-_{\rm s}
    }{q_{\rm B}},
    \qquad
    F^-_{\rm out}
    =
    \frac{
        F^+_{\rm t}F^-_{\rm s}
        +
        F^-_{\rm t}F^+_{\rm s}
    }{q_{\rm B}}.
\]
The term $R_{\rm t}R_{\rm s}$ approximates the contribution from matching nonzero bit-error strings by the probability that both inputs have nonzero bit-error strings. Those individual strings are not distinguished by the reduced label. This term contributes only to the success-probability estimate, since neither retained output component has a nonzero bit-error string.

For full phase purification, define the conditional phase quality for $B>0$ as
\[
    \phi=\frac{F^+}{B}.
\]
Here $\phi$ is the probability of the correct phase label conditioned on the state having zero bit-error string. The reduced success-probability estimate is
\[
    q_{\rm P}
    =
    B_{\rm s}
    \left(
        \phi_{\rm t}\phi_{\rm s}
        +
        (1-\phi_{\rm t})(1-\phi_{\rm s})
    \right).
\]
The factor $B_{\rm s}$ requires the sacrificial input to have zero bit-error string, while the expression in parentheses is the probability that the target and sacrifice agree on the phase label under this reduced description. Conditioned on acceptance, the phase quality becomes
\[
    \phi_{\rm out}
    =
    \frac{
        \phi_{\rm t}\phi_{\rm s}
    }{
        \phi_{\rm t}\phi_{\rm s}
        +
        (1-\phi_{\rm t})(1-\phi_{\rm s})
    }.
\]
The retained two-component label is therefore
\[
    F^+_{\rm out}=B_{\rm t}\phi_{\rm out},
    \qquad
    F^-_{\rm out}=B_{\rm t}(1-\phi_{\rm out}).
\]
Thus, the reduced phase check improves the relative phase quality among the zero-bit-error components, while preserving the target zero-bit-error mass $B_{\rm t}$.
 
\section{\bf Local Realization of Graph-State Growth Operations}
\label{app:graph-local-fusion}

\begin{figure}[t]
    \centering
    \includegraphics[width=0.45\textwidth]{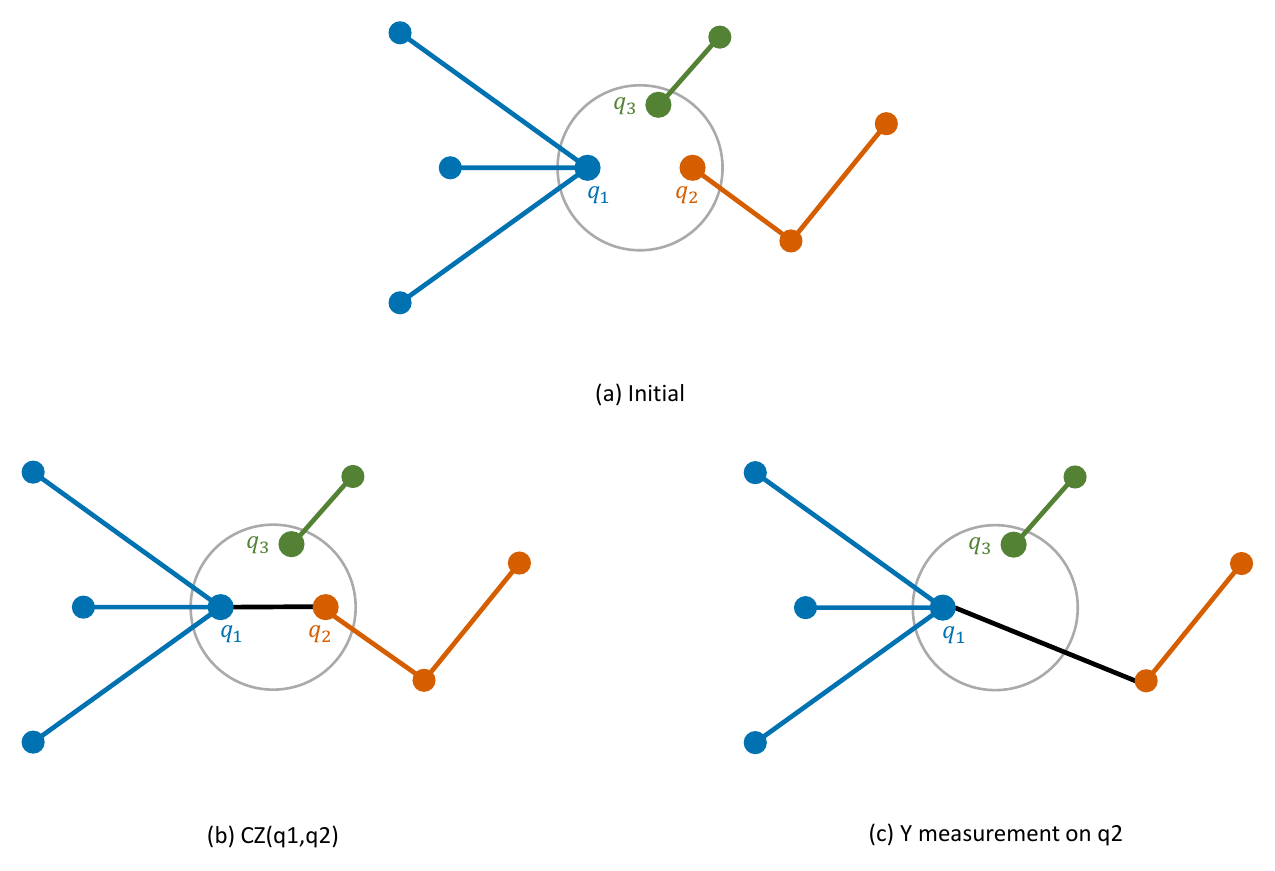}
    \caption{Local realization of the fusion step used in graph-state growth. (a) Initial configuration at a network node: $q_1$ is the designated star-center qubit, $q_2$ is the local qubit currently being merged, and $q_3$ denotes another local qubit that may be fused into the same star center later. (b) A local CZ gate is applied between $q_1$ and $q_2$. (c) A Y-basis measurement on $q_2$, together with the corresponding local complementation update, removes $q_2$ while transferring its external adjacency to the surviving star center $q_1$. Repeating this operation incrementally allows multiple locally held qubits to be fused into the same star center.}
    \vspace*{-0.1in}
    \label{fig:graph_fusion}
\end{figure}

In this appendix, we briefly describe how the graph-state growth operations used in \S\ref{sec:gen-graphs} can be realized using only local operations. Our goal here is to justify the fact that the graph-level primitives in the main text admit a natural implementation using standard graph-state transformations.

The key local ingredients are CZ gates, Y-basis measurements, and local complementation. These are standard graph-state operations: a local CZ gate adds or removes an edge between two co-located qubits, while a Y-basis measurement can be interpreted graphically as a local complementation followed by deletion of the measured vertex. Together, these operations allow one to merge locally held graph-state qubits while preserving the intended external adjacency structure of the growing target graph state.

The main idea is to organize the qubits at each network node around a designated {\em star center}. For a partially constructed target subgraph, if multiple target edges are incident on the same network node, then one locally held graph-state qubit at that node can be chosen as the star center. This qubit is the one that may participate in multiple incident target edges. Any other local qubit at that node arising from a target-edge EP is then fused into the star center using only local operations. For this local fusion step, the relevant property is that the qubit to be measured is incident on only the newly introduced target edge, regardless of whether the opposite endpoint of that edge already belongs to a larger intermediate graph state. Intuitively, this lets the node accumulate the connectivity required by the final target graph state while keeping a single representative qubit as the attachment point for future growth.

Suppose that two graph-state qubits $q_1$ and $q_2$ reside at the same network node, as illustrated in Fig.~\ref{fig:graph_fusion}. Let $q_1$ denote the current star-center qubit, and let $q_2$ be the local qubit to be merged into it. We first apply a local CZ gate between $q_1$ and $q_2$, and then perform a Y-basis measurement on $q_2$, together with the resulting local-complementation updates; see Fig.~\ref{fig:graph_fusion}(a)--(c). At the graph level, this removes $q_2$ while modifying the local neighborhood so that $q_1$ remains the attachment point. Repeating this procedure allows multiple local qubits at the same node to be fused one at a time into the same star center, thereby providing the local mechanism underlying the graph-growth operations in the main text.

Using the above local construction, the three graph-level primitives in \S\ref{sec:gen-graphs} are realized as follows. In {\tt Fuse-Two-Components}, the EP for the new target edge brings one endpoint qubit to each of the two incident nodes, and the relevant endpoint qubit is locally fused into the current star center at each endpoint. In {\tt Extend-by-New-Vertex}, one endpoint is fused into the existing component while the other initializes the new leaf attachment. In {\tt Add-Internal-Edge}, the EP is used to create an additional missing edge between two already existing parts of the partially constructed graph state. Thus, the main text reasons only at the graph level, while the actual qubit-level realization is handled locally via the above star-center construction.   
\section{\bf Graph-Basis Vector Model for Graph State Fidelity}
\label{app:graph-fidelity}

For general graph states, a single fidelity value is not closed under graph-state purification.  Two graph states can have the same overlap with the ideal graph state but different graph-basis error populations, and a color-check purification step can accept them with different probabilities and produce different output fidelities.  We therefore use a graph-basis probability vector, in direct analogy with the GHZ-basis vector used in Appendix~\ref{app:purification-detail}.

\para{Graph-Basis Vector Representation.}
Let $G=(V,E)$ be an $n$-qubit graph, and let $\ket{\psi_G}$ denote the ideal graph state.  For each vertex $v\in V$, the stabilizer generator is
\[
    K_v
    =
    X_v\prod_{u\in N_G(v)} Z_u,
\]
where $N_G(v)$ denotes the neighbors of $v$. The ideal graph state is the unique simultaneous $+1$ eigenstate of all $K_v$.  For a binary label
$\boldsymbol{\mu}\in\{0,1\}^{V}$, define the graph-basis state
\[
    \ket{\psi^{(G)}_{\boldsymbol{\mu}}}
    =
    \left(
        \prod_{v\in V} Z_v^{\mu_v}
    \right)
    \ket{\psi_G}.
\]
Equivalently, $\boldsymbol{\mu}$ records the stabilizer-sign pattern:
\[
    K_v\ket{\psi^{(G)}_{\boldsymbol{\mu}}}
    =
    (-1)^{\mu_v}
    \ket{\psi^{(G)}_{\boldsymbol{\mu}}}.
\]
Thus $\mu_v=0$ means that the $v$-th stabilizer sign is unchanged, while $\mu_v=1$ means that it is flipped.  The graph-basis states form an orthonormal basis indexed by the $2^n$ binary labels.

A graph-diagonal state is represented by the probability vector
\[
    \mathbf{P}^{(G)}
    =
    \left\{
        P^{(G)}[\boldsymbol{\mu}]
    \right\}_{\boldsymbol{\mu}\in\{0,1\}^{V}},
    \qquad
    \sum_{\boldsymbol{\mu}}P^{(G)}[\boldsymbol{\mu}]=1,
\]
meaning
\[
    \rho_G
    =
    \sum_{\boldsymbol{\mu}}
        P^{(G)}[\boldsymbol{\mu}]
        \ket{\psi^{(G)}_{\boldsymbol{\mu}}}
        \!\bra{\psi^{(G)}_{\boldsymbol{\mu}}}.
\]
The scalar fidelity with respect to the ideal graph state is the distinguished component
\[
    F\!\left(\mathbf{P}^{(G)}\right)
    =
    P^{(G)}[\mathbf{0}].
\]
As in the GHZ case, this scalar remains the quantity compared with the final fidelity threshold, but the intermediate operation updates are computed from the full vector $\mathbf{P}^{(G)}$.

\para{Pauli Labels in the Graph Basis.}
The graph-basis label is especially convenient for Clifford circuits.  On $\ket{\psi_G}$, the stabilizer relation gives
\[
    X_v
    \equiv
    \prod_{u\in N_G(v)} Z_u,
\]
up to multiplication by the stabilizer $K_v$.  Hence, ignoring irrelevant global phases, the single-qubit Pauli errors act on graph-basis labels as
\[
    Z_v:\boldsymbol{\mu}\mapsto \boldsymbol{\mu}\oplus \mathbf{e}_v,
\]
\[
    X_v:\boldsymbol{\mu}\mapsto
    \boldsymbol{\mu}\oplus \mathbf{n}_v,
\]
\[
    Y_v:\boldsymbol{\mu}\mapsto
    \boldsymbol{\mu}\oplus \mathbf{e}_v\oplus \mathbf{n}_v,
\]
where $\mathbf{e}_v$ is the unit vector at $v$ and $\mathbf{n}_v$ is the neighborhood indicator vector of $v$ in $G$. These relations allow every Pauli error to be represented by a graph-basis label and therefore enable label propagation through the Clifford circuits used for graph-state growth.

\para{Graph-State Growth Primitives.}
The second-stage construction builds the target graph by repeatedly consuming existing graph-state resources together with two-qubit graph-edge resources. We use three growth primitives: \emph{fuse-two-components}, which joins two connected components by a target edge; \emph{add-internal-edge}, which adds a missing edge inside an existing connected component; and \emph{extend-by-new-vertex}, which extends a connected component by one new vertex. Each primitive is implemented by a Clifford circuit, followed by Pauli feed-forward conditioned on the local measurement outcomes.

The important point for the vector model is that, once a successful measurement branch and its feed-forward rule are fixed, a graph-growth primitive induces a deterministic map on graph-basis labels. Let $\mathcal{O}$ be one such primitive. Its inputs consist of one or more graph-state resources and a two-qubit edge resource. Write their graph-basis labels collectively as
\[
    \boldsymbol{\eta}
    =
    (\boldsymbol{\mu}_1,\boldsymbol{\mu}_2,\boldsymbol{\epsilon}),
\]
with the unused components omitted for primitives that have fewer inputs.

For a given input-label tuple $\boldsymbol{\eta}$, the Clifford realization of $\mathcal{O}$ described in Appendix~\ref{app:graph-local-fusion} maps the corresponding input graph-basis component to an output graph-basis component. After conditioning on the successful measurement branch and applying the prescribed feed-forward corrections, the retained qubits are expressed in the graph basis of the output graph $G_{\rm out}$. We denote the resulting output label by
\[
    \boldsymbol{\gamma}
    =
    \tau_{\mathcal{O}}(\boldsymbol{\eta}),
    \qquad
    \boldsymbol{\gamma}\in\{0,1\}^{V_{\rm out}}.
\]

Thus, $\tau_{\mathcal{O}}$ is the label transition induced by the graph-growth primitive. Since the primitive is Clifford, this map can be computed once for each primitive and then used as a transition table.

For independent inputs, each input-label tuple $\boldsymbol{\eta}$ has a probability weight given by the product of the corresponding input-vector entries. This weight is assigned to the output component labeled by $\tau_{\mathcal{O}}(\boldsymbol{\eta})$. If several input-label tuples produce the same output label, their weights are added. The resulting vector updates for the three graph-growth primitives are given explicitly below.

\softpara{Fuse-Two-Components.}
Let $G_1=(V_1,E_1)$ and $G_2=(V_2,E_2)$ be two disjoint connected target subgraphs, and let $e=(a,b)$ be a target edge with $a\in V_1$ and $b\in V_2$. Write $G_e=(\{a,b\},\{e\})$ for the two-qubit edge graph. The fuse-two-components primitive consumes graph-state resources on $G_1$ and $G_2$ and a two-qubit edge resource on $G_e$, and produces the graph state on
\[
    G_{\rm out}
    =
    (V_1\cup V_2,\,
    E_1\cup E_2\cup\{e\}).
\]
Let the corresponding input vectors be
$\mathbf{P}_1$, $\mathbf{P}_2$, and $\mathbf{P}_e$, with labels
$\boldsymbol{\mu}_1\in\{0,1\}^{V_1}$,
$\boldsymbol{\mu}_2\in\{0,1\}^{V_2}$, and
$\boldsymbol{\epsilon}\in\{0,1\}^{\{a,b\}}$.
After the successful measurement branch and its feed-forward corrections, Pauli-frame propagation gives a deterministic binary label map
\[
    \boldsymbol{\gamma}_{\rm fuse}
    =
    \tau_{\rm fuse}^{G_1,G_2,e}
    (\boldsymbol{\mu}_1,\boldsymbol{\mu}_2,\boldsymbol{\epsilon})
    \in
    \{0,1\}^{V_1\cup V_2}.
\]
That is, the input component
\[
    \ket{\psi^{(G_1)}_{\boldsymbol{\mu}_1}}
    \otimes
    \ket{\psi^{(G_2)}_{\boldsymbol{\mu}_2}}
    \otimes
    \ket{\psi^{(G_e)}_{\boldsymbol{\epsilon}}}
\]
is mapped to the graph-basis component
\[
    \ket{\psi^{(G_{\rm out})}_{\boldsymbol{\gamma}_{\rm fuse}}}.
\]
For a fixed output label $\boldsymbol{\gamma}$, define the set of input labels that produce it by
\[
    \mathcal{I}_{\rm fuse}(\boldsymbol{\gamma})
    =
    \left\{
    (\boldsymbol{\mu}_1,\boldsymbol{\mu}_2,\boldsymbol{\epsilon})
    :
    \tau_{\rm fuse}^{G_1,G_2,e}
    (\boldsymbol{\mu}_1,\boldsymbol{\mu}_2,\boldsymbol{\epsilon})
    =
    \boldsymbol{\gamma}
    \right\}.
\]
Therefore, conditioned on the heralded success of the physical fusion operation, the exact output vector is obtained by summing the probabilities of all such input-label tuples:
\[
    P_{\rm fuse}[\boldsymbol{\gamma}]
    =
    \sum_{
        (\boldsymbol{\mu}_1,\boldsymbol{\mu}_2,\boldsymbol{\epsilon})
        \in
        \mathcal{I}_{\rm fuse}(\boldsymbol{\gamma})
    }
        P_1[\boldsymbol{\mu}_1]\,
        P_2[\boldsymbol{\mu}_2]\,
        P_e[\boldsymbol{\epsilon}].
\]

\softpara{Add-Internal-Edge.}
Let $G=(V,E)$ be a connected target subgraph, and let $e=(a,b)$ be a missing target edge with $a,b\in V$. Write $G_e=(\{a,b\},\{e\})$ for the two-qubit edge graph. The add-internal-edge primitive consumes a graph state on $G$ and an edge resource on $G_e$, and produces the graph state on
\[
    G_{\rm out}
    =
    (V,E\cup\{e\}).
\]
For input labels
$\boldsymbol{\mu}\in\{0,1\}^{V}$ and
$\boldsymbol{\epsilon}\in\{0,1\}^{\{a,b\}}$, the successful Clifford implementation induces a binary transition
\[
    \boldsymbol{\gamma}_{\rm add}
    =
    \tau_{\rm add}^{G,e}
    (\boldsymbol{\mu},\boldsymbol{\epsilon})
    \in
    \{0,1\}^{V}.
\]
Thus the component
$\ket{\psi^{(G)}_{\boldsymbol{\mu}}}
\otimes
\ket{\psi^{(G_e)}_{\boldsymbol{\epsilon}}}$
is mapped to
$\ket{\psi^{(G_{\rm out})}_{\boldsymbol{\gamma}_{\rm add}}}$.
For a fixed output label $\boldsymbol{\gamma}$, let
\[
    \mathcal{I}_{\rm add}(\boldsymbol{\gamma})
    =
    \left\{
    (\boldsymbol{\mu},\boldsymbol{\epsilon})
    :
    \tau_{\rm add}^{G,e}(\boldsymbol{\mu},\boldsymbol{\epsilon})
    =
    \boldsymbol{\gamma}
    \right\}.
\]
The exact vector update, conditioned on success, is
\[
    P_{\rm add}[\boldsymbol{\gamma}]
    =
    \sum_{
        (\boldsymbol{\mu},\boldsymbol{\epsilon})
        \in
        \mathcal{I}_{\rm add}(\boldsymbol{\gamma})
    }
        P_G[\boldsymbol{\mu}]\,
        P_e[\boldsymbol{\epsilon}].
\]

\softpara{Extend-by-New-Vertex.}
Let $G=(V,E)$ be a connected target subgraph, and let $e=(a,b)$ be a target edge with $a\in V$ and $b\notin V$. Write $G_e=(\{a,b\},\{e\})$ for the two-qubit edge graph. The extend-by-new-vertex primitive consumes a graph state on $G$ and an edge resource on $G_e$, and produces
\[
    G_{\rm out}
    =
    (V\cup\{b\},E\cup\{e\}).
\]
For input labels
$\boldsymbol{\mu}\in\{0,1\}^{V}$ and
$\boldsymbol{\epsilon}\in\{0,1\}^{\{a,b\}}$, the successful Clifford circuit induces
\[
    \boldsymbol{\gamma}_{\rm ext}
    =
    \tau_{\rm ext}^{G,e}
    (\boldsymbol{\mu},\boldsymbol{\epsilon})
    \in
    \{0,1\}^{V\cup\{b\}}.
\]
Thus
$\ket{\psi^{(G)}_{\boldsymbol{\mu}}}
\otimes
\ket{\psi^{(G_e)}_{\boldsymbol{\epsilon}}}$
is mapped to
$\ket{\psi^{(G_{\rm out})}_{\boldsymbol{\gamma}_{\rm ext}}}$.
For a fixed output label $\boldsymbol{\gamma}$, let
\[
    \mathcal{I}_{\rm ext}(\boldsymbol{\gamma})
    =
    \left\{
    (\boldsymbol{\mu},\boldsymbol{\epsilon})
    :
    \tau_{\rm ext}^{G,e}(\boldsymbol{\mu},\boldsymbol{\epsilon})
    =
    \boldsymbol{\gamma}
    \right\}.
\]
The exact output vector is
\[
    P_{\rm ext}[\boldsymbol{\gamma}]
    =
    \sum_{
        (\boldsymbol{\mu},\boldsymbol{\epsilon})
        \in
        \mathcal{I}_{\rm ext}(\boldsymbol{\gamma})
    }
        P_G[\boldsymbol{\mu}]\,
        P_e[\boldsymbol{\epsilon}].
\]

In all three growth primitives, the hardware success probability of the fusion operation is not part of this normalized state update; it appears separately as the rate-scaling coefficient.

\para{One Color-Check Purification Step.}
We now give the explicit vector update for one graph-state purification color check. Consider a connected graph state on graph $G=(V,E)$, where $G$ may be either an intermediate graph produced during construction or the final target graph. Fix a proper vertex coloring
\[
    V=A_1\cup A_2\cup\cdots\cup A_r,
    \qquad
    A_i\cap A_j=\emptyset \text{ for } i\neq j,
\]
where no edge has both endpoints in the same set $A_j$. We call each $A_j$ a color class. A one-color check purifies one selected color class $A=A_j$. Let $\bar A=V\setminus A$. For a graph-basis label $\boldsymbol{\mu}\in\{0,1\}^{V}$, write
\[
    \boldsymbol{\mu}
    =
    (\boldsymbol{\mu}_A,\boldsymbol{\mu}_{\bar A}).
\]
Let the target copy have vector $\mathbf{P}_{\rm t}$ on $V$. To check color $A$, the auxiliary sacrificial graph is built only from the vertices and edges needed by the corresponding color-check subprotocol. More precisely, define
\[
    N_G(A)
    =
    \{u\in V\setminus A:\exists a\in A
    \text{ with } (a,u)\in E\},
\]
and set
\[
    W=A\cup N_G(A),
    \quad
    E_A
    =
    \{(a,u)\in E:a\in A,\ u\in V\setminus A\}.
\]
Here $N_G(A)$ is the set of vertices outside $A$ that are adjacent to at least one vertex in $A$; $W$ is the vertex set used by the sacrificial resource; and $E_A$ keeps exactly the edges crossing from $A$ to its neighbors. The auxiliary sacrifice state is the graph state on $(W,E_A)$. Thus it contains the color class to be checked and the neighboring vertices needed to test that color; edges not incident to $A$ are not part of this auxiliary resource. Let $\mathbf{Q}_{\rm s}$ be the native graph-basis vector of the auxiliary sacrifice state on $(W,E_A)$. For the color-check update, we first embed it into the target label space by fixing all absent vertices to have zero label:
\[
    P_{\rm s}[\boldsymbol{\mu}]
    =
    \begin{cases}
    Q_{\rm s}[\boldsymbol{\mu}_W],
        & \text{if }\boldsymbol{\mu}_{V\setminus W}=\mathbf{0},\\
    0,
        & \text{otherwise.}
    \end{cases}
\]
After this embedding, both $\mathbf{P}_{\rm t}$ and $\mathbf{P}_{\rm s}$ are indexed by $\{0,1\}^{V}$. Operationally, the color-check subprotocol applies CNOT gates between the target and sacrifice qubits that share the same vertex label. For vertices in $A$, the CNOT direction is sacrifice to target; for vertices in $W\setminus A$, the direction is target to sacrifice. The sacrificial qubits in $A$ are then measured in the $X$ basis, and the sacrificial qubits in $W\setminus A$ are measured in the $Z$ basis. 
If $\xi_a\in\{0,1\}$ is the $X$-measurement outcome for $a\in A$ and $\zeta_b\in\{0,1\}$ is the $Z$-measurement outcome for $b\in W\setminus A$, then the measured check syndrome for $a\in A$ is
\[
    s_a
    =
    \xi_a
    \bigoplus_{\substack{b\in V\setminus A\\ (a,b)\in E}}
    \zeta_b .
\]
This syndrome is exactly the parity check between the target and sacrifice graph-basis labels on the checked color. To see this, use superscripts ${\rm t}$ and ${\rm s}$ for target and sacrificial qubits. For $a\in A$, the CNOT has direction ${\rm s}\rightarrow{\rm t}$. Therefore, in the Heisenberg picture, the post-CNOT $X$ measurement on the sacrificial qubit at $a$ is equivalent to measuring $X^{\rm s}_aX^{\rm t}_a$ before the CNOT. For a neighbor $b\in V\setminus A$ of $a$, the CNOT has direction ${\rm t}\rightarrow{\rm s}$. Thus the post-CNOT $Z$ measurement on the sacrificial qubit at $b$ is equivalent to measuring $Z^{\rm t}_bZ^{\rm s}_b$ before the CNOT. Consequently, the parity $s_a$ is the outcome bit of the pre-CNOT observable
\[
    \left(
        X^{\rm t}_a
        \prod_{\substack{b\in V\setminus A\\ (a,b)\in E}}
        Z^{\rm t}_b
    \right)
    \left(
        X^{\rm s}_a
        \prod_{\substack{b\in V\setminus A\\ (a,b)\in E}}
        Z^{\rm s}_b
    \right)
    =
    K^{\rm t}_a K^{\rm s}_a .
\]
Here $K^{\rm t}_a$ is the stabilizer generator of the target graph $G$, and $K^{\rm s}_a$ is the corresponding stabilizer generator of the auxiliary sacrifice graph $(W,E_A)$. Now take pure graph-basis inputs with labels
$\boldsymbol{\mu}^{\rm t}$ and $\boldsymbol{\mu}^{\rm s}$. Since
\[
    K^{\rm t}_a
    \ket{\psi^{(G)}_{\boldsymbol{\mu}^{\rm t}}}
    =
    (-1)^{\mu^{\rm t}_a}
    \ket{\psi^{(G)}_{\boldsymbol{\mu}^{\rm t}}},
    K^{\rm s}_a
    \ket{\psi^{(W,E_A)}_{\boldsymbol{\mu}^{\rm s}_W}}
    =
    (-1)^{\mu^{\rm s}_a}
    \ket{\psi^{(W,E_A)}_{\boldsymbol{\mu}^{\rm s}_W}},
\]
the product observable $K^{\rm t}_aK^{\rm s}_a$ has eigenvalue $(-1)^{\mu^{\rm t}_a\oplus\mu^{\rm s}_a}$. Hence
\[
    s_a
    =
    \mu^{\rm t}_a\oplus\mu^{\rm s}_a .
\]
The target is kept when all these syndrome values are zero; otherwise it is discarded. Thus the acceptance condition is equality on the checked color class, $\boldsymbol{\mu}^{\rm t}_A=\boldsymbol{\mu}^{\rm s}_A$; the remaining labels need not agree. For clarity, write an accepted target--sacrifice label pair as
\[
    \text{target label }(\boldsymbol{x},\boldsymbol{y}),
    \qquad
    \text{sacrifice label }(\boldsymbol{x},\boldsymbol{z}),
\]
where $\boldsymbol{x}\in\{0,1\}^{A}$ is the checked-color label and $\boldsymbol{y},\boldsymbol{z}\in\{0,1\}^{\bar A}$ are the remaining labels. The same $\boldsymbol{x}$ appears in both labels because acceptance has already enforced equality on the checked color. It remains to determine the output label on $\bar A$. Recall that although the graph-basis label is written as a pattern of $Z$ labels, the bit $\mu_v$ is equivalently the stabilizer-sign bit of $K_v$. Let $v\in\bar A$. The $v$-th output label bit is the eigenvalue bit of the retained target stabilizer
\[
    K^{\rm t}_v
    =
    X^{\rm t}_v
    \prod_{\substack{a\in A\\(a,v)\in E}}
        Z^{\rm t}_a
    \prod_{\substack{w\in\bar A\\(w,v)\in E}}
        Z^{\rm t}_w .
\]
We consider two cases:
\begin{enumerate}
\item If $v\in W\setminus A$, then the auxiliary sacrifice has a qubit at $v$, and the CNOT at $v$ has direction ${\rm t}\rightarrow{\rm s}$. Using the CNOT propagation rule for an $X$ operator on the control qubit,
\[
    X^{\rm t}_v
    \longmapsto
    X^{\rm t}_vX^{\rm s}_v .
\]
For each neighboring $a\in A$, the CNOT at $a$ has direction ${\rm s}\rightarrow{\rm t}$. Using the CNOT propagation rule for a $Z$ operator on the target qubit,
\[
    Z^{\rm t}_a
    \longmapsto
    Z^{\rm t}_aZ^{\rm s}_a .
\]
For each neighboring $w\in\bar A$, the operator $Z^{\rm t}_w$ is unchanged:
\[
    Z^{\rm t}_w
    \longmapsto
    Z^{\rm t}_w .
\]
Multiplying these pulled-back factors gives
\[
    K^{\rm t}_v
    \longmapsto
    \left(
        X^{\rm t}_v
        \prod_{\substack{a\in A\\(a,v)\in E}}
            Z^{\rm t}_a
        \prod_{\substack{w\in\bar A\\(w,v)\in E}}
            Z^{\rm t}_w
    \right)
    \left(
        X^{\rm s}_v
        \prod_{\substack{a\in A\\(a,v)\in E}}
            Z^{\rm s}_a
    \right)
    =
    K^{\rm t}_vK^{\rm s}_v .
\]
Here $K^{\rm s}_v$ is the stabilizer generator of $v$ in the auxiliary graph $(W,E_A)$; it contains only the neighbors of $v$ in $A$ because the auxiliary graph contains only the crossing edges between $A$ and $\bar A$. Therefore
\[
    (-1)^{\mu^{\rm out}_v}
    =
    (-1)^{\mu^{\rm t}_v}
    (-1)^{\mu^{\rm s}_v},
\]
and hence
\[
    \mu^{\rm out}_v
    =
    \mu^{\rm t}_v\oplus\mu^{\rm s}_v .
\]

\item If $v\in\bar A\setminus W$, then $v$ is not adjacent to the checked color class $A$, because $W=A\cup N_G(A)$. Hence the auxiliary sacrifice has no qubit at $v$, no CNOT acts at $v$, and $K^{\rm t}_v$ has no neighboring $Z^{\rm t}_a$ factor with $a\in A$. Therefore $K^{\rm t}_v$ pulls back to itself, so
\[
    \mu^{\rm out}_v
    =
    \mu^{\rm t}_v .
\]
By the embedding convention, the absent sacrificial label is $\mu^{\rm s}_v=0$, so the same equation can be written as
\[
    \mu^{\rm out}_v
    =
    \mu^{\rm t}_v\oplus\mu^{\rm s}_v .
\]
\end{enumerate}
Thus, in both cases,
\[
    \boldsymbol{\mu}^{\rm out}_{\bar A}
    =
    \boldsymbol{\mu}^{\rm t}_{\bar A}
    \oplus
    \boldsymbol{\mu}^{\rm s}_{\bar A}.
\]
Consequently, conditioned on acceptance, the checked part of the retained target label remains $\boldsymbol{x}$ and the unchecked part is $\boldsymbol{y}\oplus\boldsymbol{z}$. Hence the retained target label is
\[
    (\boldsymbol{x},\boldsymbol{y}\oplus\boldsymbol{z}).
\]
Now fix an output label $(\boldsymbol{x},\boldsymbol{d})$. It can be produced by any sacrificial unchecked label $\boldsymbol{z}$, provided the target unchecked label is
\[
    \boldsymbol{y}
    =
    \boldsymbol{d}\oplus\boldsymbol{z}.
\]
Therefore, the unnormalized accepted output vector is
\[
    \widetilde P_A[\boldsymbol{x},\boldsymbol{d}]
    =
    \sum_{\boldsymbol{z}\in\{0,1\}^{\bar A}}
        P_{\rm t}[\boldsymbol{x},
        \boldsymbol{d}\oplus\boldsymbol{z}]
        P_{\rm s}[\boldsymbol{x},\boldsymbol{z}].
\]
Summing all unnormalized output components gives the success probability, and normalizing gives the conditional output vector:
\[
    q_A
    =
    \sum_{\boldsymbol{x},\boldsymbol{d}}
        \widetilde P_A[\boldsymbol{x},\boldsymbol{d}],
    \qquad
    P_A[\boldsymbol{x},\boldsymbol{d}]
    =
    \frac{
        \widetilde P_A[\boldsymbol{x},\boldsymbol{d}]
    }{q_A}.
\]
The output fidelity after this color check is therefore
\[
    F_A
    =
    P_A[\mathbf{0}]
    =
    \frac{\widetilde P_A[\mathbf{0}]}{q_A}.
\]

\para{Multiple Color Checks.}
A full color-check round can be represented as a sequence of one-color checks. If the color classes are
$A_1,\ldots,A_r$, then
\[
    \mathbf{P}^{(0)}
    =
    \mathbf{P}_{\rm t},
    \qquad
    \mathbf{P}^{(j)}
    =
    \mathcal{C}_{A_j}
    \left(
        \mathbf{P}^{(j-1)},
        \mathbf{P}_{{\rm s},j}
    \right),
    \quad
    j=1,\ldots,r,
\]
where $\mathcal{C}_{A_j}$ denotes the normalized vector update for the $j$-th color check. The total success probability is the product of the conditional success probabilities produced at each step. Applying one check to each color class checks the whole graph state once.

\color{black}
\section{\bf Additional DP Details for General Graph States}
\label{app:graph-dp-details}

In this appendix, we provide additional details for the DP formulation of the \graphdp problem introduced in \S\ref{sec:graph-dp}.

\para{Optimal DP Formulation.}
Recall that $\mathcal{T}(X,k,\overrightarrow{EP})$ is the frontier of annotated construction trees that generate the graph state corresponding to the connected target subgraph $X$ at a rate of at least $k$ while consuming an EP-rate vector bounded by $\overrightarrow{EP}$. The DP combines feasible child trees through the three graph-growth primitives and one-color purification, retaining one annotated tree for each distinct reachable root vector.

\softpara{Base Case.}
The base case uses the two-vertex edge graph state $X_e$ associated with a target edge $e\in E(G)$. The first stage provides the corresponding EP at rate $r_e$ with source vector $\mathbf{P}^{(2)}_e$, which is the root vector of the single-node construction tree $T_e$. If the first stage provides only the scalar Bell-pair fidelity $f_e$, we initialize this source vector as
\[
    P^{(2)}_e[\mathbf{0}]=f_e,
    \qquad
    P^{(2)}_e[\boldsymbol{\mu}]
    =
    \frac{1-f_e}{3},
    \quad
    \boldsymbol{\mu}\neq\mathbf{0},
\]
with entries ordered by the graph-basis labels of $X_e$. The base-case frontier is initialized as
\[
\mathcal{T}(X_e,k,\overrightarrow{EP})
=
\begin{cases}
\{T_e\},&
k\le \floor{r_e}\ \text{and}\
\langle 0,\ldots,k,\ldots,0\rangle\leq\overrightarrow{EP},\\
\emptyset,&
\text{otherwise,}
\end{cases}
\]
where the entry $k$ corresponds to the target edge $e$.

\softpara{Recursive Step.}
For any construction tree with more than one node, the root operation must be one of the following four types:
\begin{enumerate}
    \item {\tt Fuse-Two-Components},
    \item {\tt Extend-by-New-Vertex},
    \item {\tt Add-Internal-Edge},
    \item One-color graph-state purification using an auxiliary sacrificial graph.
\end{enumerate}
Consequently, $\mathcal{T}(X,k,\overrightarrow{EP})$ is obtained by collecting the annotated trees produced by all feasible instantiations of the above four cases and merging duplicate root-vector representations in the retained frontier.

\begin{enumerate}
    \item \emph{{\tt Fuse-Two-Components}.}
    Suppose $X$ can be decomposed into two disjoint connected target subgraphs $X_1$ and $X_2$ together with a target edge $e=(a,b)$, where $a\in V(X_1)$, $b\in V(X_2)$, and
    \[
    X=\bigl(V(X_1)\cup V(X_2),\;E(X_1)\cup E(X_2)\cup\{e\}\bigr).
    \]
    Let $k_1,k_2,k_e$ be discretized rates, and let $\overrightarrow{EP}_1,\overrightarrow{EP}_2,\overrightarrow{EP}_e$ be the corresponding EP-consumption vectors such that
    \[
    k \le \min(k_1,k_2,k_e)\cdot c_{\rm fuse},
    \quad
    \overrightarrow{EP}_1+\overrightarrow{EP}_2+
    \overrightarrow{EP}_e
    \leq
    \overrightarrow{EP}
    \]
    where $c_{\rm fuse}$ denotes the effective rate-scaling factor of the primitive\footnote{The rate-scaling factors come from two sources: the success probability of the corresponding operation and the expected waiting time to collect all required input resources, assuming independent Poisson generation processes; see Appendix~\ref{app:waiting-model}. In particular, synchronizing two equal-rate inputs yields the factor $2/3$, while synchronizing three equal-rate inputs yields the factor $6/11$, corresponding to the reciprocals of the expected maxima of two and three i.i.d.\ exponential arrival times, respectively.}. For every choice of input trees
    \begin{footnotesize}
    \[
    T_1\in\mathcal{T}(X_1,k_1,\overrightarrow{EP}_1),
    T_2\in\mathcal{T}(X_2,k_2,\overrightarrow{EP}_2),
    T_e\in\mathcal{T}(X_e,k_e,\overrightarrow{EP}_e),
    \]
    \end{footnotesize}
    this decomposition yields an annotated tree whose root vector is
    \[
    \mathsf{Fuse}_{X_1,X_2,e}
    \left(
    \mathbf{P}_{T_1}^{(X_1)},\mathbf{P}_{T_2}^{(X_2)},\mathbf{P}_{T_e}^{(X_e)}
    \right),
    \]
    where $\mathsf{Fuse}_{X_1,X_2,e}$ denotes the vector update induced by the label transition $\tau_{\rm fuse}^{X_1,X_2,e}$ in Appendix~\ref{app:graph-fidelity}. The DP inserts all such annotated candidates into the frontier for $(X,k,\overrightarrow{EP})$.
    
    \item \emph{{\tt Extend-by-New-Vertex}.}
    Suppose $X$ is obtained from a smaller connected target subgraph $X'$ by attaching a new vertex $b$ via a target edge $e=(a,b)$, where $a\in V(X')$ and $b\notin V(X')$. Thus,
    \[
    X=\bigl(V(X')\cup\{b\},\;E(X')\cup\{e\}\bigr).
    \]
    Let $k',k_e$ be discretized rates and let $\overrightarrow{EP}',\overrightarrow{EP}_e$ be the corresponding EP-consumption vectors such that
    \[
    k \le \min(k',k_e)\cdot c_{\rm ext},
    \quad
    \overrightarrow{EP}'+\overrightarrow{EP}_e
    \leq
    \overrightarrow{EP},
    \]
    where $c_{\rm ext}$ is the rate-scaling factor of the primitive. For every choice of input trees
    \begin{footnotesize}
    \[
    T'\in\mathcal{T}(X',k',\overrightarrow{EP}'),
    \quad
    T_e\in\mathcal{T}(X_e,k_e,\overrightarrow{EP}_e),
    \]
    \end{footnotesize}
    this decomposition yields an annotated tree whose root vector is
    \[
    \mathsf{Extend}_{X',e}
    \left(
    \mathbf{P}_{T'}^{(X')},\mathbf{P}_{T_e}^{(X_e)}
    \right),
    \]
    where $\mathsf{Extend}_{X',e}$ denotes the vector update induced by the label transition $\tau_{\rm ext}^{X',e}$ in Appendix~\ref{app:graph-fidelity}. Again, all feasible annotated candidates are inserted into the frontier.
    
    \item \emph{{\tt Add-Internal-Edge}.}
    Suppose $X$ is obtained from a connected target subgraph $X'$ by adding a missing internal edge $e=(a,b)$ whose endpoints already lie in $V(X')$. Then
    \[
    V(X)=V(X'),\qquad E(X)=E(X')\cup\{e\}.
    \]
    Let $k',k_e$ be discretized rates and let $\overrightarrow{EP}',\overrightarrow{EP}_e$ be the corresponding EP-consumption vectors such that
    \[
    k \le \min(k',k_e)\cdot c_{\rm int},
    \quad
    \overrightarrow{EP}'+\overrightarrow{EP}_e
    \leq
    \overrightarrow{EP},
    \]
    where $c_{\rm int}$ is the rate-scaling factor of the primitive. For every choice of input trees
    \begin{footnotesize}
    \[
    T'\in\mathcal{T}(X',k',\overrightarrow{EP}'),
    \quad
    T_e\in\mathcal{T}(X_e,k_e,\overrightarrow{EP}_e),
    \]
    \end{footnotesize}
    this decomposition yields an annotated tree whose root vector is
    \[
    \mathsf{AddEdge}_{X',e}
    \left(
    \mathbf{P}_{T'}^{(X')},\mathbf{P}_{T_e}^{(X_e)}
    \right),
    \]
   where $\mathsf{AddEdge}_{X',e}$ denotes the vector update induced by the label transition $\tau_{\rm add}^{X',e}$ in Appendix~\ref{app:graph-fidelity}. The DP again inserts all feasible annotated candidates into the frontier.

   \item \emph{One-Color Graph State Purification.}
    Finally, the graph state corresponding to $X$ can be purified by applying one color-check subprotocol to a selected color class $A$ of $X$. Following Appendix~\ref{app:graph-fidelity}, let $(W,E_A)$ be the auxiliary sacrificial graph for this color check. The DP includes this check only when $(W,E_A)$ is connected and thus belongs to its state family. One input is the target graph state on $X$, and the other input is the sacrificial graph state on $(W,E_A)$. For notational simplicity, $\mathcal{T}((W,E_A),\cdot,\cdot)$ denotes the frontier for this auxiliary graph.
    For purification using two independently generated inputs, let $k_{\rm t},k_{\rm s}$ be discretized rates and let $\overrightarrow{EP}_{\rm t},\overrightarrow{EP}_{\rm s}$ be the corresponding EP-consumption vectors such that
    \[
    k \le \frac{2q_A}{3}\min(k_{\rm t},k_{\rm s}),
    \quad
    \overrightarrow{EP}_{\rm t}+
    \overrightarrow{EP}_{\rm s}
    \leq
    \overrightarrow{EP},
    \]
    where $q_A$ is the color-check success probability for the chosen input vectors, as defined in Appendix~\ref{app:graph-fidelity}. For every choice of input trees
    \begin{footnotesize}
    \[
    T_{\rm t}\in\mathcal{T}(X,k_{\rm t},\overrightarrow{EP}_{\rm t}),
    \qquad
    T_{\rm s}\in\mathcal{T}((W,E_A),k_{\rm s},\overrightarrow{EP}_{\rm s})
    \]
    \end{footnotesize}
    this decomposition yields an annotated tree whose root vector is
    \[
    \mathcal{C}_A
    \left(
    \mathbf{P}_{T_{\rm t}}^{(X)},\mathbf{P}_{T_{\rm s}}^{(W,E_A)}
    \right),
    \]
    where $\mathcal{C}_A$ denotes the one-color vector update defined in Appendix~\ref{app:graph-fidelity}.

    If $(W,E_A)=X$, we also consider same-stream purification using two copies generated by one child tree $T_0\in\mathcal{T}(X,k_0,\overrightarrow{EP}_0)$. This yields a candidate when
    \[
    k\leq\frac{q_A}{2}k_0,
    \qquad
    \overrightarrow{EP}_0\leq\overrightarrow{EP},
    \]
    with root vector $\mathcal{C}_A(\mathbf{P}_{T_0}^{(X)},\mathbf{P}_{T_0}^{(X)})$ and the corresponding success probability $q_A$. The DP inserts all feasible annotated candidates from both cases into the frontier.
\end{enumerate}
These four cases define the candidate sets
$\mathcal{C}_{\rm fuse}$, $\mathcal{C}_{\rm ext}$,
$\mathcal{C}_{\rm int}$, and $\mathcal{C}_{\rm pur}$ used in the frontier recurrence for $X$ with more than one edge:
\[
\mathcal{T}(X,k,\overrightarrow{EP})
=
\operatorname{Prune}
\left(
\mathcal{C}_{\rm fuse}
\cup
\mathcal{C}_{\rm ext}
\cup
\mathcal{C}_{\rm int}
\cup
\mathcal{C}_{\rm pur}
\right),
\]
where $\operatorname{Prune}$ removes only duplicate root vectors, keeping one annotated tree for each distinct vector. For an edge graph $X_e$, the frontier includes feasible trees that use the source EP directly or purify copies of it.

\section{\bf Additional LP Details for General Graph States}
\label{app:graph-lp-details}

In this appendix, we provide additional details for the LP formulation of the \graphlp problem introduced in \S\ref{sec:graph-lp}. As in the main text, the LP operates only on the second stage of the overall pipeline: for each target edge $e\in E(G)$, the first stage has already produced an end-to-end EP resource with generation rate $r_e$ and graph-basis source vector $\mathbf{P}^{(2)}_e$. The LP then determines how to combine these target edge EP resources using the adopted graph state growth and purification operations. The hypergraph construction below uses graph-basis vector labels.

\para{Hypergraph Vertices.}
Let $\mathcal{X}$ denote the family of connected target subgraphs of $G$, including the $2$-vertex edge graph states corresponding to the target edges. Let $\mathcal{D}$ denote the collection of graph-basis vectors used as labels in the hypergraph. Whenever a vector label is paired with a graph $X$, it is understood to have the graph-basis dimension for $X$. We build a directed hypergraph $H_{G,\mathcal{D}}$ as follows. The vertex set $V(H_{G,\mathcal{D}})$ contains:

\begin{enumerate}
    \item Two distinguished vertices, \emph{start} and \emph{term}.
    \item For each connected target subgraph $X\in\mathcal{X}$ and each compatible vector $\mathbf{P}\in\mathcal{D}$, the following hypervertices:
    \begin{enumerate}
    \item $\id{avail}(X,\mathbf{P})$, representing a copy of $X$ with graph-basis vector $\mathbf{P}$ that is available for further operations;
    \item $\id{fuse}(X,\mathbf{P})$, representing a copy of $X$ produced by {\tt Fuse-Two-Components};
    \item $\id{ext}(X,\mathbf{P})$, representing a copy of $X$ produced by {\tt Extend-by-New-Vertex};
    \item $\id{int}(X,\mathbf{P})$, representing a copy of $X$ produced by {\tt Add-Internal-Edge};
    \item $\id{purify}(X,\mathbf{P})$, representing a copy of $X$ produced by a color-check purification operation.
    \end{enumerate}
\end{enumerate}

\para{Hyperarcs.}
The hyperarcs of $H_{G,\mathcal{D}}$ correspond to the growth and purification operations of the second-stage graph state. 

\begin{enumerate}
\item[\textbf{[Start]}]
    For each target edge $e\in E(G)$, let $X_e$ denote the corresponding $2$-vertex edge graph state. If the first stage provides $e$ with rate $r_e$ and source vector $\mathbf{P}^{(2)}_e$, then we include the hyperarc
    \[
        \bigl(\{\emph{start}\},
        \{\id{avail}(X_e,\mathbf{P}^{(2)}_e)\}\bigr).
    \]

\item[\textbf{[Fuse]}]
Let $X_1$ and $X_2$ be disjoint connected target subgraphs, and let $e=(a,b)\in E(G)$ be a target edge such that $a\in V(X_1)$ and $b\in V(X_2)$. Let $X_e$ denote the corresponding $2$-vertex edge graph state, and define
\[
X=\bigl(V(X_1)\cup V(X_2),\; E(X_1)\cup E(X_2)\cup\{e\}\bigr).
\]
For vectors
$\mathbf{P}_1,\mathbf{P}_2,\mathbf{P}_e,\mathbf{P}_{\rm out}\in\mathcal{D}$, where 
$\mathbf{P}_{\rm out}$ is determined by the vector update rule
$\mathsf{Fuse}_{X_1,X_2,e}(\cdot)$, we add the hyperarc
\begin{small}
\[
    \bigl(
    \{\id{avail}(X_1,\mathbf{P}_1),
    \id{avail}(X_2,\mathbf{P}_2),
    \id{avail}(X_e,\mathbf{P}_e)\},
    \{\id{fuse}(X,\mathbf{P}_{\rm out})\}
    \bigr),
\]
\end{small}
followed by the consolidation hyperarc
\[
    \bigl(\{\id{fuse}(X,\mathbf{P})\},
    \{\id{avail}(X,\mathbf{P})\}\bigr).
\]

\item[\textbf{[Extend]}] Let $X_0$ be a connected target subgraph, and let $e=(a,b)\in E(G)$ be a target edge such that $a\in V(X_0)$ and $b\notin V(X_0)$. Let $X_e$ denote the corresponding $2$-vertex edge graph state, and define
\[
X=\bigl(V(X_0)\cup\{b\},\; E(X_0)\cup\{e\}\bigr).
\]
For vectors
$\mathbf{P}_0,\mathbf{P}_e,\mathbf{P}_{\rm out}\in\mathcal{D}$,
where $\mathbf{P}_{\rm out}$ is determined by the vector update rule
$\mathsf{Extend}_{X_0,e}(\cdot)$, we add the hyperarc
\[
    \bigl(
    \{\id{avail}(X_0,\mathbf{P}_0),
    \id{avail}(X_e,\mathbf{P}_e)\},
    \{\id{ext}(X,\mathbf{P}_{\rm out})\}
    \bigr),
\]
followed by the consolidation hyperarc
\[
    \bigl(\{\id{ext}(X,\mathbf{P})\},
    \{\id{avail}(X,\mathbf{P})\}\bigr).
\]

\item[\textbf{[Internal]}] Let $X_0$ be a connected target subgraph, and let $e=(a,b)\in E(G)$ be a target edge such that $a,b\in V(X_0)$ and $e\notin E(X_0)$. Let $X_e$ denote the corresponding $2$-vertex edge graph state, and define
\[
X=\bigl(V(X_0),\; E(X_0)\cup\{e\}\bigr).
\]
For vectors
$\mathbf{P}_0,\mathbf{P}_e,\mathbf{P}_{\rm out}\in\mathcal{D}$,
where $\mathbf{P}_{\rm out}$ is determined by the vector update rule
$\mathsf{AddEdge}_{X_0,e}(\cdot)$, we add the hyperarc
\[
    \bigl(
    \{\id{avail}(X_0,\mathbf{P}_0),
    \id{avail}(X_e,\mathbf{P}_e)\},
    \{\id{int}(X,\mathbf{P}_{\rm out})\}
    \bigr),
\]
followed by the consolidation hyperarc
\[
    \bigl(\{\id{int}(X,\mathbf{P})\},
    \{\id{avail}(X,\mathbf{P})\}\bigr).
\]

\item[\textbf{[Purify]}] 
Let $A$ be a color class of the graph state on $X$. Following Appendix~\ref{app:graph-fidelity}, let $(W,E_A)$ be the auxiliary graph for this one-color check. As in the DP, we include this check only when $(W,E_A)$ is connected and thus belongs to $\mathcal{X}$. Consider vectors
$\mathbf{P}_{\rm t},\mathbf{P}_{\rm s},\mathbf{P}_{\rm out}\in\mathcal{D}$,
where $\mathbf{P}_{\rm t}$ is on $X$, $\mathbf{P}_{\rm s}$ is on $(W,E_A)$, and $\mathbf{P}_{\rm out}$ is determined by the color-check update rule $\mathcal{C}_{A}(\cdot)$. For purification using two independently generated inputs with $(X,\mathbf{P}_{\rm t})\neq((W,E_A),\mathbf{P}_{\rm s})$, we add
the following hyperarc:
\begin{small}
\[
    \bigl(
    \{\id{avail}(X,\mathbf{P}_{\rm t}),
    \id{avail}((W,E_A),\mathbf{P}_{\rm s})\},
    \{\id{purify}(X,\mathbf{P}_{\rm out})\}
    \bigr).
\]
\end{small}
If $(W,E_A)=X$ and the two copies are drawn from the same stream, we instead use $\mathbf{P}_{\rm s}=\mathbf{P}_{\rm t}$ and the single-tail hyperarc
\[
    \bigl(
    \{\id{avail}(X,\mathbf{P}_{\rm t})\},
    \{\id{purify}(X,\mathbf{P}_{\rm out})\}
    \bigr).
\]
The synchronization factor is $1/2$ for this single-tail form and $2/3$ for the two-tail form.
Each purification output is followed by
\[
    \bigl(\{\id{purify}(X,\mathbf{P})\},
    \{\id{avail}(X,\mathbf{P})\}\bigr).
\]

\item[\textbf{[Term]}] 
For each vector $\mathbf{P}\in\mathcal{D}$ on the full target graph $G$ whose ideal component satisfies $P[\mathbf{0}]\ge \fth$, we add the hyperarc
\[
    \bigl(\{\id{avail}(G,\mathbf{P})\},\{\emph{term}\}\bigr).
\]
\end{enumerate}

\para{LP Variables and Constraints.}
We introduce one nonnegative variable $z_h$ for each hyperarc $h$ in $H_{G,\mathcal{D}}$; intuitively, $z_h$ denotes the generation rate assigned to hyperarc $h$. Let $in(v)$ and $out(v)$ denote the sets of incoming and outgoing hyperarcs of a hypervertex $v$, respectively. The LP constraints are as follows.

\begin{itemize}
\item \underline{Flow conservation at available-state vertices.} For every vertex $v=\id{avail}(\cdot)$,
\[
\sum_{h\in in(v)} z_h = \sum_{h'\in out(v)} z_{h'}.
\]

\item \underline{Operation constraints.} For every operation vertex
$v\in\{\id{fuse}(\cdot),\id{ext}(\cdot), \id{int}(\cdot),\id{purify}(\cdot)\}$, the outgoing rate is obtained from the incoming rate through the corresponding effective rate coefficient:
\[
\sum_{h\in in(v)} c_h z_h
=
\sum_{h'\in out(v)} z_{h'}.
\]
Here, $c_h$ is the effective rate coefficient of the operation represented by hyperarc $h$; it incorporates the synchronization factor and the success probability of that operation.

\item \underline{Capacity constraints on first-stage resources.} For each Start hyperarc $h_e$ corresponding to the target edge $e$,
\[
z_{h_e}\le r_e.
\]

\item \underline{Objective.} Maximize the total rate of target graph states above the fidelity threshold:
\[
\max \sum_{h\in in(\emph{term})} z_h.
\]
\end{itemize}

\section{\bf Impact of Discretization Gap and Bounded Iterative Purification}
\label{app:disc-gap}

\subsection{\textbf{Discretization Gap}}
Here, we bound the effect of discretization on the computed fidelity and generation rate of a fixed construction. Both fidelity discretization and rate discretization aim to maximize the generation rate subject to the same final fidelity threshold. They differ in which quantity is discretized.

\para{Setup and Discretization Rules.}
Fix an instance of the \fed, \gfed, \gdp, \ggdp, \graphdp, or \graphlp problem, together with the same source resources, waiting model, and operation rules as in the main text. For GHZ and graph states, we use the corresponding fidelity vector $\mathbf{P}$ and write $F(\mathbf{P})$ for its ideal component: $P[0,\mathbf{0}]$ for GHZ states and $P[\mathbf{0}]$ for graph states. 
Below, $f^{cont}$ and $f^{disc}$ denote the final fidelities under the continuous and discretized models, respectively; $r^{cont}$ and $r^{disc}$ denote the corresponding generation rates.

Fix a purification bound $m\geq1$. In the EP formulation, $m$ bounds the iterations within each iterated-purification node, as in the EP DP recurrence. 
For GHZ and graph states, $m$ bounds consecutive purification steps between successive fusion or graph-growth operations, including the blocks before the first such operation and after the last. In a construction tree, a dependency path is a root-to-leaf path; in a level-based structure, the same restriction applies to every source-to-output dependency path. An iterated purification or a sequence of color checks is counted by its constituent purification steps.

\softpara{Fidelity Discretization.}
As in the LP appendices, let $\mathcal{D}$ denote the finite collection of vector labels. For an intermediate resource $X$, let $Q_X$ map its fidelity vector in the continuous space to a compatible representative in $\mathcal{D}$. The representatives must remain valid probability vectors of the same dimension. Let $\delta_F$ bound the replacement error for every state considered:
\[
    \|Q_X(\mathbf{P})-\mathbf{P}\|_1\leq\delta_F.
\]

Since $F(\mathbf{P})$ is one component of the vector, this bound also gives
\[
    |F(Q_X(\mathbf{P}))-F(\mathbf{P})|
    \leq\|Q_X(\mathbf{P})-\mathbf{P}\|_1
    \leq\delta_F.
\]
Thus, controlling the whole-vector error also controls the scalar fidelity error.

For the scalar EP formulation, the replacement error is an absolute fidelity error. Let $\discretefid=\{f_1<\cdots<f_d\}$ cover the scalar fidelity domain, and define
\[
    \Delta_{\discretefid}
    =\max_{1\leq i<d}(f_{i+1}-f_i).
\]
If $f_i\leq f<f_{i+1}$, downward rounding replaces $f$ by $f_i$, so
\[
    0\leq f-f_i<f_{i+1}-f_i\leq\Delta_{\discretefid}.
\]
Hence, in the scalar case, we may take $\delta_F=\Delta_{\discretefid}$.

Unlike scalar downward rounding, vector replacement need not decrease the ideal component. Moreover, a lower ideal component before an operation does not imply a lower one after the operation. We therefore bound the magnitude of the vector error without assuming that rounding always underestimates fidelity.

\softpara{Rate Discretization.}
Use the rate grid $\discreterate=\{K_1<\cdots<K_q\}$ from \S\ref{sec:ghz-dp}, covering the positive rates considered, and define
\[
    \Delta_{\discreterate}
    =\max_{1\leq i<q}(K_{i+1}-K_i).
\]
Flooring an achieved rate, or ceiling a required input rate within this range, changes it by at most $\Delta_{\discreterate}$. Rate rounding can also affect fidelity, since the estimated storage times depend on the input generation rates. We therefore bound both rate and fidelity errors in the analysis below.

\para{Construction Depth.}
Let $B$ denote an upper bound on the total number of operation updates along any path from a source resource to the final output. We give explicit depth bounds for EP, GHZ, and graph-state generation.

\begin{lem-prf}
\label{lemma:disc-depth}
For a fixed instance and purification bound $m$, the EP DP constructions and the GHZ and graph-state constructions have a depth bound $B$ independent of the discretization grid. For the EP LP, a grid-independent depth bound is obtained by allowing a prescribed loss $\epsilon>0$ in total output rate, without changing the fidelity labels of the retained outputs.
\end{lem-prf}
\begin{prf}
For EP generation, let $n$ be the number of network nodes. The EP DP uses a tree-height limit of $2n$, as established in Lemma~\ref{lemma:nocycle}, counting an iterated-purification node as one node. Each swapping node contributes one update, and each iterated-purification node contributes at most $m$ one-step updates. Thus, we may take
\[
    B=2nm.
\]
For the EP LP, Theorem~\ref{thm:ep-lp} gives a finite level-based swapping structure, which can be unfolded into a collection of purification-augmented swapping trees. The original source-rate allocations are divided among these trees, so their combined consumption on each source link does not exceed the LP allocation. As in \S\ref{sec:ep-lp}, each operation retains at most $2/3$ of the rate allocated to each input stream. Let $R_{\rm src}$ be the sum of the source-rate capacities. A swapping tree with output rate $r$ and more than $d$ operation nodes along one path requires at least $(3/2)^d r$ of source rate along that path. Since the source allocations across all trees sum to at most $R_{\rm src}$, discarding swapping trees deeper than $d$ loses at most
\[
    \epsilon_{\rm depth}\leq R_{\rm src}(2/3)^d
\]
in total output rate, without changing the fidelity labels of the retained outputs. The same bound applies to any finite level-based swapping structure with continuous fidelity labels.

For a prescribed rate error $0<\epsilon<R_{\rm src}$, choose
\[
    d=\left\lceil
        \frac{\log(R_{\rm src}/\epsilon)}{\log(3/2)}
      \right\rceil,
    \qquad B=md.
\]
The discarded swapping trees contribute at most $\epsilon$ in total output rate, and every retained tree has at most $B$ individual operation updates along each path, since each iterated-purification node contains at most $m$ steps. For fixed source capacities, $m$, and $\epsilon$, this approximation depth is independent of the grid.

For an $n$-qubit GHZ target, each fusion produces a state on a strictly larger qubit set than either input, whereas purification never decreases the qubit set size. Starting from a two-qubit source, a path therefore contains at most $n-2$ fusions, and we may take
\[
    B=(n-2)+(n-1)m.
\]

For a target graph $G$ with $n$ vertices, every input graph is a subgraph of the output graph, including the auxiliary sacrificial input of a color check. Each graph-growth operation produces a graph with strictly more edges than any of its input graphs. Since an edge source has one edge and a simple undirected graph has at most $n(n-1)/2$ edges, a path contains at most $n(n-1)/2-1$ growth operations. Consequently, we may take
\[
    B=\frac{n(n-1)}{2}(m+1)-1.
\]
Thus, for fixed $n$ and $m$, the number of operations along any path from a source resource to the final output is bounded independently of the discretization grid.
\end{prf}

\para{Regularity of the Update Maps.}
For corresponding continuous and discretized resources, measure the vector error in the $\ell_1$ norm and divide each generation-rate or EP-consumption error by the fixed reference rate $r_*=1\,\mathrm{s}^{-1}$ to make it dimensionless. For scalar EP fidelities, use the absolute fidelity error in place of the vector error.

\begin{lem-prf}
\label{lemma:disc-lipschitz}
Fix a bounded domain containing the compared inputs and outputs, where positive input rates and purification acceptance probabilities have fixed positive lower bounds independent of the grid. The adopted operation updates, including waiting and classical-communication delays, admit a common grid-independent Lipschitz constant $L\geq1$: whenever every input has vector error and normalized rate errors at most $\epsilon$, the corresponding output errors before rounding are at most $L\epsilon$.
\end{lem-prf}
\begin{prf}
\begin{enumerate}
    \item \emph{Swapping, Fusion, and Graph Growth.}
    The fidelity updates for EP swapping, GHZ fusion, and the graph-state growth primitives are polynomial in their input fidelities or vector components. Since these inputs lie in $[0,1]$, their derivatives are bounded.

    \item \emph{Waiting and Rate Updates.}
    Waiting-time expressions include terms such as $1/r$ and $r_1/[r_2(r_1+r_2)]$, where $r,r_1,r_2$ are input generation rates. Their derivatives are bounded when these rates have a fixed positive lower bound. Rate updates also involve synchronization expressions such as $\min(r_1,r_2)$. Although this minimum is not differentiable when the two rates coincide, changing each input rate by at most $\epsilon$ changes the minimum by at most $\epsilon$, so it is Lipschitz.

    \item \emph{Purification.}
    Let $\mathbf{a}$ and $\widetilde{\mathbf{a}}$ be the unnormalized accepted output vectors calculated from the continuous and discretized inputs, respectively. Each component gives the probability that purification succeeds and produces the corresponding output label, and is polynomial in the input-vector components after waiting. Their sums
    \[
        q=\|\mathbf{a}\|_1,
        \qquad
        \widetilde q=\|\widetilde{\mathbf{a}}\|_1
    \]
    are therefore the purification success probabilities. Conditioned on success, the output vectors are $\mathbf{a}/q$ and $\widetilde{\mathbf{a}}/\widetilde q$.

    Assume both success probabilities are at least a fixed $p_{\min}>0$. Adding and subtracting $\widetilde{\mathbf{a}}/q$ and applying the triangle inequality gives
    \begin{align*}
        \left\|\frac{\mathbf{a}}{q}
        -\frac{\widetilde{\mathbf{a}}}{\widetilde q}\right\|_1
        &\leq
        \frac{\|\mathbf{a}-\widetilde{\mathbf{a}}\|_1
        +|q-\widetilde q|}{q}\\
        &\leq
        \frac{2}{p_{\min}}
        \|\mathbf{a}-\widetilde{\mathbf{a}}\|_1,
    \end{align*}
    where the second inequality uses
    $|q-\widetilde q|\leq\|\mathbf{a}-\widetilde{\mathbf{a}}\|_1$.
    Thus, normalization amplifies the vector error by at most the fixed factor $2/p_{\min}$ and is Lipschitz on this domain.
\end{enumerate}

A common grid-independent Lipschitz constant $L$ can therefore be chosen for all primitive maps on this domain.
\end{prf}

\para{Error Bounds and Convergence.}
Using the construction-depth bounds in Lemma~\ref{lemma:disc-depth} and the per-operation error bound in Lemma~\ref{lemma:disc-lipschitz}, we bound the accumulated discretization error for a fixed construction.

\begin{thm-prf}
\label{thm:disc-loss}
Consider paired continuous and discretized constructions with the same operations and depth at most $B$, satisfying the regularity conditions of Lemma~\ref{lemma:disc-lipschitz}. Let $e_0$ bound the vector errors and normalized rate errors at the sources, and let $\delta$ bound the local rounding error at each operation. For fidelity discretization, the output vector is replaced by a grid representative, giving local error $\delta=\delta_F$. For rate discretization, the vector is not rounded, while each rate label is rounded once, giving local error $\delta=\Delta_{\discreterate}/r_*$. We may take
\[
    \delta=\max\left\{\delta_F,
                    \frac{\Delta_{\discreterate}}{r_*}\right\}.
\]
Define
\[
    C_B=\sum_{j=0}^{B-1}L^j,
    \qquad E_B=L^B e_0+C_B\delta.
\]
At each terminal output,
\begin{align*}
    \|\mathbf{P}^{disc}-\mathbf{P}^{cont}\|_1
    &\leq E_B
    &&\text{(vector fidelity error)},\\
    |f^{disc}-f^{cont}|
    &\leq E_B
    &&\text{(scalar fidelity error)},\\
    |r^{disc}-r^{cont}|
    &\leq r_*E_B
    &&\text{(rate error)}.
\end{align*}
The same rate-error bound holds for each EP-consumption component.
\end{thm-prf}
\begin{prf}
Let $e_b$ bound both the vector error and every normalized rate error at all nodes reached after at most $b$ operation updates. Before rounding, the Lipschitz inequality bounds the next output error by $Le_b$. Rounding introduces at most $\delta$ more error. The triangle inequality therefore gives
\[
    e_{b+1}\leq Le_b+\delta.
\]
Induction yields
\[
    e_b\leq L^b e_0+\delta\sum_{j=0}^{b-1}L^j.
\]
Taking $b=B$ gives $e_B\leq E_B$, establishing the stated vector fidelity, scalar fidelity, and rate error bounds. For both trees and acyclic level-based structures, we process each operation only after bounding the errors of all its inputs.
\end{prf}

The error bound also yields convergence under grid refinement. We refine the vector-label collection so that $\delta_F\to0$ for fidelity discretization, or the rate grid so that $\Delta_{\discreterate}\to0$ for rate discretization. In either case, any source-rounding errors must also tend to zero. The following result compares evaluations of the same construction.

\begin{thm-prf}
\label{thm:disc-convergence}
Fix a finite construction and progressively refine the discretization grid. Assume the conditions of Theorem~\ref{thm:disc-loss} hold throughout, with a common grid-independent $L$, source error $e_0\to0$, and local rounding error $\delta\to0$. Then, at each terminal output,
\begin{align*}
    \lim_{e_0,\delta\to0}\mathbf{P}^{disc}
    &=\mathbf{P}^{cont},\\
    \lim_{e_0,\delta\to0}f^{disc}
    &=f^{cont},\\
    \lim_{e_0,\delta\to0}r^{disc}
    &=r^{cont}.
\end{align*}
For a fixed finite level-based structure, its total output rate also converges to the continuously evaluated total rate.
\end{thm-prf}
\begin{prf}
The construction has fixed finite depth $B$. Since $L$ is independent of the grid, so is $C_B=\sum_{j=0}^{B-1}L^j$. Consequently,
\[
    \lim_{e_0,\delta\to0}E_B
    =\lim_{e_0,\delta\to0}(L^B e_0+C_B\delta)
    =0.
\]
The three limits follow from the corresponding error bounds in Theorem~\ref{thm:disc-loss}. A fixed finite level-based structure has finitely many terminal streams, so summing their convergent rates gives convergence of its total output rate.
\end{prf}

Thus, under the stated regularity condition, both discretization choices recover the continuous fidelity and rate of a fixed construction in the fine-grid limit.

\para{EP Example for the Regularity Condition.}
For the \fed formulation, the update rules are the EP swapping rule in Eqn.~\ref{eqn:EP_swapfidelity}, the one-step EP purification rule in Eqn.~\ref{eqn:EP_One_Purify_Fidelity}, and the iterated-purification rules in Eqns.~\ref{eqn:EP_Purify_Fidelity} and~\ref{eqn:EP_Purify_Prob}.
The DP treats an iterated purification with $i\leq m$ rounds as a single candidate operation; see Eqn.~\ref{eqn:iterated-purify-latency}. 
Each purification round is counted as a separate update, and each iterated-purification node contributes at most $m$ updates. The height bound in Lemma~\ref{lemma:nocycle} therefore gives
$B=2nm$.

To illustrate the fidelity part of the regularity condition, first consider the scalar updates before applying the waiting channel. For the EP swapping rule
\[
\swapfidelity(f_1,f_2)=\frac14\!\left(1+\frac13(4f_1-1)(4f_2-1)\right),
\]
we have
\[
\frac{\partial \swapfidelity}{\partial f_1}=\frac{4f_2-1}{3},
\qquad
\frac{\partial \swapfidelity}{\partial f_2}=\frac{4f_1-1}{3}.
\]
Thus, over $f_1,f_2\in[\frac12,1]$, both partial derivatives are at most $1$. The swapping fidelity map is therefore Lipschitz with constant $2$ with respect to the maximum norm.
The one-step EP purification rule in Eqn.~\ref{eqn:EP_One_Purify_Fidelity}, with target fidelity $x$ and sacrificial fidelity $y$, can be written as
\[
\purifyfidelity(x,y)=\frac{10xy-x-y+1}{8xy-2x-2y+5}.
\]
For $x,y\in[\frac12,1]$, differentiation gives
\[
0<\frac{\partial\purifyfidelity(x,y)}{\partial x}
=\frac{3(-4y^2+14y-1)}{(8xy-2x-2y+5)^2}
\leq\frac34.
\]
To obtain the upper bound, note that the denominator before squaring is at least $2y+4$; the remaining inequality is equivalent to $15(y-1)^2\geq0$. By symmetry, the same bound holds for the derivative with respect to $y$. Hence, the purification fidelity map is Lipschitz with constant $\frac32$ with respect to the maximum norm.

These scalar swapping and purification maps therefore admit a common per-update Lipschitz bound of $2$. For the complete operation updates, including waiting, classical-communication delays, and rate updates, Lemma~\ref{lemma:disc-lipschitz} gives a finite Lipschitz bound for each map on the specified domain. We take $L$ to be the maximum of $2$ and these bounds; this choice is independent of the discretization grid because the maps and domain are fixed. With this common $L$ and $B=2nm$ for the EP DP, Theorems~\ref{thm:disc-loss} and~\ref{thm:disc-convergence} bound the fidelity and rate errors and establish convergence for a fixed purification-augmented swapping tree. The same argument applies to a fixed finite level-based swapping structure.

\subsection{\textbf{Effect of the Purification-Iteration Bound}}

We now examine the effect of increasing the purification bound $m$ defined above. All other operation restrictions and source resources are held fixed; in the discretized formulation, the grid is also fixed. Increasing $m$ allows longer purification sequences while preserving all previously feasible constructions.

Let $OPT^{cont}_m(f^\tau)$ and $OPT^{disc}_m(f^\tau)$ denote the least upper bounds on the generation rates achievable at fidelity threshold $f^\tau$ with purification bound $m$, in the continuous and discretized formulations, respectively. Define $OPT^{cont}_\infty(f^\tau)$ and $OPT^{disc}_\infty(f^\tau)$ analogously over finite constructions with no prescribed purification bound. The optimal value is defined to be zero when no positive-rate solution is feasible.

\medskip
\begin{thm-prf}
\label{thm:iter-bound-monotone}
For every fidelity threshold $f^\tau$,
{\footnotesize
\[
OPT^{cont}_1(f^\tau)\le OPT^{cont}_2(f^\tau)\le \cdots \le OPT^{cont}_m(f^\tau)\le OPT^{cont}_\infty(f^\tau),
\]
}
and likewise,
{\footnotesize
\[
OPT^{disc}_1(f^\tau)\le OPT^{disc}_2(f^\tau)\le \cdots \le OPT^{disc}_m(f^\tau)\le OPT^{disc}_\infty(f^\tau).
\]
}
\end{thm-prf}
\begin{prf}
Increasing the purification-iteration bound can only add feasible purification strategies; it never removes any strategy available under a smaller bound. Hence, the feasible set is nested in $m$, and the optimal generation rate is monotone non-decreasing in both the continuous and discretized formulations.
\end{prf}

\para{EP Example for the Purification-Iteration Bound.}
This example shows how a moderate purification-iteration bound can capture most of the fidelity improvement available from repeated pumping. Consider the Bennett-style EP purification model in Eqn.~\ref{eqn:EP_One_Purify_Fidelity}, without additional waiting-induced decoherence. For a fixed sacrificial-copy fidelity $f\in(\frac12,1)$, define
\[
H_f(x)=\purifyfidelity(x,f)=\frac{10fx-f-x+1}{8fx-2f-2x+5}.
\]
Then, the left-deep iterated purification rule in Eqn.~\ref{eqn:EP_Purify_Fidelity} can be written as
\[
\iterpurifyfidelity{0}(f)=f,
\qquad
\iterpurifyfidelity{i+1}(f)=H_f(\iterpurifyfidelity{i}(f)).
\]

Differentiating $H_f(x)$ gives
\[
0<\frac{\partial H_f(x)}{\partial x}
=\frac{3(-4f^2+14f-1)}{(8fx-2f-2x+5)^2}
\leq\frac34,
\]
and a further differentiation gives
\[
\frac{\partial^2 H_f(x)}{\partial x^2}
=
\frac{12(4f-1)(4f^2-14f+1)}{(8fx-2f-2x+5)^3}<0,
\]
for $x,f\in(\frac12,1)$. Thus, for each fixed $f$, the map $H_f$ is increasing and concave in $x$.

Moreover, the one-step gain from purification is
\[
\iterpurifyfidelity{1}(f)-f
=
\frac{(1-f)(2f-1)(4f-1)}{8f^2-4f+5},
\]
which is positive for $f>\frac12$ but tends to $0$ as $f\to 1$. Hence, the marginal gain of purification is already intrinsically diminishing near high fidelities.

Since $H_f(f)>f$ and $H_f(1)<1$, the map has a fixed point $x_f^\star\in(f,1)$. The derivative bound makes this fixed point unique and gives $H_f(x)>x$ for $x\in[f,x_f^\star)$. Therefore, the sequence

\[
f,\;\iterpurifyfidelity{1}(f),\;\iterpurifyfidelity{2}(f),\;\ldots
\]
is monotone non-decreasing with decreasing marginal improvements, and converges to the unique fixed point $x_f^\star\in(f,1)$ of $H_f$, namely
\[
x_f^\star
=
\frac{6f-3+\sqrt{28f^2-26f+7}}{2(4f-1)}.
\]
Since $H_f(x_f^\star)=x_f^\star$ and $0<H_f'(x)\leq3/4$, by the mean value theorem, consecutive fidelity gaps satisfy
\begin{align*}
0\leq x_f^\star-\iterpurifyfidelity{i+1}(f)
&=H_f(x_f^\star)-H_f(\iterpurifyfidelity{i}(f))\\
&\leq\frac34\left(x_f^\star-\iterpurifyfidelity{i}(f)\right).
\end{align*}
Thus, each round leaves at most $3/4$ of the preceding gap. Applying this inequality $i$ times, starting from $\iterpurifyfidelity{0}(f)=f$, gives
\[
0\leq x_f^\star-\iterpurifyfidelity{i}(f)
\leq\left(\frac34\right)^i(x_f^\star-f).
\]
Thus, repeated EP pumping approaches $x_f^\star$ geometrically.

For example, if the input EP fidelity is $f=0.9$, then
\[
\iterpurifyfidelity{1}(0.9)\simeq0.9264,
\iterpurifyfidelity{2}(0.9)\simeq0.9369,
\iterpurifyfidelity{3}(0.9)\simeq0.9409,
\]
while the limiting value is
\[
x_{0.9}^\star\simeq0.9435.
\]
Hence, choosing $m=3$ leaves a fidelity gap below $0.0026$ from the limiting value, while $m=5$ reduces this gap below $3.7\times 10^{-4}$. For this input fidelity, these moderate bounds capture most of the available fidelity improvement. This motivates limiting the candidate purification sequences to moderate lengths, rather than exploring much longer sequences with little additional fidelity gain. The appropriate choice of $m$ depends on the input fidelity and the required fidelity threshold.

\end{document}